\documentclass[twocolumn]{aastex62}
\usepackage[utf8]{inputenc}
\usepackage{longtable}
\usepackage{graphicx}
\usepackage{amsmath,amssymb}
\usepackage{color, colortbl}
\usepackage{units}
\usepackage{epstopdf}
\usepackage{hyperref}
\usepackage{multirow}
\usepackage{url}
\usepackage{subfigure}
\usepackage{rotating}
\usepackage{enumitem}\setlist[description]{font=\textendash\enskip\scshape\bfseries}
\definecolor{amber}{rgb}{1.0, 0.75, 0.0}

\newcommand{\rp}{\emph{r}-process}
\newcommand{\mej}{\ensuremath{M_{\rm ej}}}
\newcommand{\mrp}{\ensuremath{M_{r\mathrm{p}}}}
\newcommand{\xmix}{\ensuremath{M_{r\mathrm{p}}}}
\newcommand{\mni}{\ensuremath{M_{\rm 56}}}
\newcommand{\vej}{\ensuremath{\beta_{\rm ej}}}
\newcommand{\iso}[2]{\ensuremath{^{#2}\text{#1}}}

\begin{document}
% \linenumbers
\title{Collapsars as Sites of \rp{} Nucleosynthesis: Systematic Photometric Near-Infrared Follow-up of Type Ic-BL Supernovae}
\correspondingauthor{Shreya Anand}
\email{sanand@caltech.edu}
\author[0000-0003-3768-7515]{Shreya Anand}
\affil{Cahill Center for Astrophysics, California Institute of Technology, Pasadena CA 91125, USA}
\author[0000-0003-3340-4784]{Jennifer Barnes}
\affil{Kavli Institute for Theoretical Physics, Kohn Hall, University of California, Santa Barbara, CA 93106, USA}
\author{Sheng Yang}
\affil{The Oskar Klein Centre, Department of Astronomy, Stockholm University, AlbaNova, SE-10691, Stockholm, Sweden}
\affil{Henan Academy of Sciences, Zhengzhou 450046, Henan, China}
\author[0000-0002-5619-4938]{Mansi M. Kasliwal}
\affil{Cahill Center for Astrophysics, California Institute of Technology, Pasadena CA 91125, USA}
\author[0000-0002-8262-2924]{Michael W. Coughlin}
\affil{School of Physics and Astronomy, University of Minnesota, Minneapolis, Minnesota 55455, USA}
\author
[0000-0003-1546-6615]{Jesper Sollerman}
\affil{The Oskar Klein Centre, Department of Astronomy, Stockholm University, AlbaNova, SE-10691, Stockholm, Sweden}
\author[0000-0002-0786-7307]{Kishalay De}
\affiliation{MIT-Kavli Institute for Astrophysics and Space Research, 77 Massachusetts Ave., Cambridge, MA 02139, USA}
\author[0000-0002-4223-103X]{Christoffer Fremling}
\affil{Cahill Center for Astrophysics, California Institute of Technology, Pasadena CA 91125, USA}
\author[0000-0001-8104-3536]{Alessandra Corsi}
\affil{Department of Physics and Astronomy, Texas Tech University, Box 1051, Lubbock, TX 79409-1051, USA}
\author[0000-0002-9017-3567]{Anna Y. Q.~Ho}
\affiliation{Department of Astronomy, Cornell University, Ithaca, NY 14850, USA}
\author[0000-0003-0477-7645]{Arvind Balasubramanian}
\affil{Department of Astronomy and Astrophysics, Tata Institute of Fundamental Research, Mumbai, 400005, India}
\author[0000-0002-9646-8710]{Conor Omand}
\affil{The Oskar Klein Centre, Department of Astronomy, Stockholm University, AlbaNova, SE-10691, Stockholm, Sweden}
\author{Gokul P. Srinivasaragavan}
\affil{Department of Astronomy, University of Maryland, College Park, MD 20742, USA}
\author[0000-0003-1673-970X]{S. Bradley Cenko}
\affiliation{Astrophysics Science Division, NASA Goddard Space Flight Center, MC 661, Greenbelt, MD 20771, USA}
\affiliation{Joint Space-Science Institute, University of Maryland, College Park, MD 20742, USA}
\author[0000-0002-2184-6430]{Tom{\'a}s Ahumada}
\affil{Cahill Center for Astrophysics, California Institute of Technology, Pasadena CA 91125, USA}
\author[0000-0002-8977-1498]{Igor Andreoni}
\affil{Department of Astronomy, University of Maryland, College Park, MD 20742, USA}
\affiliation{Astrophysics Science Division, NASA Goddard Space Flight Center, MC 661, Greenbelt, MD 20771, USA}
\affiliation{Joint Space-Science Institute, University of Maryland, College Park, MD 20742, USA}
\author{Aishwarya Dahiwale}
\affil{Cahill Center for Astrophysics, California Institute of Technology, Pasadena CA 91125, USA}
\author[0000-0001-8372-997X]{Kaustav Kashyap Das}
\affil{Cahill Center for Astrophysics, California Institute of Technology, Pasadena CA 91125, USA}
\author[0000-0001-5754-4007]{Jacob Jencson}
\affil{Department of Physics and Astronomy, The Johns Hopkins University, Baltimore, MD 21218, USA}
\author[0000-0003-2758-159X]{Viraj Karambelkar}
\affil{Cahill Center for Astrophysics, California Institute of Technology, Pasadena CA 91125, USA}
\author[0000-0003-0871-4641]{Harsh Kumar}	
\affil{Center for Astrophysics, Harvard University, 60 Garden St. Cambridge 02158 MA, USA} 
\affil{LSSTC Data Science Fellow 2018}
\author[0000-0002-4670-7509]{Brian D. Metzger}
\affil{Department of Physics and Columbia Astrophysics Laboratory, Columbia University, Pupin Hall, New York, NY 10027, USA}
\affil{Center for Computational Astrophysics, Flatiron Institute, 162 5th Ave, New York, NY 10010, USA} 
\author[0000-0001-8472-1996]{Daniel Perley}	
\affiliation{Astrophysics Research Institute, Liverpool John Moores University, IC2, Liverpool Science Park, 146 Brownlow Hill, Liverpool L3 5RF, UK}
\author[0000-0003-2700-1030]{Nikhil Sarin}
\affil{Nordita, Stockholm University and KTH Royal Institute of Technology Hannes Alfv\'ens v\"ag 12, SE-106 91 Stockholm, Sweden}
\affil{The Oskar Klein Centre, Department of Physics, Stockholm University, AlbaNova, SE-106 91 Stockholm, Sweden}
\author{Tassilo Schweyer}
\affil{The Oskar Klein Centre, Department of Astronomy, Stockholm University, AlbaNova, SE-10691, Stockholm, Sweden}
\author[0000-0001-6797-1889]{Steve Schulze}
\affil{Center for Interdisciplinary Exploration and Research in Astrophysics and Department of Physics and Astronomy, Northwestern University, 2145 Sheridan Road, Evanston, 60208, IL, USA}
\author[0000-0003-4531-1745]{Yashvi Sharma}
\affil{Cahill Center for Astrophysics, California Institute of Technology, Pasadena CA 91125, USA}
\author{Tawny Sit}
\affil{Department of Astronomy, The Ohio State University, Columbus, OH 43210, USA}
\author[0000-0003-2434-0387]{Robert Stein}
\affil{Cahill Center for Astrophysics, California Institute of Technology, Pasadena CA 91125, USA}
\author[0000-0003-3433-1492]{Leonardo Tartaglia}
\affil{Istituto Nazionale di Astrofisica, Osservatorio Astronomico d'Abruzzo, Via Mentore Maggini s.n.c. 64100 Teramo, Italy}
\author[0000-0002-1481-4676]{Samaporn Tinyanont}
\affil{National Astronomical Research Institute of Thailand, 260 Moo 4, Donkaew, Maerim, Chiang Mai, 50180, Thailand}
\author[0000-0003-0484-3331]{Anastasios Tzanidakis}
\affil{Department of Astronomy, University of Washington, Seattle, WA 98195, USA}
\author[0000-0002-2626-2872]{Jan van Roestel}
\affil{Anton Pannekoek Institute for Astronomy, University of Amsterdam, 1090 GE Amsterdam, The Netherlands}
\author[0000-0001-6747-8509]{Yuhan Yao}
\affil{Department of Astronomy, University of California, Berkeley, CA 94720-3411, USA}
\author[0000-0002-7777-216X]{Joshua S. Bloom}
\affil{Department of Astronomy, University of California, Berkeley, CA 94720-3411, USA; Physics, Lawrence Berkeley National Laboratory, 1 Cyclotron Road, MS50B-4206, Berkeley, CA 94720, USA}
\author[0000-0002-6877-7655]{David O.~Cook}
\affiliation{IPAC, California Institute of Technology, 1200 E. California
             Blvd, Pasadena, CA 91125, USA}
\author[0000-0002-5884-7867]{Richard Dekany}
\affil{Caltech Optical Observatories, California Institute of Technology, Pasadena, CA  91125}
\author[0000-0002-3168-0139]{Matthew J. Graham}
\affil{Cahill Center for Astrophysics, California Institute of Technology, Pasadena CA 91125, USA}
\author[0000-0001-5668-3507]{Steven L. Groom}
\affiliation{IPAC, California Institute of Technology, 1200 E. California
             Blvd, Pasadena, CA 91125, USA}
\author[0000-0001-6295-2881]{David~L.~Kaplan}
\affiliation{Center for Gravitation, Cosmology, and Astrophysics,
  Department of Physics, University of Wisconsin-Milwaukee, PO Box
  413, Milwaukee, WI, 53201, USA}
\author[0000-0002-8532-9395]{Frank J. Masci}
\affiliation{IPAC, California Institute of Technology, 1200 E. California
             Blvd, Pasadena, CA 91125, USA}
\author[0000-0002-7226-0659] {Michael S. Medford}
\affiliation{Department of Astronomy, University of California, Berkeley, Berkeley, CA 94720}
\affiliation{Lawrence Berkeley National Laboratory, 1 Cyclotron Rd., Berkeley, CA 94720}
\author[0000-0002-0387-370X]{Reed Riddle}
\affil{Caltech Optical Observatories, California Institute of Technology, Pasadena, CA  91125}
\author[0000-0002-0331-6727]{Chaoran Zhang}
\affiliation{Center for Gravitation, Cosmology, and Astrophysics,
  Department of Physics, University of Wisconsin-Milwaukee, PO Box
  413, Milwaukee, WI, 53201, USA}

% \author{Friends}
% \date{October 2020}

\begin{abstract}
    One of the open questions following the discovery of GW170817 is whether neutron star mergers are the only astrophysical sites capable of producing \rp{} elements. Simulations have shown that 0.01-0.1M$_\odot$ of \rp{} material could be generated in the outflows originating from the accretion disk surrounding the rapidly rotating black hole that forms as a remnant to both neutron star mergers and collapsing massive stars associated with long-duration gamma-ray bursts (collapsars). The hallmark signature of \rp{} nucleosynthesis in the binary neutron star merger GW170817 was its long-lasting near-infrared emission, thus motivating a systematic photometric study of the light curves of broadlined stripped-envelope (Ic-BL) supernovae (SNe) associated with collapsars. We present the first systematic study of 25 SNe Ic-BL---including 18 observed with the Zwicky Transient Facility and 7 from the literature---in the optical/near-infrared bands to determine what quantity of \rp{} material, if any, is synthesized in these explosions. Using semi-analytic models designed to account for \rp{} production in SNe Ic-BL, we perform light curve fitting to derive constraints on the \rp{} mass for these SNe. We also perform independent light curve fits to models without \rp{}. We find that the \rp{}-free models are a better fit to the light curves of the objects in our sample. Thus we find no compelling evidence of \rp{} enrichment in any of our objects. Further high-cadence infrared photometric studies and nebular spectroscopic analysis would be sensitive to smaller quantities of \rp{} ejecta mass or indicate whether all collapsars are completely devoid of \rp{} nucleosynthesis.
    % consistent with \rp{} nucleosynthesis, while the remaining objects are consistent with \rp{} free models.
    %We find that about one third of our objects synthesize little to no \rp{} ejecta (m$_{\rm rp} <$\,0.02 M$_{\odot}$), another third show mild to moderate \rp{} signatures (0.02 $ \lesssim \rm m_{rp} <$ 0.05\,$M_{\odot}$) and the final third exhibit evidence for significant \rp{} production (M$_{r\mathrm{p}} \gtrsim $\,0.05 M$_{\odot}$). We find hints of correlation between the \rp{} ejecta mass, mixing fraction, total ejecta mass, and average kinetic velocity of our objects. 
\end{abstract}

\section{Introduction}
% One of the most fundamental open questions in modern astronomy concerns the origin of the heaviest elements in the universe. 
% Several studies \citep[e.g.][]{Cameron1959} have shown that measured abundances of metals lighter than Rubidium can be mostly attributed to big bang nucleosynthesis, cosmic ray fission, core-collapse and Type Ia supernovae (SNe), and dying low-mass stars. 
The dominant process responsible for producing elements heavier than iron is the rapid neutron capture process, known as the \rp{} \citep{Burbidge1957, Cameron1957}, which only has a few plausible astrophysical sites. While standard core-collapse supernovae (SNe) were previously considered as candidate sites for \rp{} nucleosynthesis \citep{Woosley1994, Takahashi1994, Qian1996}, they have since been disfavored because simulations of neutrino-driven winds in core-collapse SNe fail to create conducive conditions for \rp{} production \citep{Thompson2001, Roberts2010, MartinezPinedo2012, Hotokezaka2018}. On the other hand, before 2017, many studies \citep{LattimerSchramm1974, LattimerSchramm1976, SymbalistySchramm1982} predicted that mergers of two neutron stars (NSs) or neutron stars with black holes were capable of generating \rp{} elements during the decompression of cold, neutron-rich matter ensuing from the tidal disruption of the neutron stars. \citet{LiPaczynski1998} first suggested that the signature of such \rp{} nucleosynthesis would be detectable in an ultraviolet, optical and near-infrared (NIR) transient powered by the radioactive decay of neutron-rich nuclei, termed as a ``kilonova" for its brightness, which was predicted to be 1000$\times$ that of a classical nova \citep{Metzger+10b}. Other studies proposed that \rp{}~elements could be synthesized in a rare SN subtype known as a hypernova \citep[e.g.][]{Fujimoto2007}. In this scenario, the SN explosion produces a rapidly rotating central BH surrounded by an accretion disk. Accretion onto the BH is thought to power a relativistic jet, while material in the disk may neutronize, allowing the \rp{} to occur when the newly neutron-rich material is unbound as a disk wind.

Galactic archaeological studies \citep{Ji+16a, Ji+16b}, geochemical studies \citep{Wallner+21}, and studies of the early solar system \citep{Tissot+16} offer unique insights into which astrophysical sites could plausibly explain observed \rp{} elemental abundances. A recent study of \rp{}~abundances in the Magellanic Clouds indicate that the astrophysical \rp{} site has a time-delay longer than for core-collapse SNe \citep{Reggiani2021}. Second- and third-peak abundance patterns inferred from metal-poor Galactic halo stars show consistency with the solar \rp{} abundance pattern at high atomic number, but scatter at low atomic number that could be attributed to enrichment from multiple sources, including magneto-rotational hypernovae \citep{Yong2021}. Measurements of excess $[\rm Ba/Fe]$ and $[\rm Eu/Fe]$ abundances in the dwarf galaxy Reticulum II argue for not only a rare and prolific event, but one capable of enriching the galaxy early in its history \citep{Ji+16a, Tarumi2020}, pointing towards a potential rare SN subtype whose \rp{} production would follow star formation \citep{Cote+19, Siegel2019}. Further evidence of \emph{heavy} \rp{} enrichment in the disrupted dwarf galaxy Gaia Sausage Enceladus ($\sim$3.6 Gyr star formation duration) but not in the disrupted dwarf galaxy Kraken (with $\approx$2 Gyr star formation duration) points towards multiple \rp{}~enrichment sites operating on different timescales \citep{Naidu2021}.

Overall, geological studies and studies of the early solar system and Galactic chemical evolution exemplify the need for rare and prolific astrophysical sites to explain observed abundances, and insinuate that the solar \rp{} abundance pattern could be universal. While NS mergers are compatible with many facets of the above findings \citep{Hotokezaka2018,Cote2018, Metzger_2019.LRR_kilonovae}, assuming that mergers are the \textit{sole} producers of \rp{} material presents some potential hurdles. For example, the time delay between formation and merger of NS systems must be short enough to enrich old, ultra-faint dwarf galaxies with heavy elements \citep{Roederer+16, Ji+16a, Cote+19}. Furthermore, natal merger kicks present a challenge for low-mass galaxies to retain pre-merger compact binaries \citep{Yutaka&Toshikazu_16}. The question of whether NS mergers alone can explain the relative abundances of \rp{} elements (e.g. $[\rm Eu/Fe]$ vs. $[\rm Fe/H]$) in the solar neighborhood remains unanswered \citep{Beniamini+16, Bonetti+19}.

The multi-messenger detection of the binary neutron star merger GW170817 \citep{LVCGW170817}, an associated short burst of gamma-rays GRB170817 \citep{2017gfo_GRB} and the kilonova AT2017gfo \citep{2017gfoAUS, 2017gfoGeminiS, 2017gfoSwope, 2017gfoCowperthwaite, 2017gfoDrout, 2017gfoSwift, 2017gfoKasliwal, 2017gfoSpitzer, 2017gfoKilpatrick, 2017gfoMASTER, 2017gfoMcCully, 2017gfoNicholl, 2017gfoShappee, 2017gfoDECam, 2017gfoPian, 2017gfoSmartt, 2017gfolanthanides, Villar+17, 2017gfoJGEM} relayed the first direct evidence that NS mergers are an astrophysical site of \rp{} nucleosynthesis and short GRB progenitors. Multi-band photometry and optical/NIR spectroscopy of AT2017gfo indicated that the KN ejecta was enriched with \rp{} elements \citep{2017gfoDrout, 2017gfoChornock, 2017gfoCowperthwaite, 2017gfoKasliwal, 2017gfoKilpatrick, 2017gfoPian, 2017gfoSmartt, 2017gfolanthanides, 2017gfoWatson} including heavier species occupying the second- and third-peak \citep{2017gfolanthanides, 2017gfoWatson, 2017gfoSpitzer, Gillanders2021}.

% Since the discovery of GW170817, though no additional KNe have been spectroscopically confirmed, there have been a few photometric KN candidates reported in the literature \citep{Gompertz2018, Fong2021} including a recent one associated with a long-duration gamma-ray burst (GRB; \citet{Rastinejad2022}).
% \rednote{add some text referencing state-of-the-art modeling to predict yields of kilonovae!}
Although GW170817 confirmed NS mergers as \rp{} nucleosynthesis sites, some fundamental open questions on the nature of \rp{} production still remain. Namely, can the rates of and expected yields from NS mergers explain the \textit{total} amount of \rp{} production measured in the Universe? Or, do the direct and indirect clues about \rp{} production in the Universe point towards an alternative \rp{} site, such as rare core-collapse SNe?
%and 2) do the relative abundances of elements produced in NS mergers match the measured solar, galactic, and extra-galactic \rp{} abundance patterns?

% \subsection{Classes of long-duration GRB-collapsars}
The discovery of the broadlined Type Ic supernova SN\,1998bw at 40 Mpc \citep{Galama1998}, following the long GRB 980425 was a watershed event that provided the first hints that some GRBs were connected to stellar explosions \citep{Kulkarni1998,Galama1999}. However, due to the anomalous nature of the explosion, it was not until GRB 030329 that a direct long GRB--SN connection was securely established \citep{Fynbo2004}. The spectra of these SNe exhibit broad features due to  high photospheric velocities (${\gtrsim}20,000$ km s$^{-1}$).
They have higher inferred kinetic energies than typical SNe (at $\sim$10$^{52}$ erg), and are stripped of both hydrogen and helium \citep{Modjaz2016, GalYam2017}. Since SN\,1998bw, several other SNe Ic-BL have been discovered in conjunction with long GRBs (e.g. \citealt{Kocevski2007, Olivares2012, Cano2017_SN2016jca, CorsiLazzati2021}), boosting the existing collapsar theory \citep{Woosley1993, MacFaydenWoolsey1999, MacFayden2001} as a mechanism to explain long GRBs and their associated SN counterparts. The term collapsar refers to a rapidly-rotating, massive star that collapses into a black hole, forming an accretion disk around the central black hole. Collapsars are distinct from the magnetar-powered explosions (referred to as ``magnetohydrodynamic (MHD) SNe") also proposed to be related to SNe Ic-BL \citep{Metzger2011, Kashiyama2016}. However, puzzling discoveries including that of GRB060505 and GRB060614 which lacked a clear SN counterpart to deep limits \citep{Gehrels2006} and that of GRB211211A, a long-duration GRB associated with a kilonova \citep{Rastinejad2022} have shifted the paradigm from the traditional conception that all long GRBs have a collapsar or magnetar origin. Thus some fraction of long-duration GRBs may also originate from compact binary mergers. 

% However, recent discoveries of kilonovae associated with long GRBs \citep[e.g.][]{Rastinejad2022} have shifted the paradigm from the traditional conception that all long GRBs have a collapsar or magnetar origin. Thus some fraction of long-duration GRBs may also originate from compact binary mergers. 
% A rare subclass of explosions referred to as collapsars have been proposed as an alternative dominant site of \rp{} nucleosynthesis aside from NS mergers \citep{Fujimoto2007}. 
 
% It is therefore also associated with the broadlined Type Ic (Ic-BL) SNe observed in conjunction with long GRBs.
% The collapsar scenario was originally constructed to explain
% long gamma-ray bursts.

% However, despite several thousands of classical long GRB detections, only a small fraction of them have SN Ic-BL counterparts, due to the challenge of detecting these SNe at high redshifts \citep{Cano2017, WoosleyBloom2006}. Likewise, the majority of SNe Ic-BL, typically detected in wide-field, time-domain surveys, have no known prompt associated GRB trigger \citep{Berger2003_Ibc_radio, Soderberg2006, Corsi2016, Ho2020_ZTF20aajnksq}. Low-luminosity GRBs (LLGRBs; \citealt{Liang2007}) lie in-between, as mildly relativistic explosions with isotropic peak luminosities of $10^{46}-10^{48}$ erg s$^{-1}$ and energies 2-3 orders of magnitude lower than typical long GRBs, and are often associated with radio emission and X-ray flashes \citep{Toma2007, Virgili2009, Zhang2012, Nakar2015, Cano2017}. 

Several works \citep{Fujimoto2007, Ono2012, Nakamura2015, Soker2017} have since hypothesized that the explosions that give rise to SNe Ic-BL and (in some cases) to their accompanying long GRBs (i.e. collapsars) are capable of producing 0.01-0.1M$_{\odot}$ of \rp{} material per event. Simulations suggest that in the case of a NS merger, an accretion disk forms surrounding the merger's newly-born central black hole \citep{ShibataTaniguchi2006} and \rp{} elements originate in the associated disk outflows \citep{MetzgerPiroQuataert2008, MetzgerPiroQuataert2009}. Such accretion flows are not only central to the short GRBs associated with NS mergers, but also with the long classes of GRBs associated with collapsars. However, predictions about \rp{}-production in the collapsar context are sensitive to assumptions about the magnetic field, the disk viscosity model, and the treatment of neutrinos, among other factors. \citet{Surman2006} argued that only light \rp{} elements can be synthesized in collapsar accretion disks due to neutrino-driven winds. More recently, \citet{Siegel2019} conducted 3D general-relativistic, magnetohydrodynamic simulations demonstrating sufficient \rp{} yields to explain the observed abundances in the Universe. \citet{Siegel2019} found that the disk material becomes neutron-rich through weak interactions, enabling the production of even 2nd and 3rd peak \rp{} elements in disk-wind outflows. Other works in the literature \citep{Miller2020, Just2022, Fujibayashi2022} have argued that collapsars are inefficient producers of \rp{} elements based on studies of the full radiation transport and $\alpha$-viscosity in collapsar disks. Whether or not collapsars are sites of \rp{} nucleosynthesis is still an active area of investigation, motivating detailed studies of the photometric evolution of \rp{} enriched SNe. 

Recently, \citet{Barnes2022}, motivated by \citet{Siegel2019}, created semi-analytic models of the light curves of SNe from collapsars producing \rp{} elements, yielding concrete predictions for the photometric evolution of \rp-enriched SNe Ic-BL. Our work is focused on observationally testing the models from \citet{Barnes2022}.

% More recently, \citet{Siegel2019} conducted 3D general-relativistic, magnetohydrodynamic simulations demonstrating sufficient \rp{} yield to explain the observed abundances in the Universe. Past works (e.g. \citealt{Surman2006}) have argued that only light \rp{} elements can be synthesized in collapsar accretion disks due to neutrino-driven winds, or have argued against \rp{} production in collapsar disks \citep{Miller2020, Just2022}, \citet{Siegel2019} found that the disk material becomes neutron-rich through weak interactions, enabling the production of even 2nd and 3rd peak \rp{} elements. Recently, \citet{Barnes2022}, building on the work of \citet{Siegel2019}, created semi-analytic models of collapsars producing \rp{} elements, yielding concrete photometric predictions for the light curves of SNe Ic-BL. However, thus far, direct observational evidence in the literature has been too limited to verify whether SNe Ic-BL are \rp{} sites.

In this work, we report our findings from an extensive observational campaign and compilations from the literature to determine whether collapsars powering SNe Ic-BL are capable of synthesizing \rp{} elements.
%powering mildly relativistic explosions (LLGRBs, radio-bright SNe Ic-BL) and explosions with no detected relativistic ejecta (standard SNe Ic-BL) are capable of synthesizing \rp{} elements. 
We present optical and near-infrared photometric observations and compare both color evolution and absolute light curves against the predictions from \citet{Barnes2022}. Our paper is structured as follows:  First, we detail our sample selection criteria in Sec.~\ref{sec:sample}, then Sec.~\ref{sec:observations} describes our optical and NIR observations, followed by Sec.~\ref{sec:candidates} which provides the discovery details about each candidate. Sec.~\ref{sec:literature} introduces the objects from the literature used in our study, and in Sec.~\ref{sec:models} we introduce the latest collapsar \rp{} models. In Sec.~\ref{sec:analysis}, we show how we derive explosion properties. The results of our light curve model fits are presented in Sec.~\ref{sec:results}, and finally we discuss our conclusions and future work in Sec.~\ref{sec:discussion}.

\section{Sample Selection} \label{sec:sample}
\begin{figure*}
    \centering
    \includegraphics[width=0.50\textwidth]{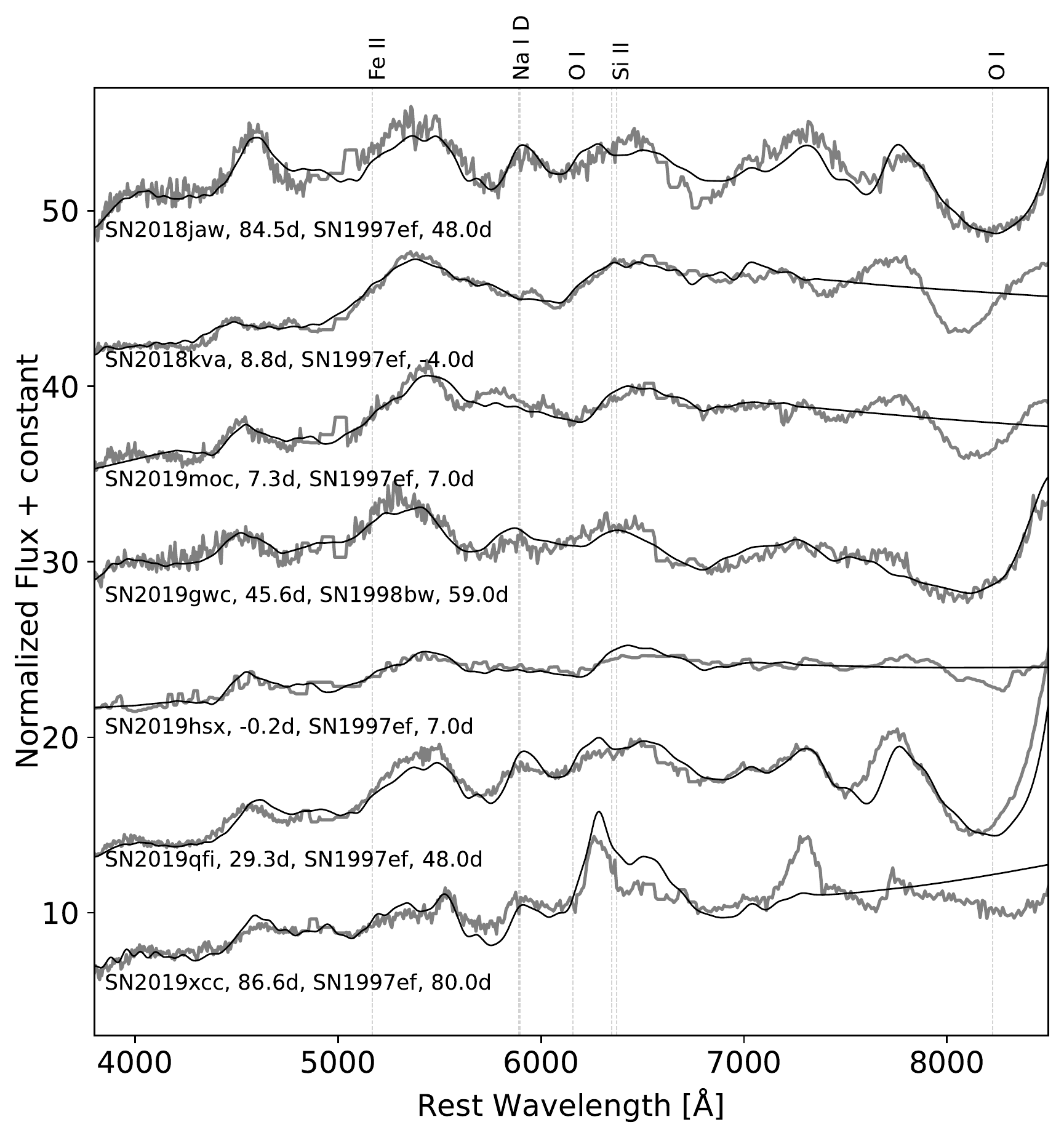}\includegraphics[width=0.50\textwidth]{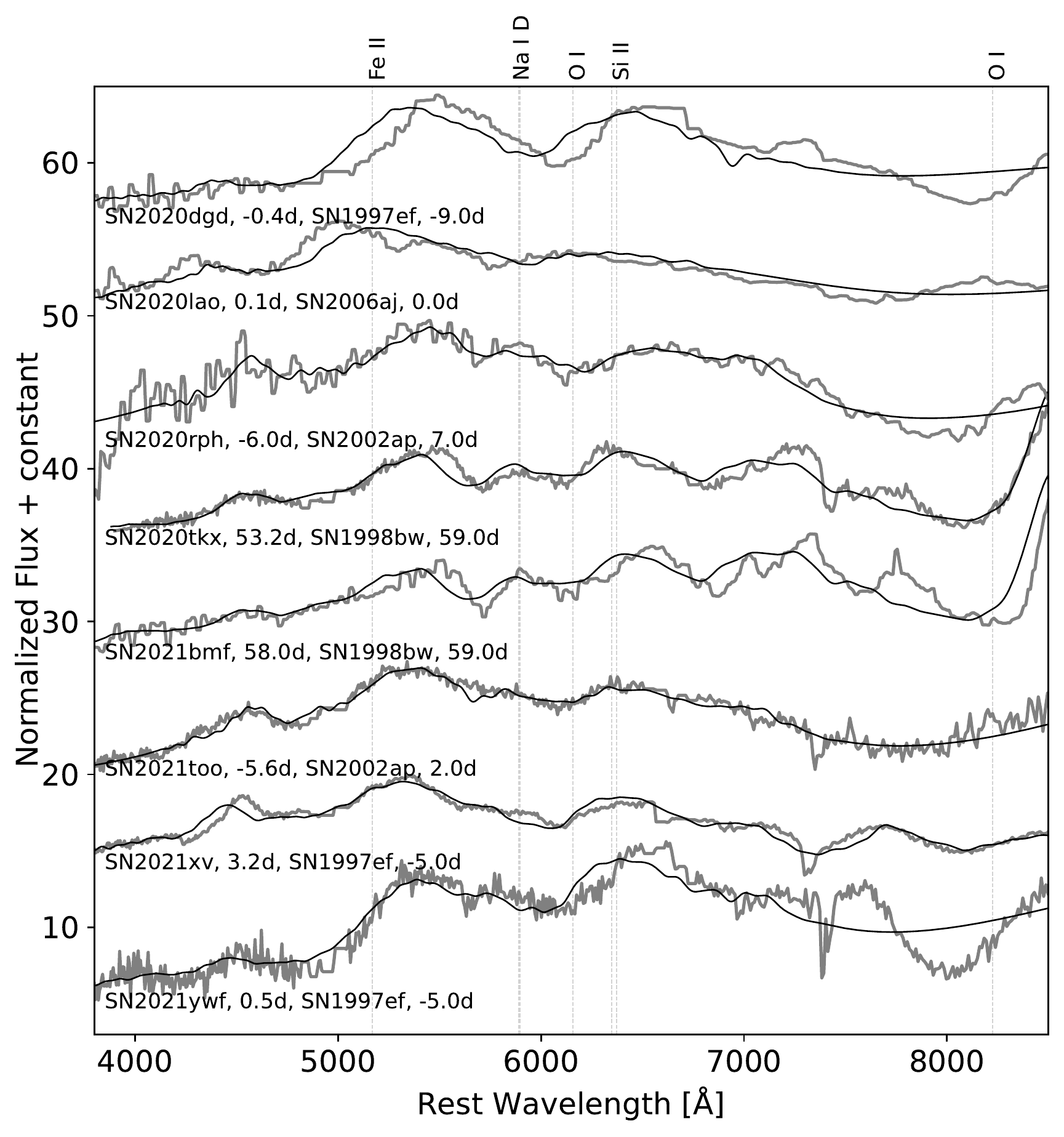}
    \caption{Classification spectra for the SNe Ic-BL in our sample, along with their SNID best-match templates, labeled by name, supernova phase relative to the peak light, and corresponding template name, and template phase from SNID. GRB190829A only has a host spectrum, which we do not display here. The spectra for SN\,2018gep and SN\,2020bvc are published in \citet{Ho2019_SN2018gep} and \citet{Ho2020_ZTF20aalxlis} so we do not show them here. The spectra show broad Fe II, Si II and O I lines. The Na I D absorption line, an indicator of host extinction \citep{Stritzinger2018_host}, is plotted for reference -- none of the SNe appear to have strong Na I D features.}
    \label{fig:spectra}
\end{figure*}

To test the hypothesis that SNe Ic-BL generate \rp{} elements, we require a statistically robust sample size of SNe with contemporaneous NIR and optical light curves. To obtain optical light curves, we use data from the Zwicky Transient Facility (ZTF; \citealt{Bellm2019, Masci2019, Graham2019}), a 47~sq.~deg. field-of-view mosaic camera with a pixel scale of 1$\arcsec$/pixel \citep{Dekany2020} installed on the Palomar 48~in. telescope. ZTF images the entire Northern sky every $\sim$2 nights in g- and $r-$bands, attaining a median 5$\sigma$ detection depth of 20.5\,m$_{\mathrm{AB}}$. Amongst the systematic efforts aimed at SN detection with ZTF, our SNe draw from two surveys in particular: ``Bright Transient Survey'' (BTS; \citealt{Fremling2020}) and the ZTF ``Census of the Local Universe'' survey (CLU; \citealt{CLU_De2020}) which are conducted as a part of ZTF's nightly operations. BTS is a magnitude-limited survey aimed at spectroscopically classifying all SNe $<18.5$\,mag at peak brightness \citep{Perley2020}. CLU, in contrast, is a volume-limited survey aimed at classifying all SNe within 150\,Mpc whose hosts belong to the CLU galaxy catalog \citep{CLU_Cook2019}. The CLU galaxy catalog is designed to provide spectroscopic redshifts of all galaxies within 200\,Mpc, and is 90\% complete (for an H$\alpha$ line flux of 4$\times 10^{-14}$ erg cm$^2$ s$^{-1}$). Hence the two surveys provide complementary methods for SN identification.     
Our sample consists of 18 spectroscopically-confirmed ZTF SNe Ic-BL within $z\lesssim0.05$. Due to our low redshift cut, we assume that the photometric K-corrections are negligible \citep{Taddia2018}. The details of the instruments and configurations used to take our classification spectra are described in Sec.~\ref{sec:observations} (see also Fig.~\ref{fig:spectra}). Where available, we use the spectroscopic redshift from the SDSS galaxy host (especially for sources falling in the CLU sample) and otherwise determine the SN redshift from spectral fitting to the narrow galaxy H$\alpha$ feature. For each spectrum, we use the Supernova Identification code (SNID; \citealt{Blondin2007}) to determine the best match template (also plotted in Fig.~\ref{fig:spectra}), fixing the redshift to the value determined using the methods described above.  We overplot the characteristic spectroscopic lines for SNe Ic-BL including O I, Fe II, and Si II in dashed lines, along with Na I D, an indicator of the amount of supernova host galaxy extinction \citep{Stritzinger2018_host}. For all of the ZTF SNe, we assume zero host attenuation; this assumption is backed by the lack of any prominent Na I D absorption features in the spectra (see Fig.~\ref{fig:spectra}). Higher host attenuation results in redder observed SN colors.

We impose a redshift cut to eliminate distant SNe that might fade rapidly below ZTF detection limits within 60\,days post-peak. ZTF yields an average rate of SNe Ic-BL discovery of $\sim$1/month, but due to visibility and weather losses, we followed-up $\sim$10 SNe per year. As a consequence of our sample selection from ZTF, probing only the local volume, we are biased against GRB-SNe. However, amongst our sample, we include one LLGRB (GRB190829A), SN\,2018gep, a published SN with fast and luminous emission \citep{Ho2019_SN2018gep}, and another published SN with a mildly-relativistic ejecta, SN\,2020bvc \citep{Ho2020_ZTF20aalxlis}, which contribute diversity to our ZTF sample. The two SNe exhibited broad features in their spectra and were classified as SNe Ic-BL, while the LLGRB was too faint for spectroscopy, and only had photometric evidence of an associated SN bump.

In the analyses in subsequent sections, we assume the following cosmological parameters: H$_{0}=63.7 \rm km\,s^{-1}\,Mpc^{-1}$, $\Omega_m=0.307$.

% We define different sample sets according to the following photometric criteria: 1) $\geq$2 epochs in a \textit{single} NIR filter (J, H, Ks or K\') 2) optical light curves lasting $\gtrsim$60\,days. The SNe satisfying these criteria are considered part of the ``gold'' sample. The remaining Ic-BL SNe for which obtained NIR follow-up are part of the ``silver'' sample. In the appendix, we show Type Ic (not broadlined) SNe for which we obtained follow-up but do not include in our main two sample sets.

\section{Observations} \label{sec:observations}

Here we describe the photometric and spectroscopic observations obtained by various facilities in our follow-up campaign.

\subsection{Photometry} \label{sec:optical}
\subsubsection{ZTF}
We use the ZTF camera on the Palomar 48-in telescope for supernova discovery and initial follow-up. ZTF's default observing mode consists of 30\,s exposures. Alerts (5$\sigma$ changes in brightness relative to the reference image) are disseminated in avro format \citep{Patterson2019} and filtered based on machine-learning real-bogus classifiers \citep{Mahabal2019}, star-galaxy classifiers \citep{Tachibana2018}, and light curve properties. Cross-matches with solar-system objects serve to reject asteroids. ZTF's survey observations automatically obtain $r-$, $g-$ and sometimes $i-$ band imaging lasting $\approx$60 days after peak, while the supernova is brighter than 20.5\,mag. \citet{Masci2019} provides more information about the data processing and image subtraction pipelines. More details about specific surveys used to obtain these data are provided in Sec~\ref{sec:sample}.

\subsubsection{LCOGT}
We performed photometric follow-up of our SNe with the Sinistro and Spectral cameras on the Las Cumbres Observatory Global Telescope (LCOGT; \citealt{LCOGT2013}) Network's 1-m and 2-m telescopes respectively. The Sinistro (Spectral) camera has a field of view of 26.5 (10.5)$\arcmin$ x 26.5 (10.5)$\arcmin$ and a pixel scale of 0.389 (0.304)$\arcsec$/pixel. The observations relied on two separate LCO programs: one aimed at supplementing ZTF light curves of Bright Transient Survey objects and the other intending to acquire late-time $r-$ and $i-$ band follow-up of stripped-envelope SNe fainter than 21\,mag. The exposure times and number of images requested varied based on filter and desired depth, ranging from 160\,s to 300\,s and from 1 to 5 images. The data are automatically flat-fielded and bias-subtracted. Though both programs use different data reduction pipelines, the methodology is nearly the same. Both pipelines extract sources using the Source Extractor package \citep{BertinArnouts2010} and calibrate magnitudes against Pan-STARRS1 (PS1) \citep{Chambers2016, Flewelling2018} objects in the vicinity. The BTS-targeted program uses the High Order Transform of Psf ANd Template Subtraction code (HOTPANTS; \citealt{Becker2015}) to subtract a PSF scaled Pan-STARRS1 template previously aligned using SCAMP \citep{Bertin2006}.  For the late-time LCOGT follow-up program, our pipeline performed image subtraction with pyzogy \citep{GuevelHosseinzadeh2017}, based on the ZOGY algorithm \citep{Zackay+16}. Both pipelines stack multiple images to to increase depth. 

\subsubsection{WASP}
We performed deep imaging with the WAfer-scale imager for Prime (WASP), mounted on the Palomar 200-in. prime focus with a 18.5$\arcmin$ x 18.5$\arcmin$ field of view and a plate scale of 0.18$\arcsec$/pixel. We obtained data from WASP for the transients at late times in the $g'-$, $r'-$ and $i'-$ filters. The data were reduced using a python based pipeline that applied standard optical reduction techniques (as described in \citealt{De2020_Gattini}), and the photometric calibration was obtained against PS1 sources in the field. Image subtraction was performed with HOTPANTS with references from PS1 and SDSS.

\subsubsection{SEDM}
We obtained additional photometric follow-up with the Spectral Energy Distribution Machine (SEDM; \citealt{Blagorodnova2018, Rigault2019, Kim2022}) on the Palomar 60-inch (P60) telescope which has a field of view of 13$\arcmin$ x 13$\arcmin$ and a plate scale of 0.378$\arcsec$/pixel. The processing is automated, and can be triggered using the Fritz marshal \citep{Kasliwal2019_GROWTH, duev2019real, skyportal2019}. Standard imaging requests involve $g-$, $r-$, and $i-$ band 300s exposures with the Rainbow Camera on SEDM. The data are later reduced using a python-based pipeline that applies standard reduction techniques and applies a customized version of FPipe (Fremling Automated Pipeline; \citealt{Fremling2016}) for image subtraction.

\subsubsection{Liverpool IO:O}
We acquired late-time, multi-band imaging with the Liverpool Telescope \citep{Steele2004} using the IO:O camera with the Sloan $griz$ filter set. The IO:O camera has a 10$\arcmin$x10$\arcmin$ field of view with a plate scale of 0.15$\arcsec$/pixel. An automatic pipeline reduces the images, performing bias subtraction, trimming of the overscan regions, and flat fielding. Once a PS1 template is aligned, the image subtraction takes place, and the final photometry comes from the analysis of the subtracted image.

\subsubsection{GROWTH-India Telescope}
We obtained photometric follow-up of our SNe with the 0.7m robotic GROWTH-India Telescope (GIT; \citealt{Kumar2022}) equipped with a 4096×4108 pixel back-illuminated Andor camera. GIT has a circular field of view of 0.86$\deg$ x 0.86 $\deg$ (corresponding to 51.6$\arcmin$ x 51.6$\arcmin$) and has a pixel scale of 0.676$\arcsec$/pixel. GIT is located at the IAO (Hanle, Ladakh).  Targeted observations were conducted in SDSS $r', \text{and}\,i'$ filters with varying exposure times. All data were downloaded in real time and processed with the automated GIT pipeline. Zero points for photometry were calculated using the PanSTARRS catalogue \citep{Flewelling2018}, downloaded from Vizier \citep{Vizier2000}. We performed image subtraction with pyzogy and PSF photometry with PSFEx \citep{Bertin2011}. 

% \subsubsection{LRIS}
% \rednote{check if LRIS imaging data was used.}
% We obtained imaging using the Low Resolution Imaging Spectrometer (LRIS; Oke et al. 1995) mounted at the Keck I telescope. Our data was taken in the g- and i-bands reaching mAB ≈ 24. The data was reduced following standard methods.

\subsubsection{WIRC}
We obtained near-infrared follow-up imaging of candidates with the Wide-field Infrared Camera (WIRC; \citealt{Wilson2003}), on the Palomar 200-inch telescope (P200) in $J-$, $H-$ and $K$-short ($Ks-$) bands. WIRC's field of view is 8.7$\arcmin$ x 8.7$\arcmin$ with a pixel scale of 0.2487$\arcsec$/pixel. The WIRC data was reduced using the same pipeline as described above for WASP, but it was additionally stacked using Swarp \citep{Bertin2010} while the calibration was done using the 2MASS point source catalog \citep{Skrutskie+06}. We obtained the WIRC data during classical observing runs on a $\sim$monthly cadence between January 2019 and December 2021. Due to the fact that the 2MASS Catalog is far shallower ($J=15.8, H=15.1, Ks=14.3$ m$_{\rm AB}$; \citealt{Skrutskie+06}) compared to WIRC's limiting magnitudes ($J=22.6, H=22.0, Ks=21.5$, in AB mag), we obtained reference images with WIRC after the SNe had faded in order to perform reference image subtraction. We perform image subtraction using the HOTPANTS algorithm and obtain aperture photometry using \textsc{photutils} \citep{photutils}. 

\subsection{Spectroscopy}
\subsubsection{SEDM}
We also used the SEDM's low-dispersion (R$\sim$100) integral field spectrograph (IFU) to obtain classification spectra for several of our objects. The field of view is 28$\arcsec$ x 28$\arcsec$ with a pixel scale of 0.125$\arcsec$/pixel. The SEDM is fully roboticized from the request submission to data acquisition to image reduction and uploading. The IFU images are reduced using the custom SEDM IFU data reduction pipeline \citep{Blagorodnova2018, Rigault2019}, which relies on the steps flat-fielding, wavelength calibration, extraction, flux calibration, and telluric correction.

\subsubsection{DBSP}
We obtained low to medium resolution (R$\sim$1000-10000) classification spectra of many of the SNe in our sample with Double Spectrograph (DBSP; \citealt{Oke1982}) on the Palomar 200-in telescope. Its plate scale is 0.293$\arcsec$/pixel (red side) and 0.389 $\arcsec$/pixel (blue side) and field of view is 120$\arcsec$ x 70$\arcsec$. The setup included a red grating of 316/7500, a blue grating of 600/400, a D55 dichroic, and slitmasks of 1$\arcsec$, 1.5$\arcsec$, and 2$\arcsec$. Some of our data was reduced using a custom PyRAF DBSP reduction pipeline \citep{Bellm2016} while the rest were reduced using a custom DBSP Data Reduction pipeline relying on Pypeit \citep{Prochaska2019, Roberson2021}.

\subsubsection{LRIS}
Some of the SNe in our sample also have spectra from the Low Resolution Imaging Spectrometer (LRIS; \citealt{Oke1995}) mounted on the 10\,m Keck I telescope. LRIS has a 6$\arcmin$ x 7.8$\arcmin$ field of view and a pixel scale of 0.135$\arcsec$/pixel. We used the 400/3400 grism on the blue arm and the 400/8500 grating on the red arm, with a central wavelength of 7830 $\rm \AA$ to cover the bandpass from 3,200-10,000 $\rm \AA$. We used longslit masks of 1.0$\arcsec$ and 1.5$\arcsec$ width. We typically used an exposure time of 600\,s to obtain our classification spectra. The spectra were reduced using LPipe \citep{Perley2019}.

\subsubsection{NOT}
We obtained low-resolution spectra with the Alhambra Faint Object Spectrograph and Camera (ALFOSC)\footnote{\href{http://www.not.iac.es/instruments/alfosc}{{http://www.not.iac.es/instruments/alfosc}}} on the 2.56\,m Nordic Optical Telescope (NOT) at the Observatorio del Roque de los Muchachos on La Palma (Spain). The ALFOSC has a field of view of 6.4$\arcmin$ x 6.4$\arcmin$ and a pixel scale of 0.2138$\arcsec$/pixel. The spectra were obtained with a 1\farcs0 wide slit and grism \#4. The data were reduced with IRAF and PypeIt. The spectra were calibrated with spectrophotometric standard stars observed during the same night and the same instrument setup.

\section{Description of ZTF candidates} \label{sec:candidates}

In the section below we include descriptions of all of the 18 candidates with ZTF data that were analyzed in this paper, including details about its discovery, coincident radio and X-ray data and any other notable characteristics about the objects. Our literature sample is described in Sec.~\ref{sec:literature}. Some of these candidates are part of a companion study \citep{Corsi2022} focusing on radio properties of SNe Ic-BL; the full ZTF sample of SNe Ic-BL will be presented in Srinivasaragavan et al., in prep. For all Swift XRT fluxes reported from the companion study, we assume a spectral model of a power-law spectrum with photon index $\Gamma=2$ corrected for Galactic absorption only. The 90\% flux upper limits for Swift XRT reported below are calculated by converting counts to flux using the same power-law model. All Swift fluxes have an energy range of 0.3-10 keV. For a more thorough discussion of whether the reported X-ray and radio emission correspond to transient or host-only emission, see Sections 3.4 and 3.5 of \citet{Corsi2022}.

The objects described here range from M$_{r}=-16.58$ to M$_{r}=-20.60$\,mag and from $z=0.017$ to $z=0.056$ (excluding the LLGRB, at $z=0.077$). All of the transients included below are ZTF SNe, but we hereafter refer to them by their IAU names. We performed forced photometry (using the MCMC method) for all of the candidates using ForcePhotZTF\footnote{https://github.com/yaoyuhan/ForcePhotZTF} \citep{Yao2019}.

We found no coincident \textit{Fermi}, \textit{Swift}, \textit{MAXI}, \textit{AGILE}, or \textit{INTEGRAL} GRB triggers or serendipitous \textit{Chandra} or \textit{XMM} X-ray coverage for these SNe based on their derived explosion dates. Though several candidate counterparts were found in temporal coincidence with \textit{KONUS} instrument on the Wind satellite, the explosion epoch uncertainties hinder our ability to make any firm association to the \textit{KONUS} sources. These objects are summarized in Tables~\ref{tab:rprocess_radio}, and their classification spectra are shown in Figure~\ref{fig:spectra}.

\subsection{SN\,2021ywf}
Our first ZTF photometry of SN\,2021ywf (ZTF21acbnfos)
was obtained on 2021 September 12 ($\mathrm{MJD}=59469.47$) with the P48.
This first detection was in the $r-$ band, with a host-subtracted magnitude of $20.03\pm0.20$, at $\alpha=05^{h}14^{m}11.00^{s}$, $\delta=+01\degr52\arcmin52.3\arcsec$ (J2000.0).
The discovery was reported to TNS on 2021 September 14
\citep{2021ywfTNSPhot}, 
with a note saying that the latest non-detection from ZTF was just 1 day prior to discovery ($r = 20.2$\,mag). 
The high cadence around discovery allows for a well constrained explosion date.
 With power-law fits to the early $g-$ and $r-$ band data, we estimate the explosion date as 
$\mathrm{MJD_{explosion}^{SN2021ywf}} = 59467.70 \pm 0.2$ (see below). 
% fitting plots in mcmc directory. ZTF21acbnfos*png
%2459486.20-18.00 (-18.57, -17.77)
%err = 0.57-0.23)/2

We classified the transient as a Type Ic-BL using a spectrum from P200+DBSP obtained on 2021 September 27 \citep{2021ywfTNSSpec}. The first spectrum was actually obtained using the P60+SEDM. However, the quality of that spectrum was not good enough to warrant a classification. SN\,2021ywf exploded in the outskirts of the spiral
galaxy CGCG 395-022 with a well established redshift of $z=0.028249$, which corresponds to a luminosity distance of 127.85\,Mpc and a distance modulus of 35.534.
This redshift is confirmed with narrow host lines in our classification spectrum.

On 2021 September 30, SN\,2021ywf was detected (3.2$\sigma$ significance) both with the \textit{Swift} XRT with $5.3_{-3.3}^{+4.9} \times 10^{-14}\rm\,erg\,cm^{-2}\,s^{-1}$ in a 7.2 ks observation, and with the VLA at $83\pm10\,\mu$Jy at 5.0 GHz (see \citealt{Corsi2022} for details).
% there is an earlier atel spec
% https://www.astronomerstelegram.org/?read=14925
% AT 2021ywf | 05:14:10.988 | +01:52:52.30 | 2021 Sep 16 | 0.0281   | SN Ibc | 

\subsection{SN\,2021xv}
Our first ZTF photometry of SN\,2021xv (ZTF21aadatfg)
was obtained on 2021 January 10 ($\mathrm{MJD}=59224.52$) with the P48. The transient was discovered in the public ZTF alert stream and reported by ALeRCE \citep{ALeRCE}.
This first detection was in the $r-$ band, with a host-subtracted magnitude of 19.93, at $\alpha=16^{h}07^{m}32.82^{s}$, $\delta=+36\degr46\arcmin46.07\arcsec$ (J2000.0).
The discovery was reported to TNS \citep{ZTF21aadatfgTNSPhot}, 
with a note saying that the last non-detection was 3\,days before discovery (on 2021 January 07 at $r=$19.52\,mag). 
We classified the transient as a Type Ic-BL using a spectrum from the NOT+ALFOSC obtained on 2021 Jan 25 \citep{ZTF21aadatfgTNSSpec}.
The transient appears to be associated with the galaxy host SDSS J160732.83+364646.1. We measure a redshift of $z=0.041$ from the narrow host lines in the NOT spectrum, corresponding to a luminosity distance of 187.29 Mpc and a distance modulus of 36.363. SN\,2021xv was marginally detected with the VLA on 2021 May 19 at F$_{\nu}=34.3\pm8.1\, \mu$Jy at 5.2 GHz, but the detection is consistent with host galaxy emission (see \citealt{Corsi2022} for details).

\subsection{SN\,2021too}
SN\,2021too (ZTF21abmjgwf) was reported first by the PS1 Young Supernova Experiment
 on 2021 July 17 ($\mathrm{MJD}=59412.60$) with the internal name PS21iap, but the first ZTF alerts are from 2021 July 16.
This first detection was in the $i-$ band, with a host-subtracted magnitude of 19.5, at $\alpha=21^{h}40^{m}54.28^{s}$, $\delta=+10\degr19\arcmin30.3\arcsec$ (J2000.0).
The discovery was reported to TNS \citep{ZTF21abmjgwfTNSPhot}. Our last non-detection with ZTF was on 2021 July 16 at $r=20.4$\,mag. 
The transient was classified as a Type Ic-BL using a spectrum from EFOSC2-NTT obtained on 2021 August 02 by ePESSTO \citep{ZTF21abmjgwfTNSSpec}. The object was positioned in the starforming galaxy SDSS J214054.29+101930.5. We measure a redshift of 0.035 from the narrow host lines in its P200+DBSP spectrum taken on 2021 Aug 07. This corresponds to a luminosity distance of 159.19 Mpc and a distance modulus of 36.01.

\subsection{SN\,2021bmf}
SN\,2021bmf (ZTF21aagtpro) was discovered by ATLAS on 2021 January 30 ($\mathrm{MJD}=59244.0$) with the internal name ATLAS\,21djt, and later by ZTF ($\mathrm{MJD}=59248.0$).
This first detection was in the $o$ band, with a host-subtracted magnitude of 18.12, at $\alpha=16^{h}33^{m}29.41^{s}$, $\delta=-06\degr22\arcmin49.53\arcsec$ (J2000.0).
The discovery was reported to TNS \citep{ZTF21aagtproTNSPhot}, 
with a note saying that the last non-detection was on 2021 January 16 at $c=18.4$\,mag. 
The transient was classified as a Type Ic-BL using a spectrum from ePESSTO obtained on 2021 February 03 \citep{ZTF21aagtproTNSSpec}. SN\,2021bmf was found in the faint host galaxy SDSS J163329.48-062249.9, which was determined to be at $z=0.0175$ based on narrow host lines in the Keck I LRIS spectrum taken on 2021 July 09, which corresponds to a luminosity distance of 78.57 Mpc and a distance modulus of 34.476.

% \subsubsection{ZTF20adadrhw}
% Our first ZTF photometry of ZTF20adadrhw (SN\,2020adow)
% was obtained on 2020 December 27 ($\mathrm{JD}=2459210.7560$) with the P48.
% This first detection was in the $g$ band, with a host-subtracted magnitude of $15.84\pm0.03$, at $\alpha=08^{h}33^{m}42.26^{s}$, $\delta=+27\degr42\arcmin43.8\arcsec$ (J2000.0).
% The transient was discovered by ASASSN already the day before, on 2020 December 26
% \citep{2020TNSTR3922....1S}.
% % With power-law fits to the early $g$ and $r-$ data, 
% We estimate the explosion date as 
% $\mathrm{JD_{explosion}^{SN2020adow}} = 2459206.98\pm3.86$ based on the first ZTF detection and the last non-detection. % (there're not enough points available for the MC fits.)

% ZTF20adadrhw was discovered in the starforming galaxy SDSS J083342.05+274246.0 with a known spectroscopic redshift of 0.007.

% The transient was classified as a Type Ic-BL on December 27 \citep{2020TNSCR3946....1Z}. 
% {\bf However, with a few days later P200 DBSP spectrum we reclassified the transient as a Type Ic !!
% --- So we got several early Ic-BL looking spectra but later reclassified, should it be part of sample?}\\

 \subsection{SN\,2020tkx}
Our first ZTF photometry of SN\,2020tkx (ZTF20abzoeiw)
was obtained on 2020 September 16 ($\mathrm{MJD}=59108.26$) with the P48. This first detection was in the $g-$ band, with a host-subtracted magnitude of $18.09\pm0.08$, at $\alpha=18^{h}40^{m}09.01^{s}$, $\delta=+34\degr06\arcmin59.5\arcsec$ (J2000.0).
The discovery was done by Gaia two days earlier
\citep{ZTF20abzoeiwTNSPhot}. The last ZTF non-detection is from 2021 September 07, a full week before discovery, and the constraints on the explosion date are therefore imprecise.

The transient was classified as a Type Ic-BL by \cite{ZTF20abzoeiwTNSSpec}
based on a spectrum from the Spectrograph for the Rapid Acquisition of Transients (SPRAT) on LT, obtained on 2020 September 18. Our sequence of P60 spectra taken in 2020 confirm this classification.

SN 2020tkx exploded in a faint host galaxy without a known redshift. Using the spectral template fitting SNID for our best NOT+ALFOSC spectrum taken on 2020 November 18, the redshift can be constrained to $z\sim0.02-0.03$, and our adopted redshift of $z=0.027$ is based on a weak, tentative H$\alpha$ line from the host galaxy in the spectrum. The adopted redshift translates to a luminosity distance of 122.09 Mpc and a distance modulus of 35.433.

The object has a upper limit of $< 3.3 \times 10^{-14} \rm erg\,cm^{-2}\,s^{-1}$ with the \textit{Swift} XRT (8.1 ks exposure) on 2020 October 03, 8.9\,days after peak light.  SN\,2020tkx was detected with the VLA at $286\pm 15\,\mu$Jy (10 GHz) on 2021 September 25 (see \citealt{Corsi2022} for more details).

\subsection{SN\,2020rph}
Our first ZTF photometry of SN\,2020rph 
(ZTF20abswdbg) was obtained on 2020 August 11 ($\mathrm{MJD}=59072.49$) with the P48. The transient was discovered in the public ZTF alert stream and reported by ALeRCE.
This first detection was in the r band, with a host-subtracted magnitude of 20.36, at $\alpha=03^{h}15^{m}17.82^{s}$, $\delta=+37\degr00\arcmin50.57\arcsec$ (J2000.0).
The discovery was reported to TNS \citep{ZTF20abswdbgTNSPhot}, 
with the last non-detection just 1\,hour before discovery at $r=$19.88\,mag. 
We classified the transient as a Type Ic-BL using a spectrum from the P60+SEDM obtained on 2020 August 24 \citep{ZTF20abswdbgTNSSpec}.
The supernova was found offset from the galaxy WISEA J031517.67+370055.3. We measure a redshift of $z=0.042$ based on a Keck I LRIS spectrum taken on 2020 October 19, which corresponds to a luminosity distance of 192.0 Mpc and a distance modulus of 36.42. SN\,2020rph has a \textit{Swift} XRT upper limit of f$< 3.6 \times 10^{-14} \text{ erg cm}^{-2} \text{ s}^{-1}$ on 2020 August 27, 3.5\,days after peak, in a 7.5 ks observation. It is detected with the VLA at $42.7\pm7.4 \mu$Jy (5.5 GHz) one day later, but the detection is consistent with host galaxy emission (see \citealt{Corsi2022} for details).

 \subsection{SN\,2020lao}
Our first ZTF photometry of SN\,2020lao (ZTF20abbplei)
was obtained on 2020 May 25 ($\mathrm{MJD}=58994.41$) with the P48.
This first detection was in the $g-$ band, with a host-subtracted magnitude of $19.69\pm0.10$, at $\alpha=17^{h}06^{m}54.61^{s}$, $\delta=+30\degr16\arcmin17.3\arcsec$ (J2000.0).
The discovery was reported to TNS on the same day
\citep{ZTF20abbpleiTNSPhot}. The field was well covered both before and after this first detection, and the P60 telescope was immediately triggered to provide $ugr$ photometry 1.4 hours after first detection.
%2458994.9710 - 2458994.9131  1.4 hours
The high cadence around discovery allows for a well constrained explosion date.
 With power-law fits to the early $g-$ and $r-$ data, we estimate the explosion date as 
$\mathrm{MJD_{explosion}^{SN2020lao}} = 58993.07\pm0.75$.

SN\,2020lao was also reported in a paper by the Transient Exoplanet Satellite Survey (TESS; \citealt{TESS2021}) with high cadence photometry. The TESS paper finds a slightly different rise time (13.5 $\pm$ 0.22 days) relative to our ZTF observations; however this can be attributed to their broad peak and bandpass that may also contain NIR flux. On the other hand, we find that our narrow $i-$ band peak is consistent with our estimated $r-$band peak.

Our first spectrum of this event was obtained with the P60+SEDM on 2020 May 26. It was mainly blue and featureless and did not warrant any classification. We obtained several more inconclusive spectra the following days, and the transient was finally classified as a Type Ic-BL by the Global SN Project on 2020 June 02 \citep{ZTF20abbpleiTNSSpec}. Our subsequent P60+SEDM and NOT+ALFOSC spectra taken in 2020 confirmed this classification based on its broad Fe II features.

SN 2020lao exploded in the face on spiral
galaxy  CGCG 169-041 with a well established redshift of $z=0.030814$, which corresponds to a luminosity distance of 141.3 Mpc and a distance modulus of 35.8.
This redshift is confirmed with narrow host lines in our later spectra.

On 2020 June 07. 3.5\,days after peak light, we obtained an upper limit on the \textit{Swift} XRT flux of f$< 2.9 \times 10^{-14} \text{ erg cm}^{-2} \text{ s}^{-1}$ (14 ks).\\

\subsection{SN\,2020dgd}
Our first ZTF photometry of SN\,2020dgd (ZTF20aapcbmc)
was obtained on 2020 February 19 ($\mathrm{MJD}=58898.52$) with the P48.
This first detection was in the $r-$ band, with a host-subtracted magnitude of 18.99, at $\alpha=15^{h}45^{m}35.57^{s}$, $\delta=+29\degr18\arcmin38.4\arcsec$ (J2000.0).
The discovery was reported to TNS \citep{ZTF20aapcbmcTNSPhot}, 
with a note saying that the last non-detection was 5\,days before discovery (on 2020 February 14 at $r=20.03$\,mag). 
We classified the transient as a Type Ic-BL using a spectrum from the P60 SEDM obtained on 2020 March 05 \citep{ZTF20aapcbmcTNSSpec}. The transient appears to be separated by 14$\arcsec$ from any visible host galaxy in the vicinity; however with a Keck I LRIS spectrum taken on 2020 June 23 in the nebular phase (not shown in Figure~\ref{fig:spectra}), we measure weak host lines at a redshift of $z = 0.032$, corresponding to a distance of 145.2 Mpc and a distance modulus of 35.8. In addition, that LRIS spectrum of the SN exhibits strong Ca II emission features.

\subsection{SN\,2020bvc}
Our first ZTF photometry of SN\,2020bvc (ZTF20aalxlis)
was obtained on 2020 February 04 ($\mathrm{MJD}=58883.0$) with the P48. This first detection was in the $i-$ band, with a host-subtracted magnitude of 17.48, at $\alpha=14^{h}33^{m}57.01^{s}$, $\delta=+40\degr14\arcmin37.5\arcsec$ (J2000.0).
SN\,2020bvc, reported originally in \citet{Ho2020_ZTF20aalxlis}, shows very similar optical, X-ray and radio properties to SN\,2006aj, which was associated with the low-luminosity GRB 060218. See \citet{Ho2020_ZTF20aalxlis} for more details about this object.

\subsection{SN\,2019xcc}
Our first ZTF photometry of SN\,2019xcc (ZTF19adaiomg)
was obtained on 2019 December 19 ($\mathrm{MJD}=58836.48$) with the P48.
This first detection was in the $r-$ band, with a host-subtracted magnitude of $19.40\pm0.13$, at $\alpha=11^{h}01^{m}12.39^{s}$, $\delta=+16\degr43\arcmin29.1\arcsec$ (J2000.0).
The discovery was reported to TNS on the same day
\citep{ZTF19adaiomgTNSPhot}, 
with a note saying that the latest non-detection from ZTF was five days prior to discovery ($r = 19.3$). This transient has very sparse light curves with only four data points from P48 in the alert stream, all in the
$r-$ band, but forced photometry also retrieved detections in the $g-$ band.
% {\bf also P60 data,  and for many of course there may also be public ATLAS data...}

The transient was classified as a Type Ic-BL by \cite{ZTF19adaiomgTNSSpec}, based on a spectrum from SPRAT on the Liverpool Telescope obtained on 2019 December 20. We could confirm this classification with a spectrum from the Keck telescope a few days later, using the LRIS instrument.

SN 2019xcc exploded close to the centre of the face on grand  spiral CGCG 095-091 with a well established redshift of $z=0.028738$, which corresponds to a luminosity distance of 129.8 Mpc, and a distance modulus of 35.6. This redshift is confirmed with narrow host H$\alpha$ in our Keck spectrum.\\

\subsection{SN\,2019qfi}
SN\,2019qfi (ZTF19abzwaen) was discovered by ATLAS
on 2019 September 07 ($\mathrm{MJD}=58743.29$) with the internal name ATLAS2019vdc, with the first ZTF alerts around the same time.
This first detection was in the $o$ band, with a host-subtracted magnitude of 18.81, at $\alpha=21^{h}51^{m}07.90^{s}$, $\delta=+12\degr25\arcmin38.5\arcsec$ (J2000.0).
The discovery was reported to TNS \citep{ZTF19abzwaenTNSPhot}, 
with a note saying that the last non-detection was 6\,days before the discovery at $o = 18.69$\,mag. 
We classified the transient as a Type Ic-BL using a spectrum from the P60+SEDM obtained on 2019 Sep 21 \citep{ZTF19abzwaenTNSSpec}.
SN\,2019qfi was identified in the starforming galaxy SDSS J215107.99+122542.5 with a known spectroscopic redshift of $z=0.028$. This corresponds to a luminosity distance of 129.0 Mpc and a distance modulus of 35.5.

\subsection{SN\,2019moc}
SN\,2019moc (ZTF19ablesob) was first reported by ATLAS
on 2019 August 04 ($\mathrm{MJD}=58699.47)$) with the internal name ATLAS2019rgu.
This first detection was in the $c$ band, with a host-subtracted magnitude of 18.54, at $\alpha=23^{h}55^{m}45.95^{s}$, $\delta=+21\degr57\arcmin19.67\arcsec$ (J2000.0). However, its first ZTF detection preceded that of ATLAS, on 2019 July 31. 
The discovery was reported to TNS \citep{ZTF19ablesobTNSPhot}, 
with a note saying that the last non-detection was 6\,days before discovery at $c = 19.44$\,mag. 
We classified the transient as a Type Ic-BL using a spectrum from the P200 DBSP obtained on 2019 Aug 10 \citep{ZTF19ablesobTNSSpec}.
The SN was found in the galaxy SDSS J235545.94+215719.7 with a known spectroscopic redshift of 0.055, corresponding to a luminosity distance of 257.6 Mpc and a distance modulus of 37.1.

\subsection{SN\,2019gwc}
Our first ZTF photometry of SN\,2019gwc (ZTF19aaxfcpq)
was obtained on 2019 June 04 ($\mathrm{MJD}=58638.28$) with the P48.
This first detection was in the $r-$ band, with a host-subtracted magnitude of 19.73, at $\alpha=16^{h}03^{m}26.88^{s}$, $\delta=+38\degr11\arcmin02.6\arcsec$ (J2000.0).
The discovery was reported to TNS \citep{ZTF19aaxfcpqTNSPhot}, 
with a note saying that the last non-detection was three days before discovery (2019 Jun 01 at $r = 20.98$\,mag). 
We classified the transient as a Type Ic-BL using a spectrum from the P60 SEDM obtained on 2019 Jun 16 \citep{ZTF19aaxfcpqTNSSpec}.
The transient was identified in the starforming host galaxy SDSS J160326.65+381057.1 at a known spectroscopic redshift of $z=0.038$, corresponding to a distance of 173.2 Mpc, and a distance modulus of 36.2.

\subsection{SN\,2019hsx}
Our first ZTF photometry of SN\,2019hsx (ZTF19aawqcgy)
was obtained on 2019 June 02 ($\mathrm{MJD}=58636.31$) with the P48.
This first detection was in the $r-$ band, with a host-subtracted magnitude of $18.62\pm0.08$, at $\alpha=18^{h}142^{m}56.22^{s}$, $\delta=+68\degr21\arcmin45.2\arcsec$ (J2000.0).
The discovery was reported to TNS %late..2 weeks after
\citep{ZTF19aawqcgyTNSPhot}, 
with a note saying that the latest non-detection from ZTF was 3 days prior to discovery (May 30; $g = 20.3$). 
We classified the transient as a Type Ic-BL using a spectrum from P60+SEDM obtained on June 14 \citep{ZTF19aawqcgyTNSSpec}. SN\,2019hsx exploded fairly close to the center of NGC 6621
with redshift $z=0.020652$. This corresponds to a distance of 92.9 Mpc and a distance modulus of 34.8.
SN\,2019hsx was detected with a \textit{Swift} XRT flux of $6.2_{-1.8}^{+2.3} \times 10^{-14}\rm\,erg\,cm^{-2}\,s^{-1}$ (at $\sim6\sigma$) in a 15 ks observation on 2019 July 20, 36.7\,days after peak.
% With our adopted cosmology and correcting for Virgo, Shapley and the Great Attractor as implemented on NED\footnote{NED; http://ned.ipac.caltech.edu/, see also \cite{Mould2000}.}
% this corresponds to a luminosity distance of 94.2 Mpc.\\

% \rednote{Host galaxy is being detected in the forced photometry both in the early and late-times, it seems...}
% {\bf From ZTF forced LC, there are lots of detections with SNR $>5$ 300 days before. We need to check on that?}

\subsection{SN\,2018kva}
Our first ZTF photometry of SN\,2018kva (ZTF18aczqzrj)
was obtained on 2018 December 23 ($\mathrm{MJD}=58475.51$) with the P48.
This first detection was in the $r-$ band, with a host-subtracted magnitude of 19.08, at $\alpha=08^{h}35^{m}16.21^{s}$, $\delta=+48\degr19\arcmin03.4\arcsec$ (J2000.0).
The discovery was reported to TNS \citep{2018kvaTNSPhot}, 
with a note saying that the latest non-detection was 3\,days before discovery, at $g = 20.33$\,mag. 
We classified the transient as a Type Ic-BL using a spectrum from the P60+SEDM obtained on 2019 Jan 03 \citep{2018kvaTNSSpec}.
The object was identified in the host galaxy WISEA J083516.34+481901.2 at a known redshift of $z=0.043$, which corresponds to a luminosity distance of 196.2 Mpc and a distance modulus of 36.5.

\subsection{SN\,2018jaw}
Our first ZTF photometry of SN\,2018jaw (ZTF18acqphpd)
was obtained on 2018 November 20 ($\mathrm{MJD}=58442.51$) with the P48.
This first detection was in the $g-$ band, with a host-subtracted magnitude of 18.39, at $\alpha=12^{h}54^{m}04.10^{s}$, $\delta=+13\degr32\arcmin47.9\arcsec$ (J2000.0).
The discovery was reported to TNS \citep{2018jawTNSPhot}, 
with a note that the object was missing ZTF non-detection limits. 
We classified the transient as a Type Ic-BL using a spectrum from the P60+SEDM obtained on 2018 Dec 12 \citep{2018jawTNSSpec}, and tentatively estimated its redshift to be $z=0.037$.
However, the narrow host lines in the Keck I-LRIS spectrum taken on 2019 April 06 indicate that the object is at a redshift of $z=0.047$. This corresponds to a luminosity distance of 168.5 Mpc and a distance modulus of 36.1.
SN\,2018jaw was identified in the galaxy host WISE J125404.15+133244.9.

\subsection{SN\,2018gep}
Our first ZTF photometry of SN\,2018gep (ZTF18abukavn)
was obtained on 2018 September 09 ($\mathrm{MJD}=58370.16$) with the P48. This first detection was in the $r-$ band, with a host-subtracted magnitude of 20.5, at $\alpha=16^{h}43^{m}48.22^{s}$, $\delta=+41\degr02\arcmin43.4\arcsec$ (J2000.0).

SN\,2018gep belongs to the class of Fast Blue Optical Transients (FBOTs) with its rapid rise time, high peak luminosity, and blue colors at peak \citep{Pritchard2021}. It was classified as a Ic-BL supernova whose early multi-wavelength data can be explained by late-stage eruptive mass loss. The transient is detected with the VLA over three epochs (9, 9.7 and 14 GHz), but the emission is likely galaxy-dominated. See \citet{Ho2019_SN2018gep} for more details on the discovery of this supernova. 

\begin{table*}[b]
    \centering
    \begin{tabular}{cccccccccc}
    \hline\hline
        IAU name & ZTF name & type & RA & Dec & z & F$_{\nu}^{a}$ & $\rm F_{0.3-10 \rm keV}^{b}$\\
         &  &  &  &  &  & ($\mu$Jy) & (10$^{-14}$ erg cm$^{-2}$ s$^{-1}$) \\
        \hline
        2018gep & ZTF18abukavn & FBOT* & 16:43:48.21 & +41:02:43.29 & 0.032 & $< 34 \pm 4$ (9.7 GHz) & $ < 9.9$ \\
        2018jaw & ZTF18acqphpd & Ic-BL & 12:54:04.10 & +13:32:47.9 & 0.047 & - & -\\
        2018kva & ZTF18aczqzrj & Ic/Ic-BL & 08:35:16.21 & +48:19:03.4 & 0.043 & - & -\\
        2019gwc & ZTF19aaxfcpq & Ic-BL & 16:03:26.88 & +38:11:02.6 & 0.038 & - & -\\
        2019hsx & ZTF19aawqcgy & Ic-BL & 18:12:56.22 & +68:21:45.2 & 0.021 & $\lesssim 19$ (6.2 GHz) & 6.2$^{+2.3}_{-1.8}$\\
        2019moc & ZTF19ablesob & Ic-BL & 23:55:45.95 & +21:57:19.67 & 0.056 & - & -\\
        2019qfi & ZTF19abzwaen & Ic-BL & 21:51:07.90 & +12:25:38.5 & 0.029 & - & -\\
        2019xcc & ZTF19adaiomg & Ic-BL & 11:01:12.39 & +16:43:29.30 & 0.029 & $< 62.7 \pm 8.7$ (6.3 GHz) & -\\
        GRB190829A & - & LLGRB & 2:58:10.580 & -8:57:29.82 & 0.077 & - & -\\
        2020bvc & ZTF20aalxlis & Ic-BL** & 14:33:57.01 & +40:14:37.5 & 0.025 & $63 \pm 6$ (10 GHz) & 9.3$^{+10.6}_{-6.1}$\\
        2020dgd & ZTF20aapcbmc & Ic-BL & 15:45:35.57 & +29:18:38.4 & 0.03 & - & -\\
        2020lao & ZTF20abbplei & Ic-BL & 17:06:54.61 & +30:16:17.3 & 0.031 & $\lesssim 33$ (5.2 GHz) & $< 2.9$\\
        2020rph & ZTF20abswdbg & Ic-BL & 03:15:17.82 & +37:00:50.57 & 0.042 & $<42.7 \pm 7.4$ (5.5 GHz) & $< 3.6$\\
        2020tkx & ZTF20abzoeiw & Ic-BL & 18:40:09.01 & +34:06:59.5 & 0.027 & $272 \pm 16$ (10 GHz) & $< 3.3$\\
        2021bmf & ZTF21aagtpro & Ic-BL & 16:33:29.41 & -06:22:49.53 & 0.017 & - & - \\
        2021xv & ZTF21aadatfg & Ic-BL & 16:07:32.82 & +36:46:46.07 & 0.041 & $< 34.3 \pm 8.1$ (5.2 GHz) & -\\
        2021ywf & ZTF21acbnfos & Ic-BL & 05:14:11.00 & +01:52:52.28 & 0.028 & $83 \pm 10$ (5.0 GHz) & 5.3$^{+4.9}_{-3.3}$\\
        2021too & ZTF21abmjgwf & Ic-BL & 21:40:54.28 & +10:19:30.33 & 0.035 & - & -\\
    \hline\hline
    \end{tabular}
    \caption{Sample summary table of Ic-BL supernova properties, estimated \rp{} ejecta mass and mixing fraction along with their 1$\sigma$ uncertainties, and first radio/X-ray detection. In the absence of any X-ray/radio detections we quote an upper limit; if the source was not observed we mark the cell with a dash. a) Flux density in $\mu$Jy with the VLA. We list only the first VLA observation at $\lesssim$50 days from the first ZTF detection as reported in \citet{Corsi2022}. b) \textit{Swift} XRT flux in units of $10^{-14}$erg cm$^{-2}$ s$^{-1}$, taken from \citet{Corsi2022}. *This SN Ic-BL is also categorized as a fast blue optical transient (FBOT), and was published in \citet{Ho2019_SN2018gep}. The quoted radio detection with the VLA could be galaxy-dominated. **This Ic-BL had a double-peaked light curve from shock-cooling; X-ray and radio measurements taken from \citet{Ho2020_ZTF20aalxlis}.}
    \label{tab:rprocess_radio}
\end{table*}

\section{Literature sample} \label{sec:literature}
In addition to the ZTF SNe in our sample we examine the Open Supernova Catalog\footnote{https://github.com/astrocatalogs/supernovae} for historical low-redshift SNe Ic-BL with $\gtrsim$3 epochs of multi-band NIR photometry concurrent with the optical coverage. Our requirement for the minimum number of epochs is to probe the color evolution over time, which then can be compared against the \rp{} models. We exclude those objects with only NIR observations of the afterglow and early ($\lesssim10$\,days from explosion) SN light curve, in the case of a GRB association. We find that SN\,1998bw \citep{Patat2001, Clochiatti2011}, SN\,2002ap \citep{Yoshii2003, Tomita2006}, SN\,2010bh \citep{Olivares2012}, and SN\,2016coi \citep{Terreran2019} match our criteria. We also find that SN\,2016jca has extensive optical and NIR follow-up \citep{Cano2017_SN2016jca, Ashall2019} but exclude it from further study because the reported NIR photometry is neither host- nor afterglow-subtracted.

SN\,2016coi uniquely shows a huge 4.5\,$\mu$m excess in the mid-infrared with archival WISE coverage in its late-time light curve. This object also has detections in the H-band past 300\,days post-peak which coincide with the mid-IR detections. Given that it also has a bright radio counterpart, the mid-IR excess could be attributed to CO formation in the ejecta \citep{Liljegren2022}, or dust formation due to adiabatic cooling \citep{Omand2019}, or metal cooling in highly mixed SN ejecta \citep{OmandJerkstrand2022}.  Though we lack model predictions in the mid-IR bands, we test whether the long-lived NIR emission could also be attributed to \rp{} production.

In addition, \citet{Bianco2014} collected optical and NIR photometry for a set of 61 stripped envelope SNe that also satisfy our low-redshift cut after conducting template-based subtraction in order to subtract host galaxy emission (for most SNe).  Amongst the SNe in that sample  classified as Type Ic-BL, only two SNe have observations in the $J$, $H$, or $K_s$ bands: SN\,2007I and SN\,2007ce. Similar to the case of our ZTF SNe, during the earlier epochs ($<$\,60\,days post-peak) these two SNe have well-sampled optical photometry, while later in time there is only NIR coverage. The second study, \citet{Stritzinger2018_supernova}, acquired optical light curves for 34 stripped-envelope SNe, 26 of which have NIR follow-up in the $YJH$ bands as a part of the Carnegie Supernova Project. Explosion and bolometric light curve properties for some of these SNe were released in a companion paper \citep{Taddia2018}. Of the 26 SNe, only one (SN\,2009bb) has adequate coverage at late times in the NIR.

\citet{Li2022} perform detailed blackbody fits to several SNe from the Open SN Catalog that have optical and NIR coverage to search for SNe that show NIR excesses in their SEDs that could be attributed to dust formation. Amongst the sample they consider, the authors find SN\,2007I and SN\,2009bb to be consistent with blackbody emission with a slight NIR excess that evolves from a photospheric temperature of $\sim$5000 ($\sim$7000) to 4300 K over the course of 51 (33) days in the case of SN\,2007I (SN\,2009bb). The same authors find that the SED of SN\,2007ce is inconsistent with a blackbody, though they use only the early-time measurements of the object (at 1.9\,days). Furthermore, \citet{Li2022} find no evidence for intrinsic dust formation or significant host extinction in order to explain their SEDs. In contrast to their study, we note that our analysis includes photometry for these SNe over a much longer baseline taken from \citet{Bianco2014} and \citet{Stritzinger2018_supernova}. 

For each of the above-mentioned SNe, we correct for Galactic extinction where extinction has not been accounted for, and convert from Vega to AB magnitudes. We also correct the light curves for host attenuation for all of these SNe except for SN\,1998bw (light curve already corrected for Galactic and host extinction) and SN\,2007ce (lacks host galaxy extinction information); the assumed host E(B-V) values are listed in Table~\ref{tab:literature_sne}. We include host extinction here as it is significant for the literature SNe. The measurements of total ejecta mass, kinetic energy, and nickel mass for each object are recorded in Table~\ref{tab:literature_sne}, along with the appropriate reference we took these estimates from. We include the following seven SNe: SN\,1998bw, SN\,2002ap, SN\,2010bh, SN\,2016coi, SN\,2009bb, SN\,2007I and SN\,2007ce in our analysis, described in Sec.~\ref{sec:results}.
% All of the light curves of these SNe from the literature are plotted in Figure~\ref{fig:mlit_lcs}. 

% \begin{figure*}
%     \centering
%     \includegraphics[width=0.5\textwidth]{figures/literature_IcBL_light curves_BVRJHK.pdf}\includegraphics[width=0.5\textwidth]{figures/literature_IcBL_light curves_griJHKs.pdf}
%     \caption{light curve collage for IcBL SNe from the literature.}
%     \label{fig:literature_sne}
% \end{figure*}

\section{Collapsar light curve models} \label{sec:models}

We model the evolution of the emission from \rp{}-enriched collapsars using a semi-analytic model of \citet{Barnes2022}.
While the details of our method are described there, we present an outline here.

The models comprise a series of concentric shells whose densities ($\rho(v)$) follow a broken power law in velocity space:
\begin{equation}
    \rho(v) \propto 
    \begin{cases}
        v^{-n}, \:\: v \leq v_{\rm t}, \\
        v^{-\delta} \:\: v > v_{\rm t},
    \end{cases}\label{eq:density_bpl}
\end{equation}
where we set the power-law index $n$ ($\delta$) equal to 1 (10). Our density profile, varying with velocity, contrasts with that of \citet{Arnett1987}, which uses a one-zone formulation. Such a density profile is necessary to enrich SNe with \rp{} elements out to a particular mixing coordinate, as we describe below.
In Eq.~\ref{eq:density_bpl}, $v_{\rm t}$ is a transition velocity chosen to produce the desired total mass \mej{} and kinetic energy 
$E_{\rm kin}$, which is parameterized via the average velocity 
$v_{\rm ej} = \sqrt{2E_{\rm kin}/\mej}$.
In addition to \mej{} and $v_{\rm ej}$, each model is characterized by its mass of \iso{Ni}{56}, $M_{56}$, which we assume is uniformly distributed throughout the ejecta \citep{Taddia2018, Yoon2019, SuzukuMaeda2021}. This choice departs from the analytical model of \citet{Arnett1982}, which assumes the nickel is centrally located. Furthermore, while the Arnett models do not allow for inefficient deposition of gamma-ray energy, these models calculate gamma-ray deposition based on a gray gamma-ray opacity. Thus these models do not match the Arnett models at maximum light. Different \iso{Ni}{56} profiles will also affect the distribution of diffusion times, altering the shape of the bolometric light curve.

We assume that some amount \mrp{} of the ejecta is composed of pure \rp{} material, and that this material is mixed
 evenly into the ejecta interior to a velocity $v_{\rm mix}$, which we define such that 
\begin{equation}
    \int\limits_0^{v_{\rm mix}} \rho(v) \: \mathrm{d} \; V =\xmix \mej{},
\label{eq:xmix}
\end{equation}
with \xmix{} a parameter of the model, and dV the volume of the ejecta.
(In other words, Eq.~\ref{eq:xmix} shows that \xmix{} is the fraction of the total ejecta mass for which the \rp{} mass fraction is non-zero.)
By distributing the \rp{}{} mass within a core of mass ${>}\mrp{}$, we account for hydrodynamic (e.g., Kelvin-Helmholtz) instabilities at the wind-ejecta boundary, which may mix the \rp{}-rich disk outflow out into the initially \rp{}-free ejecta.

The \rp{}{} elements serve as a source of radioactive energy beyond \iso{Ni/Co}{56}. 
More importantly (especially at early times---see \citealt{Siegel2019}), they impart to the enriched layers the high opacity \citep{Kasen.Badnell.Barnes_2013.ApJ_rprocess.opacs,Tanaka.Hotokezaka_2013.ApJ_rprocess.opacs} known to be a unique feature of \rp{}{} compositions.
This high opacity affects local diffusion times and the evolution of the photosphere, thereby altering SN emission relative to the \rp{}-free case.

We model the spectral energy distribution (SED) from the photospheric ejecta layers ($r \leq R_{\rm ph}$) as a black body, and integrate it to get the bolometric luminosity, given by 
\begin{equation}
L = 4\pi R^2_{\rm ph} \sigma_{\rm SB} T_{\rm ph}^4
\label{eq:Lbb}
\end{equation}
with $\sigma_{\rm SB}$ the Stefan-Boltzman constant.
The opacity in our model is gray and defined for every zone, allowing a straightforward determination of the photospheric radius $R_{\rm ph}$.
The photospheric temperature $T_{\rm ph}$ is then chosen so the RHS of Eq.~\ref{eq:Lbb} is equal to the luminosity emerging from behind the photosphere, which is an output of our calculation.

Since we are equally interested in SN signals beyond the photospheric phase, we also track emission from optically thin regions of the ejecta.
These are assumed to have an SED determined by their composition.
The \emph{r}-process free layers conform to expectations set by observed SNe \citep[e.g.][]{Hunter2009}.
For enriched layers, we rely on theoretical studies of nebular-phase \rp{}{} transients \citep{Hotokezaka.ea_2021.MNRAS_rprocess.nebular}.
The radioactive heating, opacity, and photospheric and nebular SEDs of each model are thus fully determined, allowing us to predict light curves and colors as a function of time.

\section{Analysis} \label{sec:analysis}
In the sections below, for the analysis and fitting of our light curves, we assume the central wavelengths for the optical and NIR bandpasses listed in Table~\ref{tab:filters}, ignoring any small differences due to non-standard filters.

\begin{table}[b]
    \centering
    \begin{tabular}{cc}
    \hline
    \hline
    filter & central wavelength ($\mathrm{\AA}$) \\
    \hline
        $g$ & 4770 \\
        $r$ & 6231 \\
        $i$ & 7625 \\
        $U$ & 3600 \\
        $B$ & 4380 \\
        $V$ & 5450 \\
        $R$ & 6410 \\
        $I$ & 7980 \\
        $J$ & 12350 \\
        $H$ & 16620 \\
        $K_s$ & 21590 \\
        $K$ & 21900 \\
        \hline\hline
    \end{tabular}
    \caption{Central wavelengths for the optical and NIR filters assumed during the analysis and fitting of our light curves.}
    \label{tab:filters}
\end{table}

\subsection{Estimation of explosion properties} \label{subsec:lc_properties}
The combination of using both a volume-limited and a magnitude-limited survey for SN Ic-BL discovery has yielded SNe with a diverse range of absolute magnitudes. In Table~\ref{tab:rprocess_radio} we summarize the SNe in our sample, which have redshifts ranging from 0.01 to 0.05 and peak $r-$band absolute magnitudes from M$_{\rm r}\sim-17$\,mag to M$_{\rm r}\sim-19$\,mag. For the purpose of this analysis, we consider distance uncertainty to have a negligible effect on our estimation of the explosion properties (the SNe we are fitting have a distance uncertainty of $<5$\,Mpc). Here we summarize our process for deriving explosion parameters (i.e. total ejecta mass, kinetic energy, and nickel mass) from these SN light curves.

The details of the methodology behind our analysis of the bolometric light curves in this sample are described at length in a companion paper, \citet{Corsi2022}, though only a subset of our sample is included in the companion paper. This analysis is done with the open-source code \texttt{HAFFET}\footnote{https://github.com/saberyoung/HAFFET} \citep{Yang2023}. First, we correct the light curves for Milky Way extinction, and then derive bolometric light curves from the $g-$ and $r-$ band photometry after calculating bolometric corrections from the empirical relations given in \citet{Lyman2014, Lyman2016}. In spite of the diversity in SNe Ic-BL colors and temporal evolution, \citet{Lyman2016} found that the variation in the bolometric magnitude was $<0.1$\,mag; thus we consider the Lyman+ relations to be valid for our SN sample. We estimate the explosion epoch with power law fits unless the early-time SN data are limited, in which case the explosion times are set as the midpoint between the last non-detection before discovery and the first ZTF detection. We then fit the bolometric light curves to Arnett models \citep{Arnett1982} between $-20$ and 60 days from the peak to obtain the \iso{Ni}{56} mass, \mni{} and the characteristic timescale $\tau_m$. $\tau_m$ is calculated from \mej{}, the kinetic energy $E_{\rm k}$, and the ejecta opacity $\kappa$, which is assumed to be a constant (0.07$\mathrm{cm}^2 \mathrm{g}^{-1}$; \citealt{Chugai2000, Barbarino2020, Tartaglia2021}). The uncertainties on our explosion epochs propagate into the uncertainties on $\tau_m$, \mej{}, and \mni{}. The early-time optical light curves of typical SNe Ic-BL are well-approximated by the Arnett model which describes the $^{56}$Ni-powered light curve during the supernova's photospheric phase.

\begin{figure}
    \centering
    \includegraphics[width=0.50\textwidth]{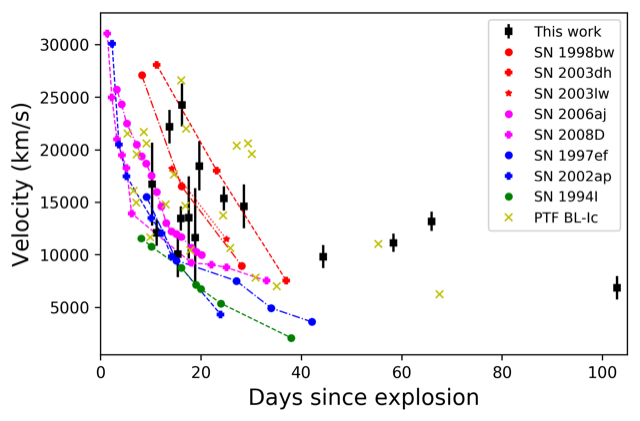}
    \caption{SN velocities measured from the Fe II 5169$\rm \AA$ line as a function of the spectroscopic phase for each supernova in our sample (black points) plotted along with the measured velocities of SNe Ic-BL from the literature and from PTF \citep{Taddia2018}. The velocities we measure here are broadly consistent with both the literature and PTF sample.}
    \label{fig:velocity}
\end{figure}

For each of the SNe we estimate the photospheric velocity ($v_{\rm ph}$) using the earliest high-quality spectrum taken of the object. We use the IDL routine \texttt{WOMBAT} to remove host galaxy lines and tellurics, and then smooth the spectrum using \texttt{SNspecFFTsmooth} \citep{Liu2016}. The broad Fe II feature at 5169\,$\rm \AA$ is considered to be a proxy for the photospheric velocity of a Type Ic-BL SN \citep{Modjaz2016}. Thus we use the open source code SESNspectraLib\footnote{https://github.com/nyusngroup/SESNspectraLib} \citep{Liu2016, Modjaz2016} to fit for the Fe II velocity by convolving with SN Ic templates. The velocities were measured at different phases for each SN, as shown in Figure~\ref{fig:velocity}.

We then estimate the kinetic energy, $E_{\rm k}$, and the total ejecta mass \mej{} of the explosion using our derived values for $\tau_\mathrm{m}$ and $v_{\rm ph}$ and the empirical relations from \citet{Lyman2016}. In some cases where $v_{\rm ph}$ was only measured $>$15\,days after the peak, we could only quote lower limits on the kinetic energy and ejecta mass of the explosion.

The explosion properties we derive are given in Table~\ref{tab:basic_properties}.

\subsection{Comparing color-color predictions to observations} \label{subsec:color}
Optical-NIR colors are a useful diagnostic to determine whether SNe Ic-BL could be potential sites of \rp{} production.
The high opacity of \rp{} elements causes emission from the enriched regions to shift to redder wavelengths. 

\begin{figure*}
    \centering
    \includegraphics[width=0.52\textwidth]{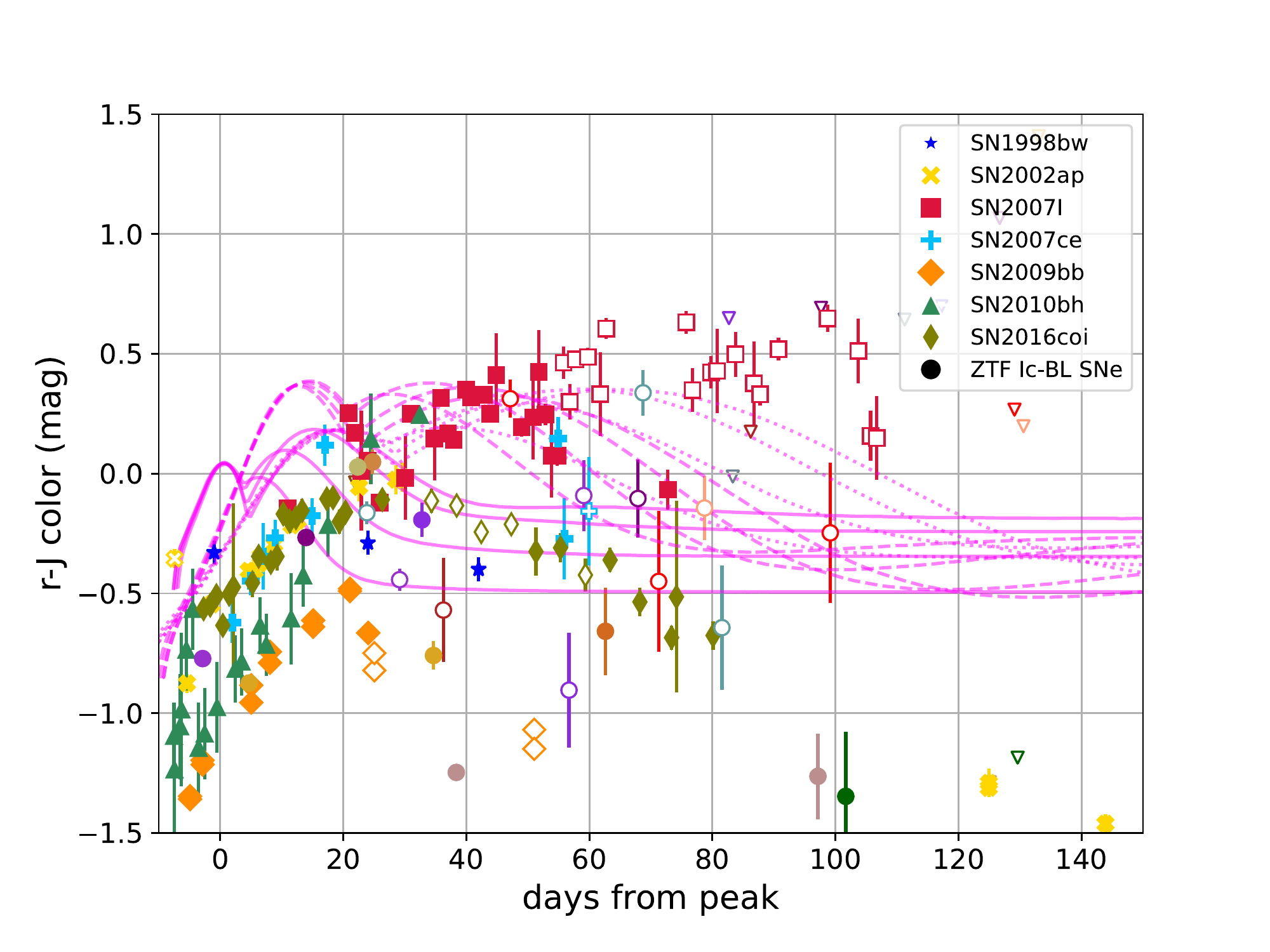}\includegraphics[width=0.52\textwidth]{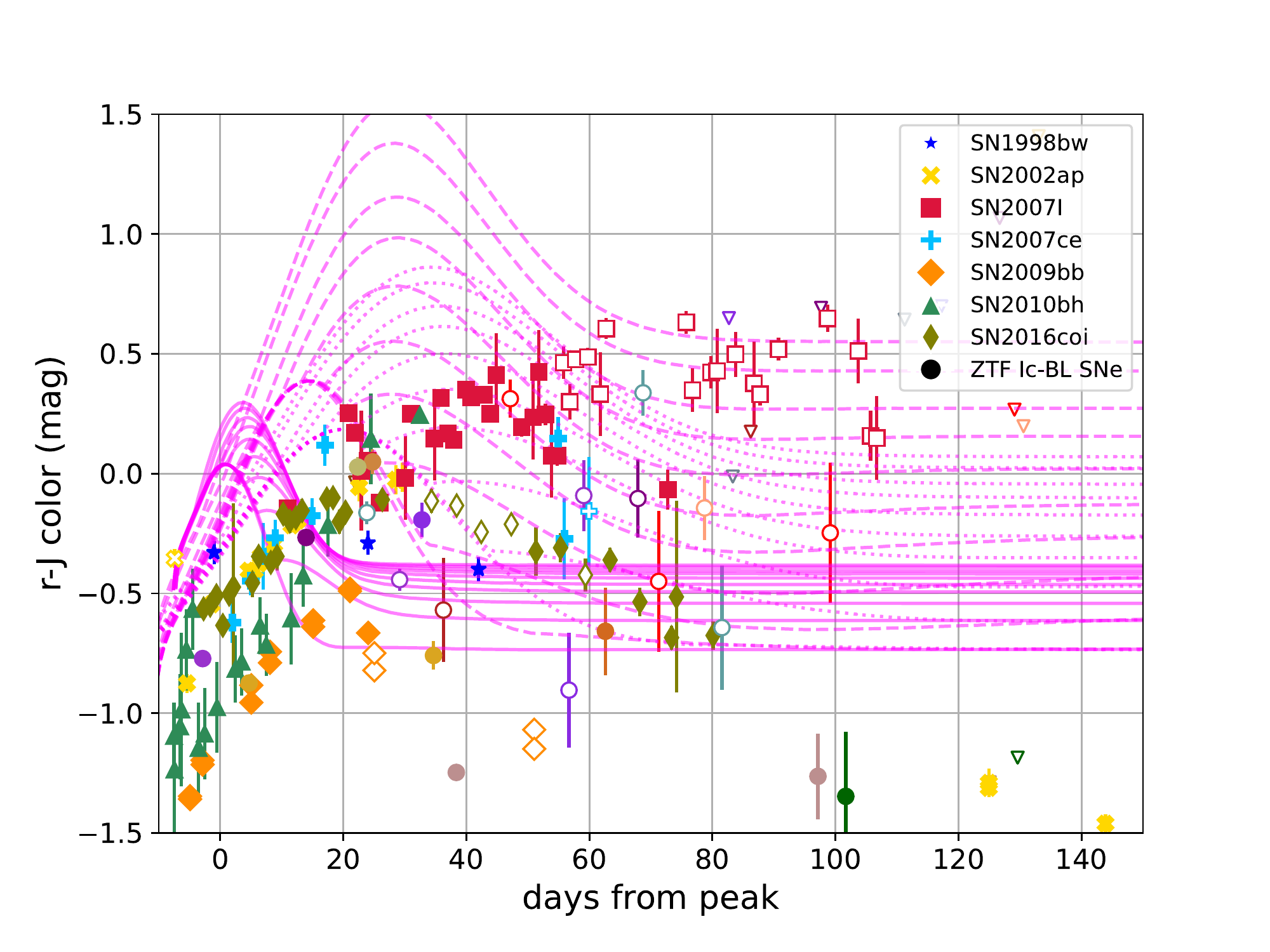}
    \includegraphics[width=0.52\textwidth]{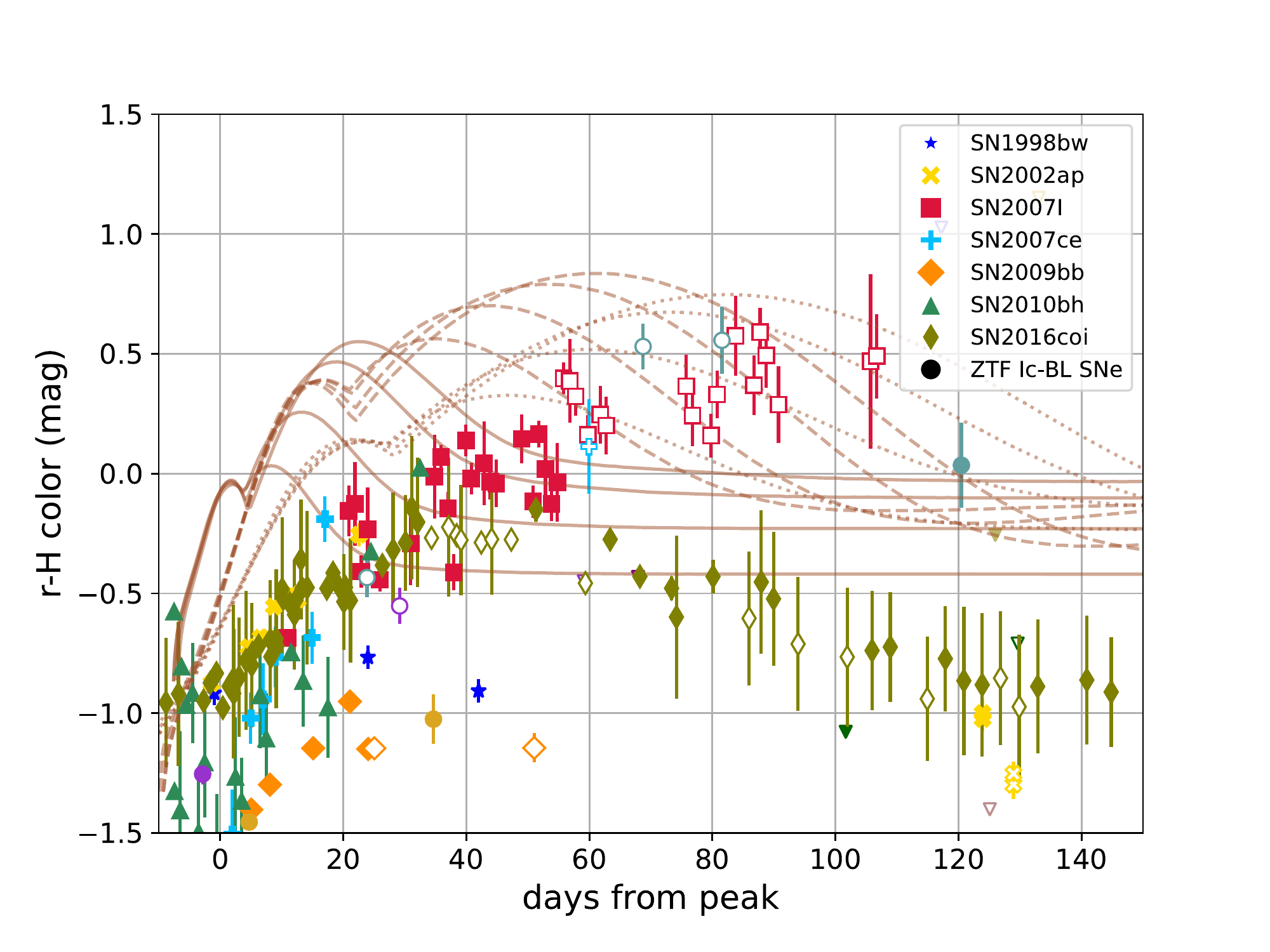}\includegraphics[width=0.52\textwidth]{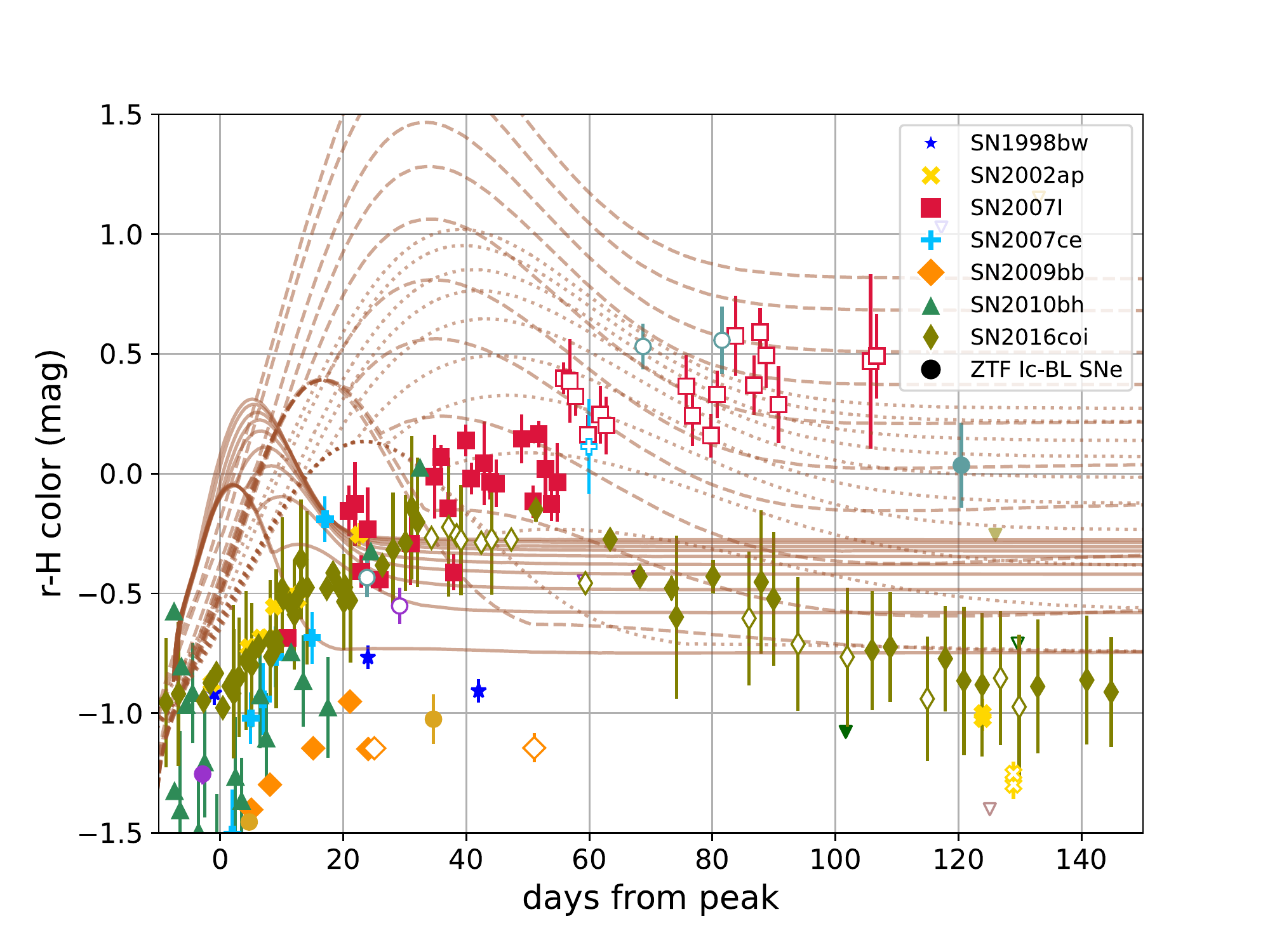}
    \includegraphics[width=0.52\textwidth]{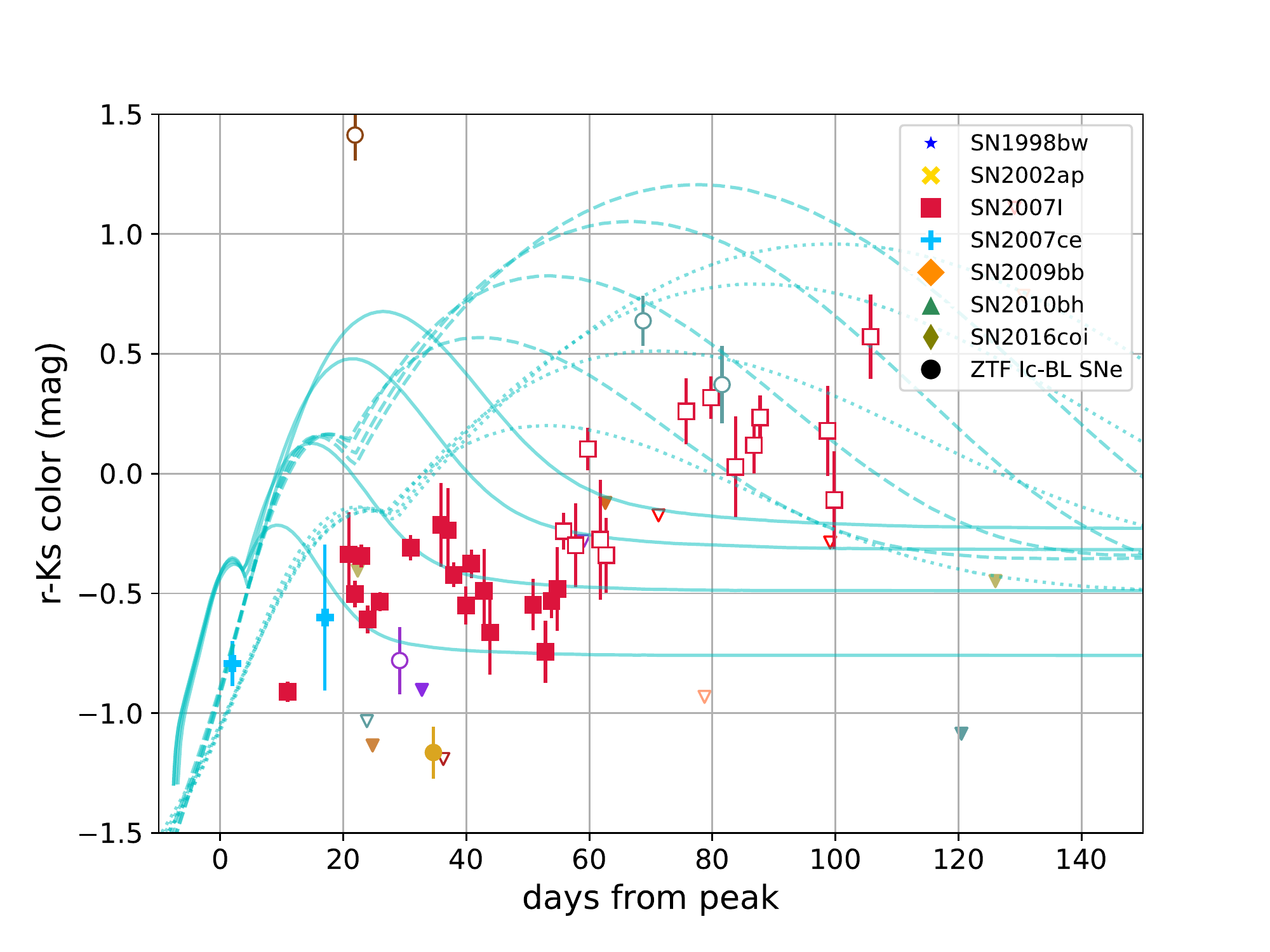}\includegraphics[width=0.52\textwidth]{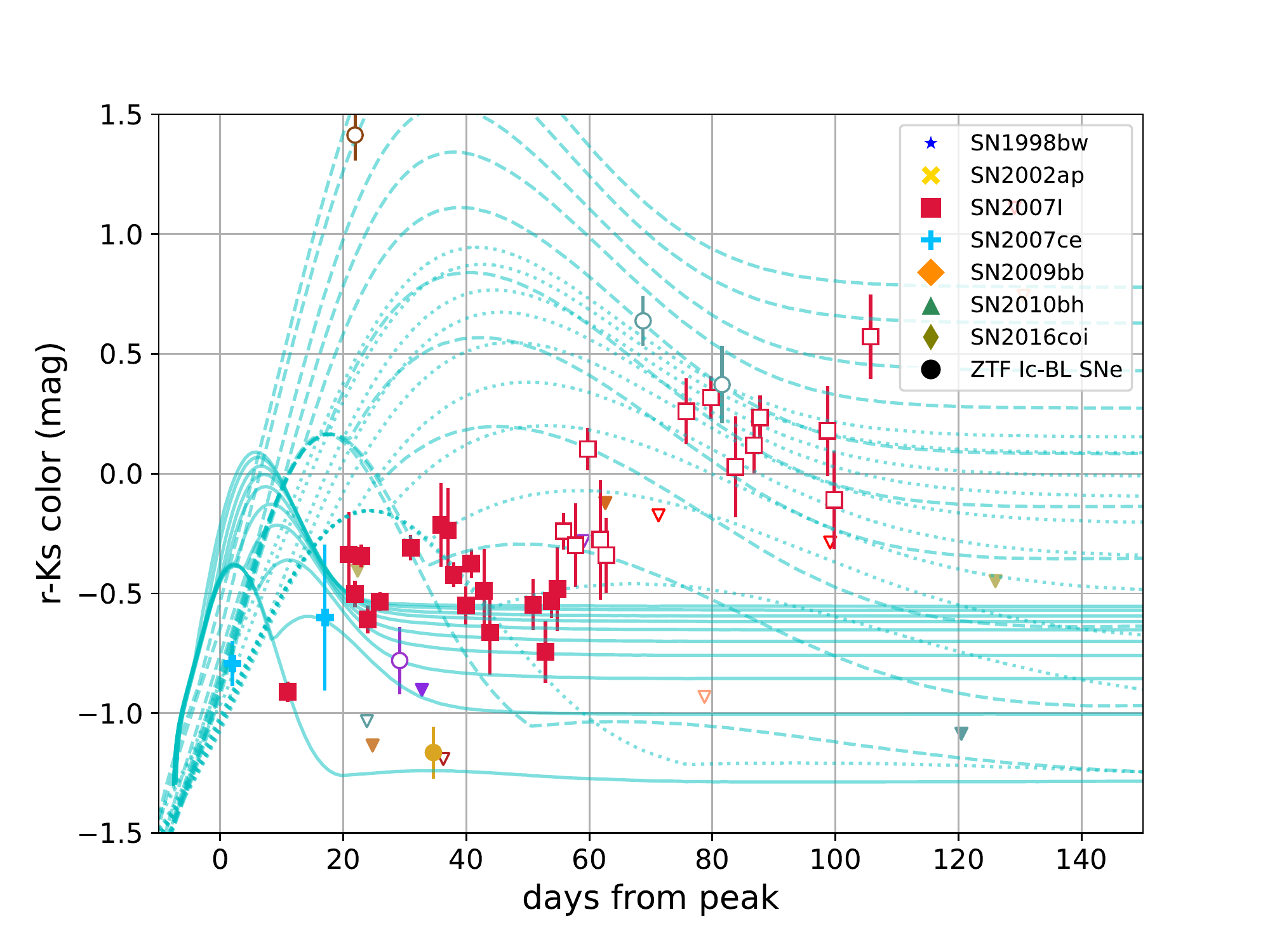}
    \caption{$r-J$, $r-H$, and $r-K_s$ color evolution plots for the \rp{} enriched models for a representative set of model parameters, compared to color measurements for the SNe in our sample. Each model is shown in a separate linestyle: ``solid": \mej{}$=7.93 \rm M_{\odot}$, \vej{}$=0.25c$, \mni{}$=0.85 \rm M_{\odot}$, ``dotted": \mej{}$=2.62 \rm M_{\odot}$, \vej{}$=0.038c$, \mni{}$=0.39 \rm M_{\odot}$, and ``dashed": 
    \mej{}$=1.00 \rm M_{\odot}$, \vej{}$=0.033c$, \mni{}$=0.07 \rm M_{\odot}$. When possible, the $r-X$ color of observed SNe was estimated using either concurrent $r-$band photometry or the closest optical photometry within 3\,days of a given NIR datapoint (filled markers).
    Otherwise, the $r-$band magnitude is extrapolated from a stretched and scaled light curve of SN\,2020bvc (unfilled markers). Left: Fixing the mixing fraction to a moderate value of 0.3, we vary \rp{} ejecta masses [0.01, 0.03, 0.08, 0.13]\,M$_{\odot}$. Right: Fixing the \rp{} mass to a conservative value of 0.01\,M$_{\odot}$, we vary the mixing coordinate from 0.1 to 0.9. In general, the objects in our sample appear to have bluer colors relative to the models (with the exception of SN\,2007I).}
    \label{fig:color}
\end{figure*}

In Figure~\ref{fig:color}, we plot colors with respect to the $r-$band as $r-X$ ($X=J,H,K_s$) for \rp{} enriched models corresponding to the following parameters: ``high mass, high velocity": \mej{}$=7.93 \rm \,M_{\odot}$, \vej{}$=0.25c$, \mni$=0.85\,\rm M_{\odot}$ (solid line), ``medium mass, medium velocity": \mej$=2.62\,\rm M_{\odot}$, \vej{}$=0.038c$, \mni$=0.39 \rm\,M_{\odot}$ (dotted line), and ``low mass, low velocity": \mej$=1.00 \rm\,M_{\odot}$, \vej{}$=0.033c$, \mni$=0.07 \rm\,M_{\odot}$ (dashed line). This set of models illustrates how different combinations of assumed parameters affect the color curves. These specific model grids were used to fit the light curves of three objects in our sample and represent the broad range of explosion parameters derived for our SNe.

% First, we compare our color measurements to \rp{}-free models corresponding to the three grids (see Figure~\ref{fig:norp_color}). 
We use these color evolution predictions from the models to compare against the optical-NIR colors of our SNe. Our $r-X$ color measurements rely on two different methods: if there is an optical data point within three days of the NIR data point, we compute the color difference directly (filled circles); otherwise, we estimate the color by subtracting the NIR photometry from a scaled and shifted optical template (open circles). We construct this template from the light curve of SN\,2020bvc, one of the SNe with the most well-sampled light curves, and then compute the shift and scale factors needed for the template to fit the data.  For the cases in which the optical model does not fit the optical light curve perfectly, there can be a systematic offset between the open and closed circles. For example, the estimated $r-K_s$ color of SN\,2019xcc (Figure~\ref{fig:color}, bottom panel) is $>1$\,mag, but this is likely attributed to the fact that there is no concurrent optical photometry along with the $Ks-$band data point, and the optical light curve fades much faster than that of SN\,2020bvc.

The predicted $r-J$ colors for \rp{} collapsar light-curve models range from $r-J\sim-$0.5\,mag to $r-J\sim$1.5\,mag. In the lefthand-side panels of Figure~\ref{fig:color}, we fix the mixing fraction to a moderate value of \xmix{}$=0.3$ and vary the amount of \rp{} ejecta mass. On the righthand-side panels, we fix the \rp{} ejecta mass to 0.01M$_{\odot}$ and vary the mixing fraction. The amount of reddening in the model light curves is more strongly affected by the amount of mixing assumed; even for the lowest value of $M_{r\mathrm{p}}$, we find prolific reddening predictions for high mixing fractions relative to models with moderate mixing fractions and high \rp{} yield. 

However, \rp{} enrichment is not the only factor affecting colors; unenriched SN models also have a range of colors, depending on their masses, velocities, and nickel production. Even amongst different models with identical \rp{} composition, color evolution can be sensitive to the explosion properties assumed. Here, the ``high mass, high velocity" model set also shows the most dramatic reddening predictions for models that have extreme mixing; in general, higher mass models tend to show larger $r-X$ colors.

Late-time interaction with the circumstellar medium is also known to affect the color evolution of SNe \citep{Ben-Ami2014, Kuncarayakti2022}. While this is a rare phenomenon in SNe Ic-BL, SN\,2022xxf showed evidence for a clear double-peaked light curve and narrow emission-line profiles in the later-phase spectra characteristic of interaction with a H/He-poor CSM \citep{Kuncarayakti2023}. SN\,2022xxf also exhibited a dramatic red-to-blue color evolution as a result of interaction. We do not observe any of the above evidences for CSM interaction in our Ic-BL SNe, and therefore consider it unlikely that interaction could account for bluer colors at later times.

When comparing our color measurements against \rp{} models, we find that several of our objects show colors similar to the \rp{} models with minimal mixing. However, after 50\,days post-peak, our detections and upper limits altogether strongly suggest that our SNe are brighter in the optical compared to the NIR. In particular, as many of our SNe are detected in the $J-$band over a wide range of phases, we can constrain the $r-J$ color to $<\,-0.5$ after 50\,days post-peak. On the other hand, only one object shows $r-J$/$H$/$K_s$ colors $\sim$\,0.5\,mag: SN\,2007I. In particular, SN\,2007I exhibits an increase in its $r-J$ color until about 60\,days. 

While these empirical color comparisons can be useful for identifying any obvious reddening signature that could be a smoking gun for \rp{} enrichment, more detailed fitting is required to establish whether or not these SNe are \rp{} enriched. Hence, in the next section, we describe our detailed model fitting aimed at determining whether there is room for an \rp{} contribution to their light curves.

\section{Results of light curve Model Fitting} \label{sec:results} 

% In Figures~\ref{fig:mhigh_lcs}, \ref{fig:mmed_lcs} and \ref{fig:mlow_lcs} we show the ZTF optical light curves of each of the SNe in our sample, supplemented with follow-up imaging from the Liverpool Telescope, LCO Sinistro and Spectral imagers, as well as SEDM.  
% Due to ZTF's imaging cadence and sensitivity and the redshift cuts in our sample selection, the optical light curves tend to have regular detections until $\sim$60\,days post maximum light. For some SNe, the light curves are missing patches of data due to observability constraints or weather. On the other hand, our \rp{} enriched and \rp{}-free models last up to 200\,days in the optical and IR. 
% Therefore we use the \rp{}-free model from the grid chosen for each SN as a proxy for its late-time $r$-band evolution. 

% To quantitatively determine whether \rp{} contribution is required to explain the light curves of SNe Ic-BL, we first only considered \rp{} free models for our light curve fitting. For each SN, we fit the light curves to the model suite, comprising a subset of the full model grid centered around the independently derived explosion parameters (i.e $M_{56}$, $M_{\rm ej}$, $\beta_{\rm ej}$) and encompassing the 1$\sigma$ parameter uncertainties. Relative to the other explosion parameters, we intentionally chose a wide parameter range for $M_{56}$ to compensate for the fact that these light curve models tend to underestimate the model flux for a given quantity of \iso{Ni}{56}.  

To quantitatively determine whether \rp{} contribution is required to explain the light curves of SNe Ic-BL, we perform nested sampling fits over multi-dimensional parameter space spanned by the \rp{} enriched models. However, in order to perform the fitting, we need a distribution over functions with a \emph{continuous} domain. Since these \rp{} models are discretely parameterized, we invoke gaussian process regression (GPR) to predict light curves from the training set (which are the \rp{} enriched models, in this case) for each linear combination over the continuous ranges of parameters.

We first considered the full grid of \rp{} enriched models from \citet{Barnes2022}. For objects for which it was possible to estimate the total ejecta mass and kinetic energy, we select grids where the parameters fall within the following bounds: $\mej{} \in (M_{\rm ej, 0}-3\sigma, M_{\rm ej, 0}+3\sigma)$, $\vej{} \in (\beta_{\rm ej,0}-3\sigma, \beta_{\rm ej,0}+3\sigma)$, and $M_{56} \in (M_{56}-3\sigma, M_{56}+10\sigma)$, where $M_{\rm ej, 0}$, $\beta_{\rm ej, 0}$, and $M_{56}$ are the independently derived explosion properties for the supernovae (see Table~\ref{tab:basic_properties}). We changed the upper bound on \mej{} (\vej{}) to $M_{\rm ej, 0}+10\sigma$ ($\beta_{\rm ej,0}+10\sigma$) for those SNe for which only a lower limit on those quantities was derived. We use the entire range of parameters in the grid for \mrp{} and \xmix{}.

% We followed a similar methodology to define the subset of the model space in which to conduct model fitting for each object. The only differences between the two searches are 1) considering a smaller range of values for explosion parameters (especially \iso{Ni}{56}) and 2) including $M_{r\mathrm{p}}$ and \xmix as additional fitting parameters. 

% Our modified constraint on explosion parameters is purely motivated by the need to reduce the grid size to conserve memory, since adding the \rp{} parameters makes the grid 40$\times$ larger, though in all cases our grids encompass the 1$\sigma$ uncertainty on explosion parameters.

We then perform singular value decomposition on each light curve in the model grid tailored to each supernova and interpolate between model parameters using scikit-learn's GPR package, sampling between $-5$ and 200 days relative to the supernova peak in a similar fashion to \citet{Coughlin2019} and \citet{Pang2022}. We allow GPR to interpolate the range of \rp{} ejecta masses and mixing fractions between \mrp{}$=$\,0.01\,M$_{\odot}$, \xmix$=$\,0.1 (which are technically the lowest values in the \rp{} enriched grid) and \mrp{}$=0.00$, \xmix{}$=0.0$, though we do not allow it to exceed the maximum values for these quantities (i.e. \mrp{}$\leq$\,0.15\,M$_{\odot}$ and \xmix$\leq$\,0.9). We limit interpolation of the remaining explosion parameters within the maximum and minimum bounds of the original grid. For a given set of explosion parameters (\mej{}, \vej{}, \mni{}), each grid also contains an \rp{}-free model. 

We compute a likelihood function based on the interpolated light curve models and our multi-band ZTF forced photometry, follow-up photometry, and WIRC photometry. Since the errors from GPR are small (i.e. they well-approximate the original model grid), we assume a systematic fitting uncertainty of 0.5\,mag in the NIR bands and a fitting uncertainty of 1.0\,mag in the optical. We converged upon a 1.0\,mag systematic uncertainty in the optical after evaluating how different assumed errors affect the fit quality. The difference in the systematic errors is motivated by the fact that the NIR bands, rather than the optical bands, are a stronger determinant of whether there is evidence for \rp{} production. Furthermore, these assumptions on the systematic error compensate for the finer sampling in the optical bands relative to the NIR.  In the likelihood calculation, we also impose a condition that rejects samples with a linear least squares fitting error worse than 1.0\,mag. For the \rp{} enriched model fits, our prior also restricts the inference of parameters within the ranges of the grid (0.0\,$\leq \rm x_{\rm mix} \leq$\,0.90; 0.0\,M$_{\odot} \, \leq \rm M_{r\mathrm{p}} \leq\,0.15 \, \rm M_{\odot}$) and within physical constraints (i.e. \mrp{}$<$ \xmix (\mej{} - \mni{})). We impose this upper limit on M$_{r\mathrm{p}}$ to satisfy the requirement that the \rp{} enriched core also contains \iso{Ni}{56} (see Figure~8 of \citealt{Barnes2022}). Finally, we employ \texttt{PyMultinest}'s \citep{Buchner2014} nested sampling algorithm to maximize the likelihood and converge on the best fit parameters and their uncertainties. 

\emph{Most of the SNe in our sample show no compelling evidence for r-process production.} In our model fits, the general trend we observe is that the best fit consistently under-predicts the peak of the optical light curve, while performing better at predicting the NIR flux. In some cases, the under-prediction is egregious, while in other cases it is more modest. In general under-prediction indicates that the optical-NIR color of the SN is actually \emph{bluer} than predicted by the models, providing stronger evidence towards favoring \rp{}-free models over the enriched models. As mentioned earlier, as M$_{r\mathrm{p}}$ increases, the NIR light curve gets brighter; as \xmix{} increases, the optical light curve peak diminishes and the optical flux is suppressed more at later times.

To quantitatively assess the fit quality, we compute $\chi^2$ values between the best-fit model and the data points. We adopt the convention that if $\chi^2 > \chi^2_{crit}$ (at the $>5$\% level), we can reject our hypothesis that these SNe are well-described by the best-fit \rp{} enriched model. Therefore, given that our fits have 4 degrees of freedom, a $\chi^2 > 9.49$ is indicative that the \rp{} enriched models are poor fits to the data. Applying this criteria suggests that SN\,2018gep, SN\,2019xcc and SN\,2020rph are very unlikely to harbor \rp{} material in their ejecta.

Similarly, we select the subset of objects for which $\chi^2 < \chi^2_{crit}$ for a p-value of 0.90 ($\chi^2_{crit}=1.06$, for 4 degrees of freedom). Based on their $\chi^2$ values, SN\,1998bw, SN\,2007ce, SN\,2018kva, SN\,2019gwc, SN\,2020lao, SN\,2020tkx, SN\,2021xv and SN\,2021bmf show the most convincing fits to the \rp{} enriched models. 
Upon visual inspection of the remainder of the light curve fits, we find that none of the other objects are well-described by the \rp{} model predictions.
We display the corner plots showing the posterior probability distributions on the derived parameters in Fig.~\ref{fig:rp_corner} for the objects passing our $\chi^2$ cut, along with the best-fit light curves in Fig.~\ref{fig:rp_lcs}.

\begin{figure*}
\centering
\begin{subfigure}
    \centering
    \includegraphics[width=0.55\textwidth]{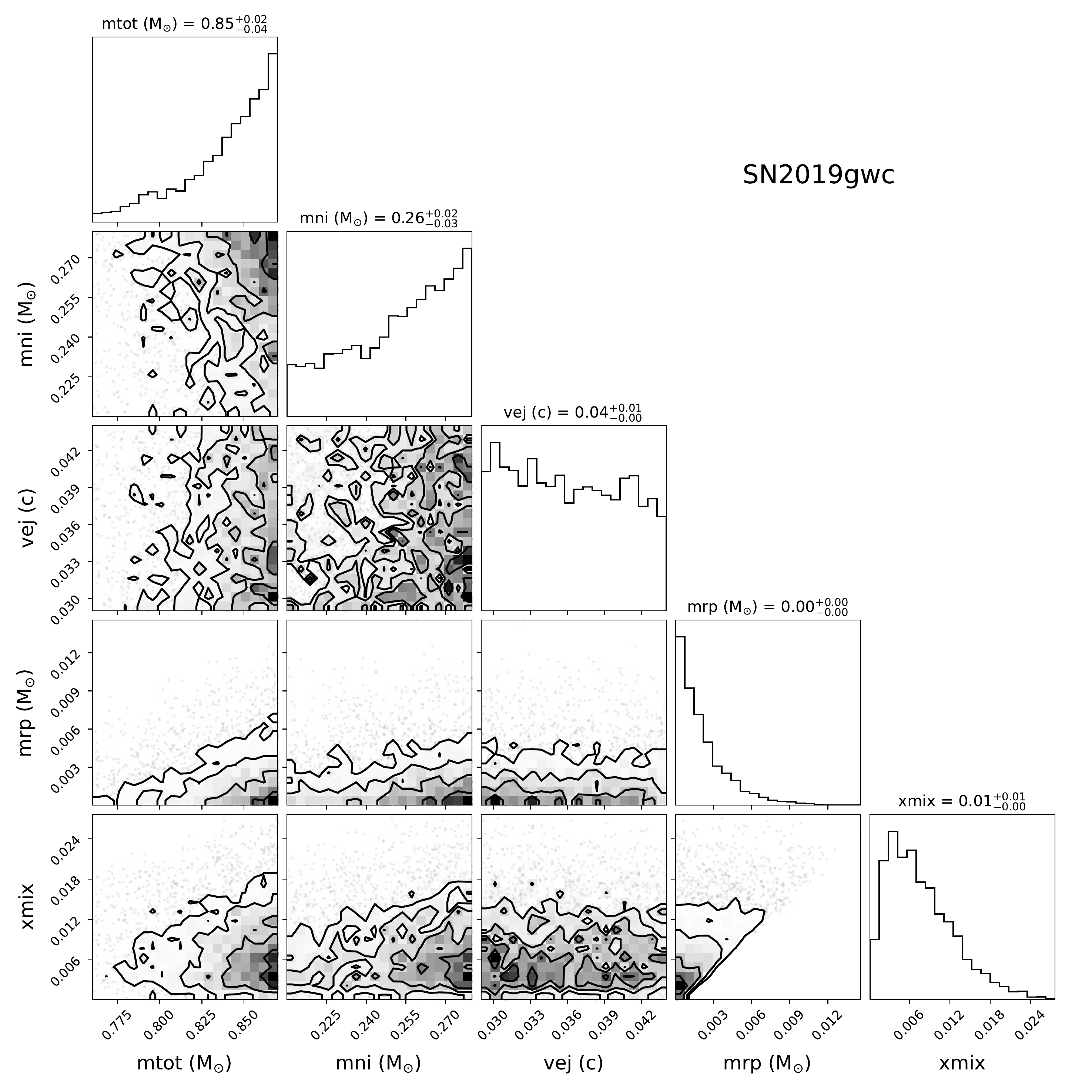}\includegraphics[width=0.55\textwidth]{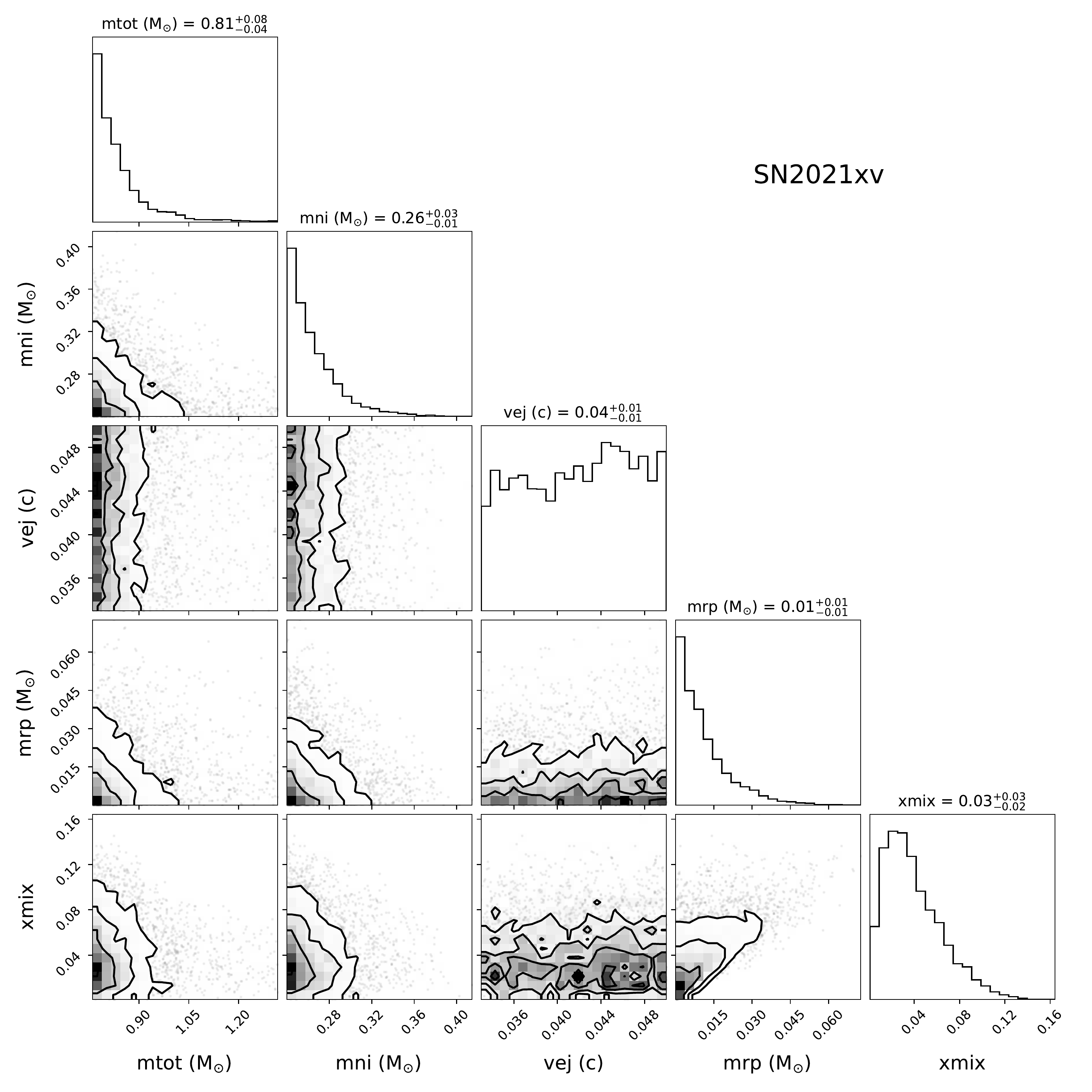}
    \end{subfigure}
    \begin{subfigure}
    \centering
    \includegraphics[width=0.55\textwidth]{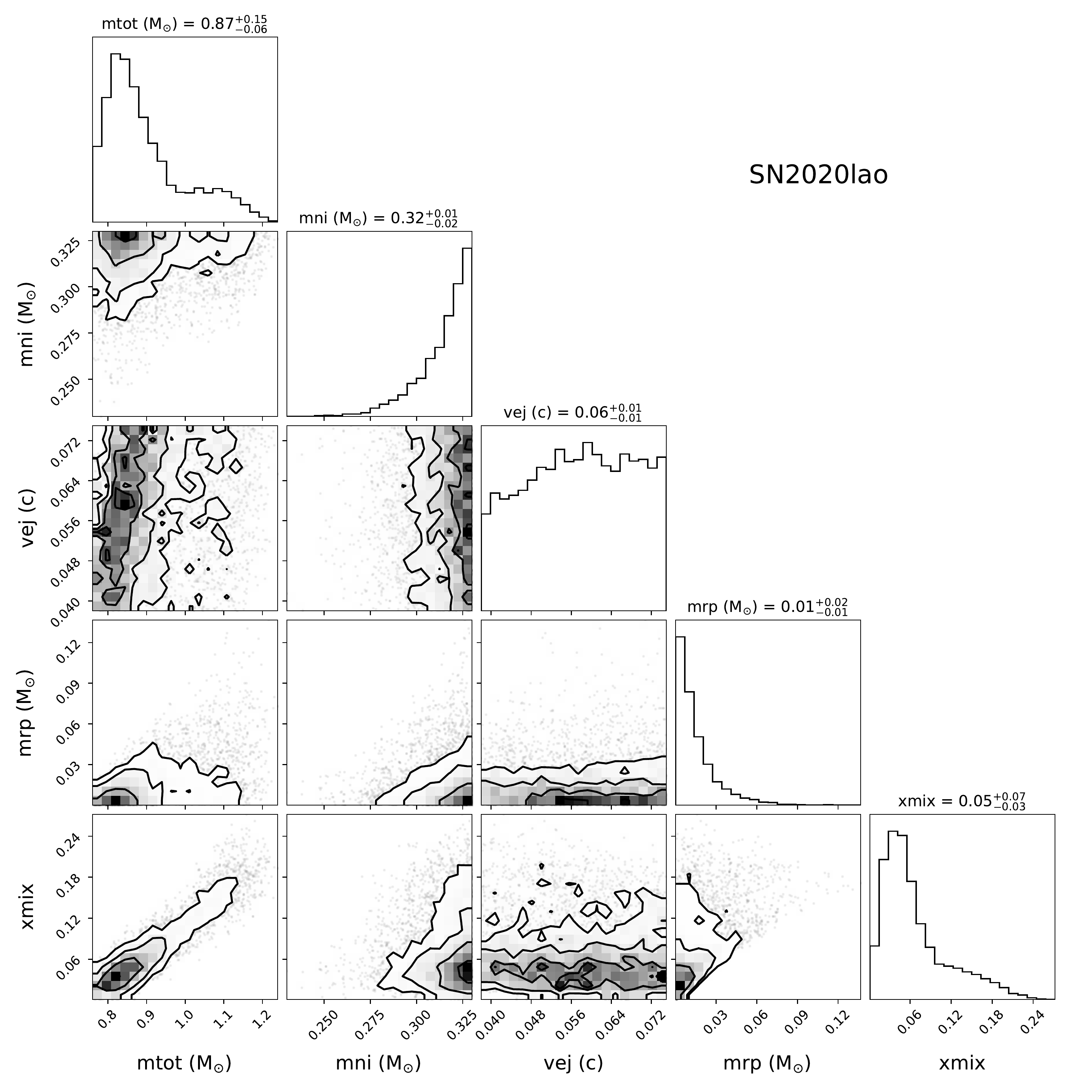}\includegraphics[width=0.55\textwidth]{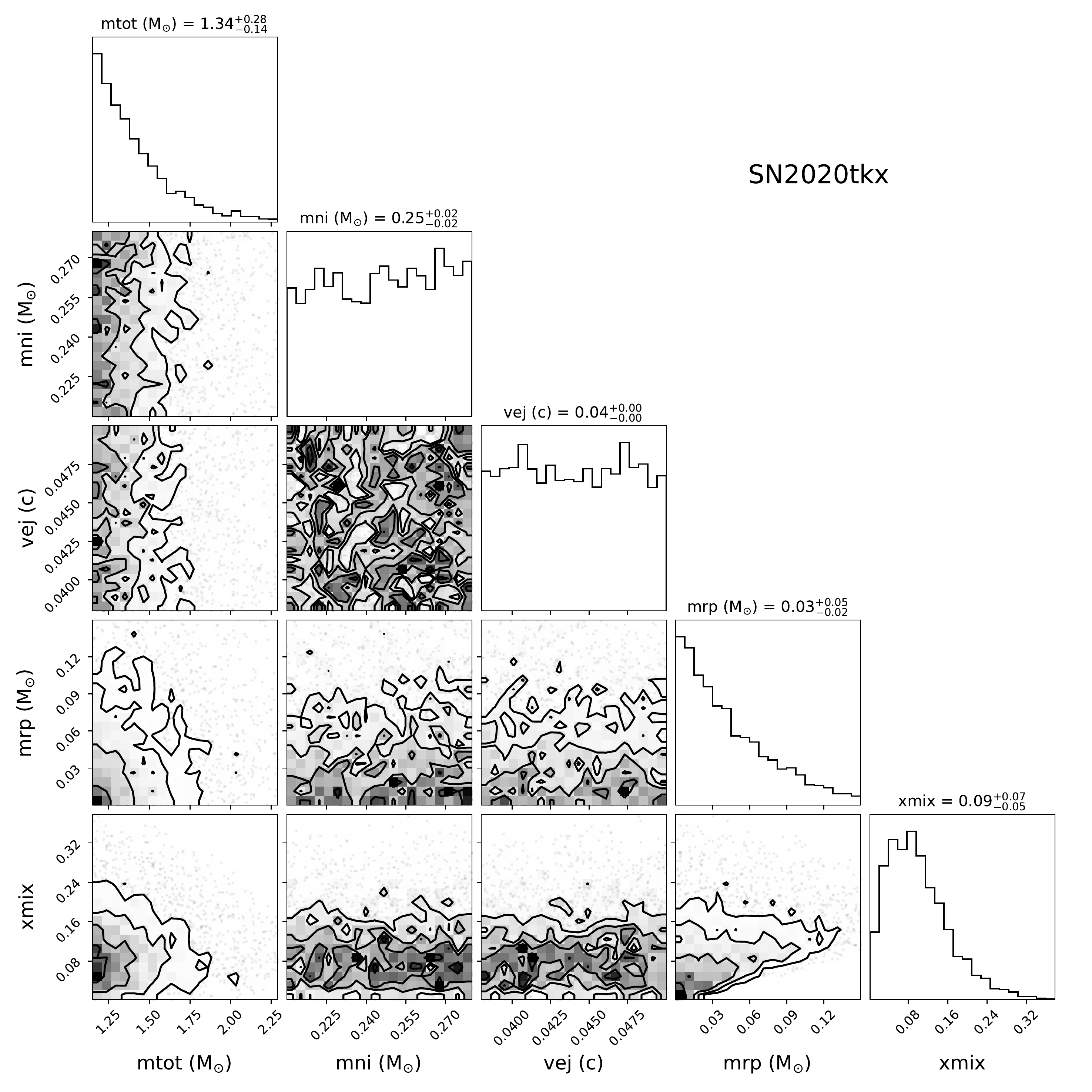}
    \end{subfigure}
\end{figure*}
\begin{figure*}
    \begin{subfigure}
    \centering
    \includegraphics[width=0.55\textwidth]{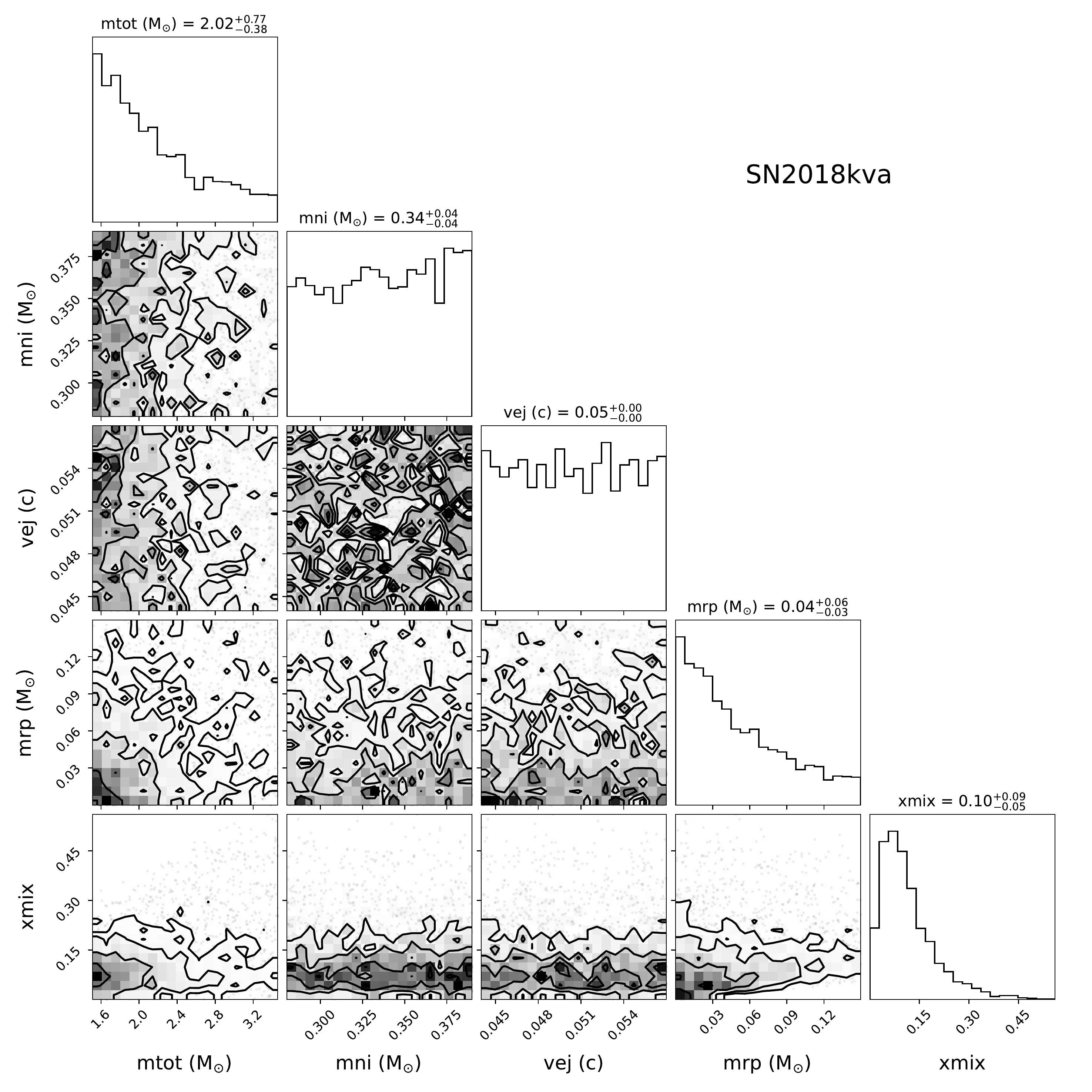}\includegraphics[width=0.55\textwidth]{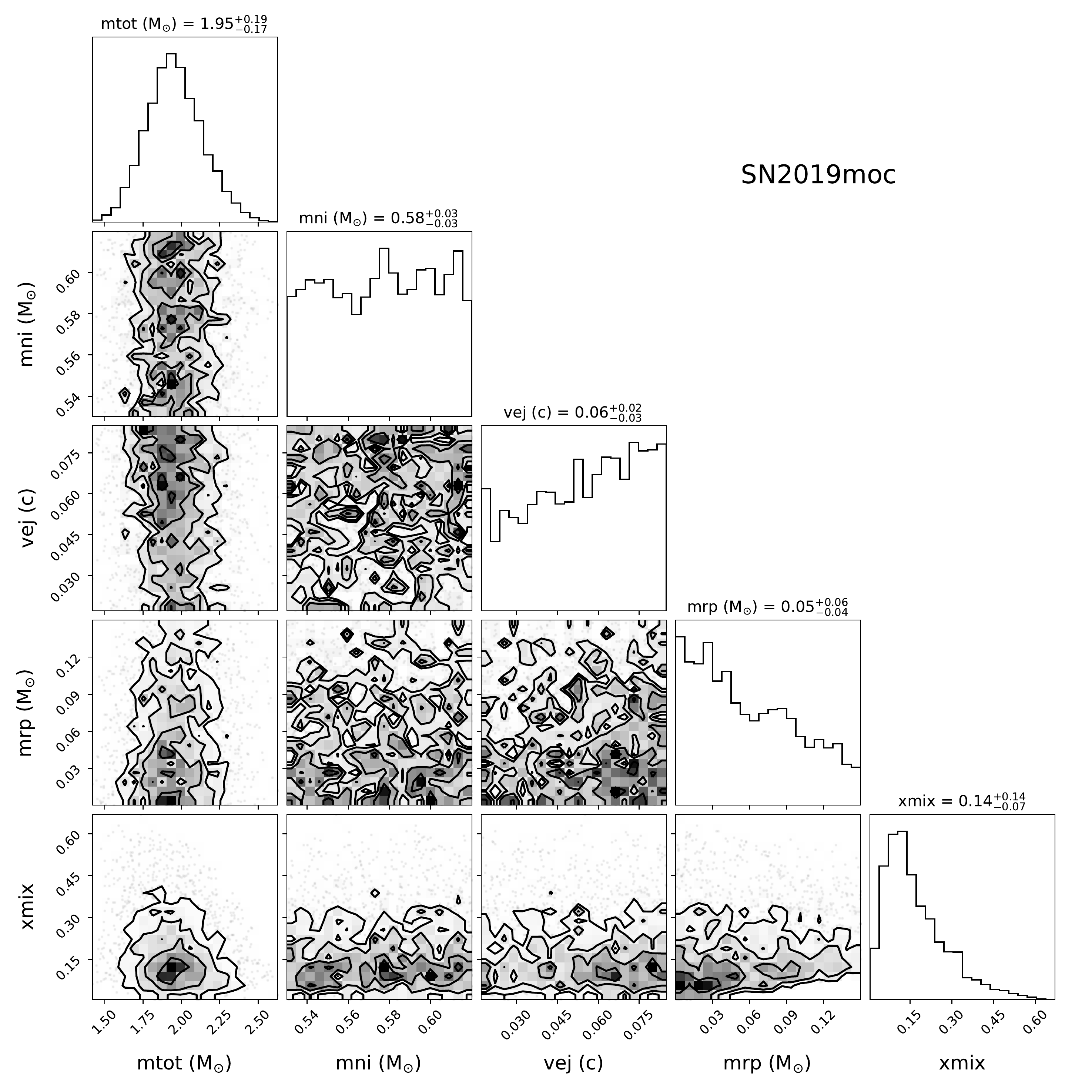}
    \end{subfigure}
    \begin{subfigure}
    \centering
    \includegraphics[width=0.55\textwidth]{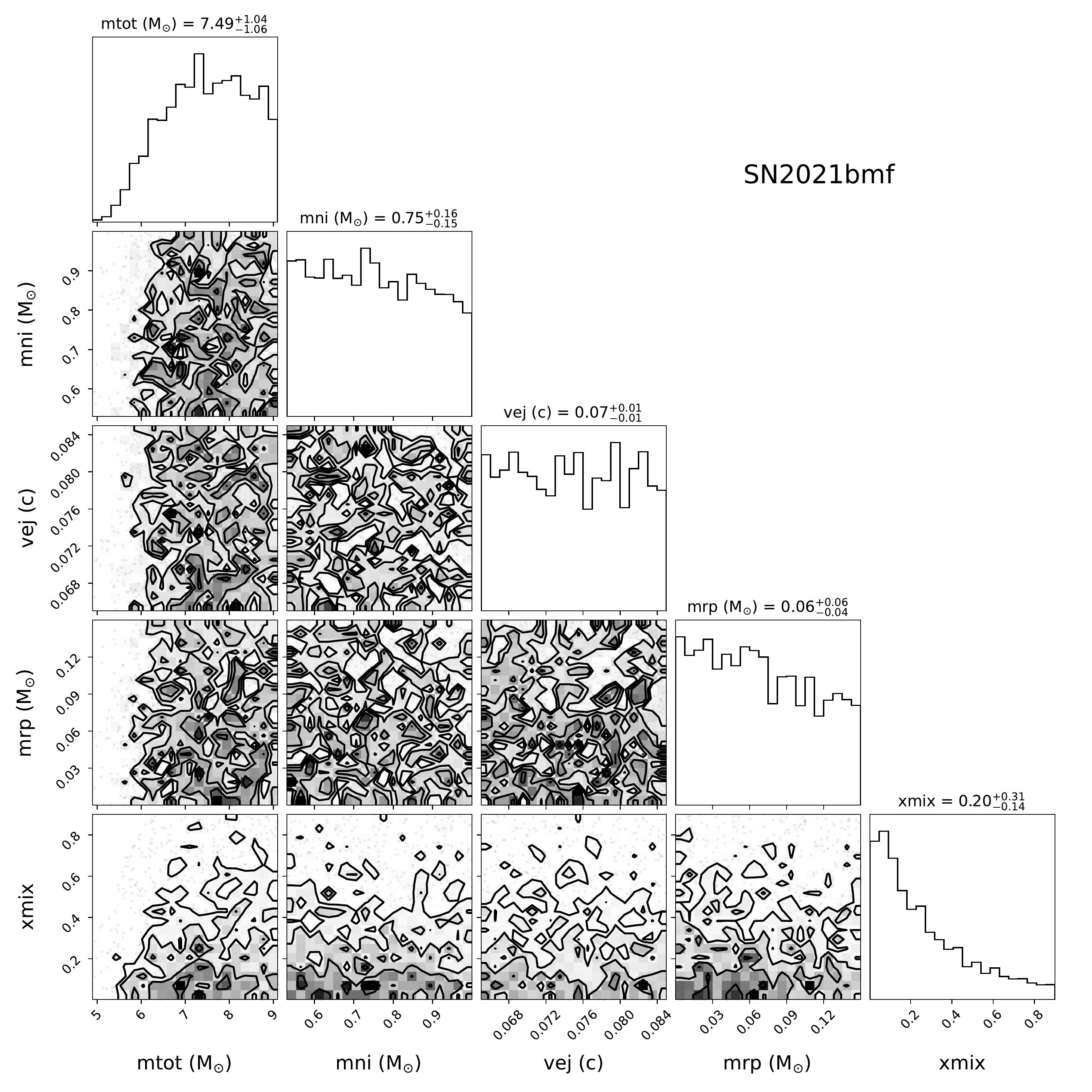}\includegraphics[width=0.55\textwidth]{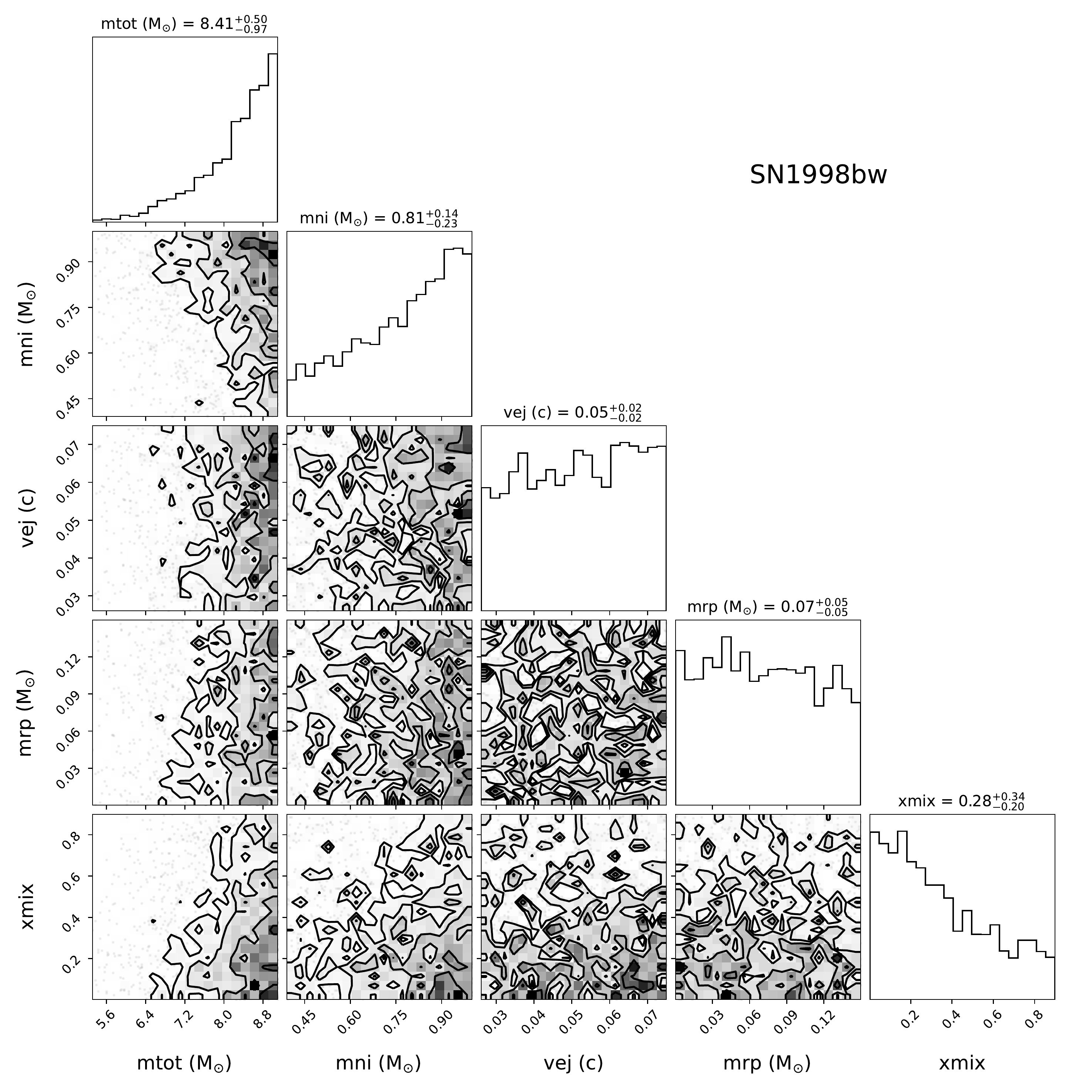}
    \end{subfigure}
\end{figure*}
\begin{figure*}
    \begin{subfigure}
    \centering
    \includegraphics[width=0.55\textwidth]{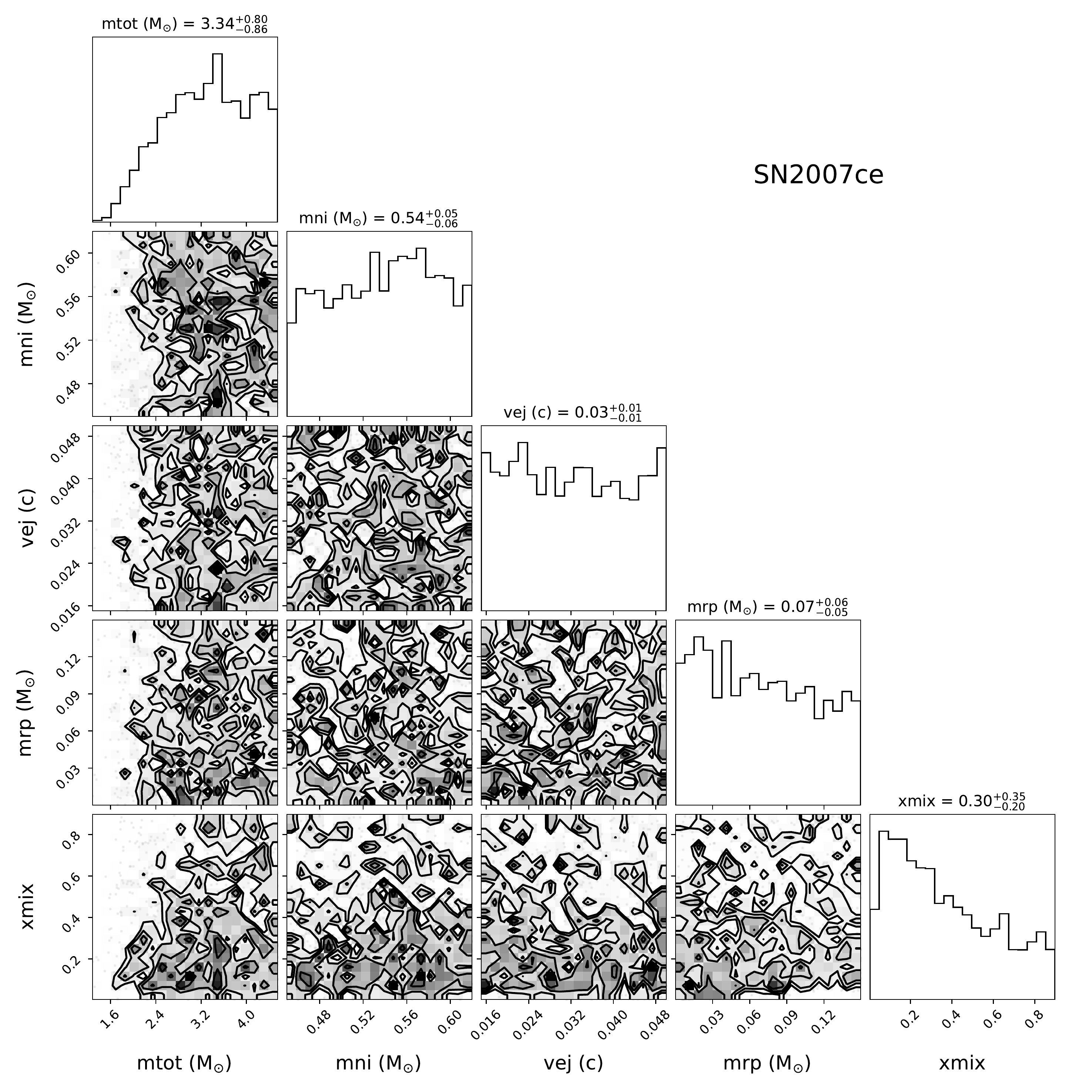}\includegraphics[width=0.55\textwidth]{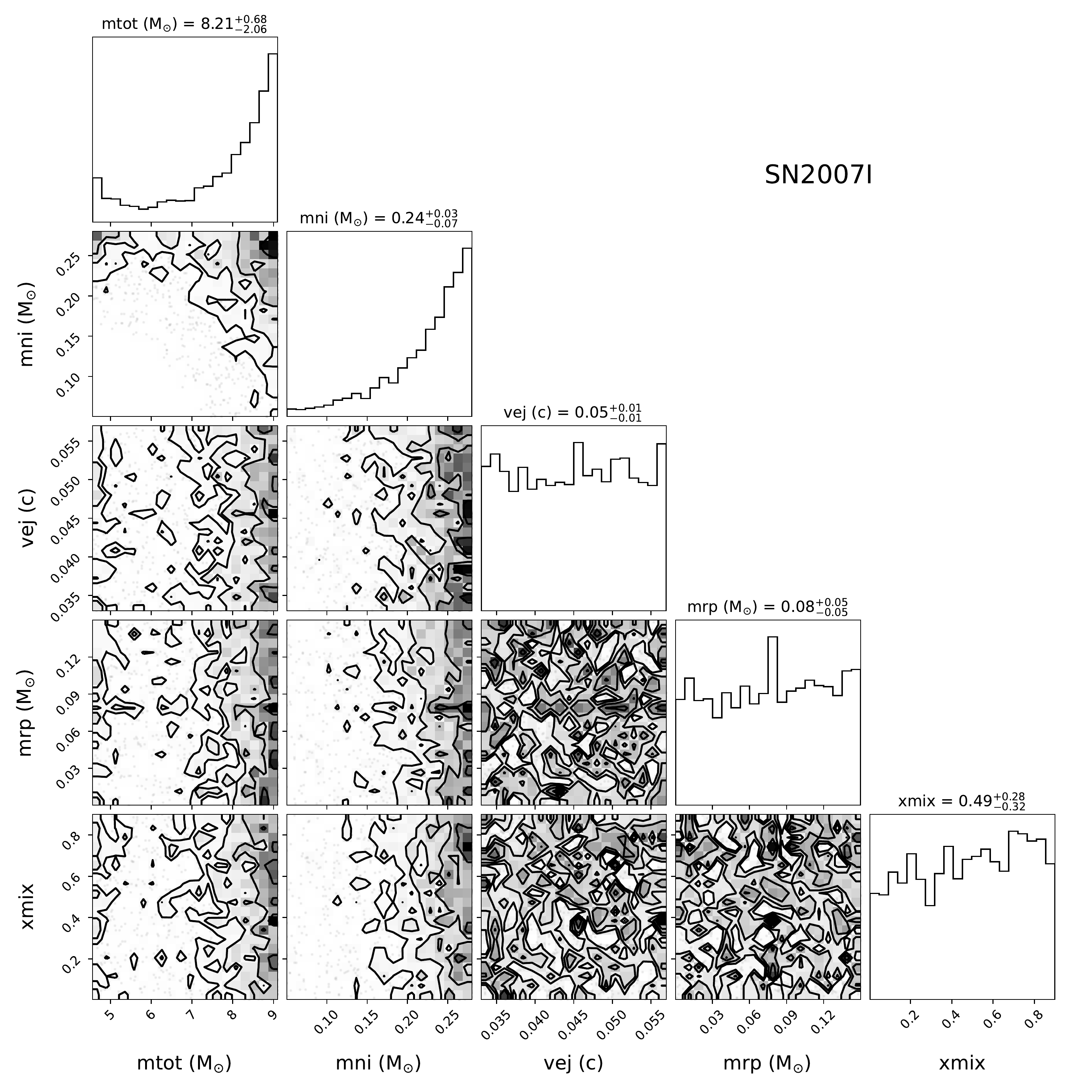}
    \end{subfigure}
    \caption{Corner plots showing the posterior probability distributions for each of the parameters in the \rp{} enriched models for the subset of objects satisfying our $\chi^2$ cut, ordered by the amount of \rp{} mass inferred. The posterior probability plots are more well-constrained for the objects with low inferred \mrp{}; in the remaining cases, the posterior distributions are poorly constrained. \mej{} and \vej{} inferred here are generally in agreement with \texttt{HAFFET}, but discrepancies exist in the amount of nickel mass inferred. }
    \label{fig:rp_corner}
\end{figure*}

\begin{figure*}
    \centering
    \begin{subfigure}
    \centering
    \includegraphics[width=0.5\textwidth]{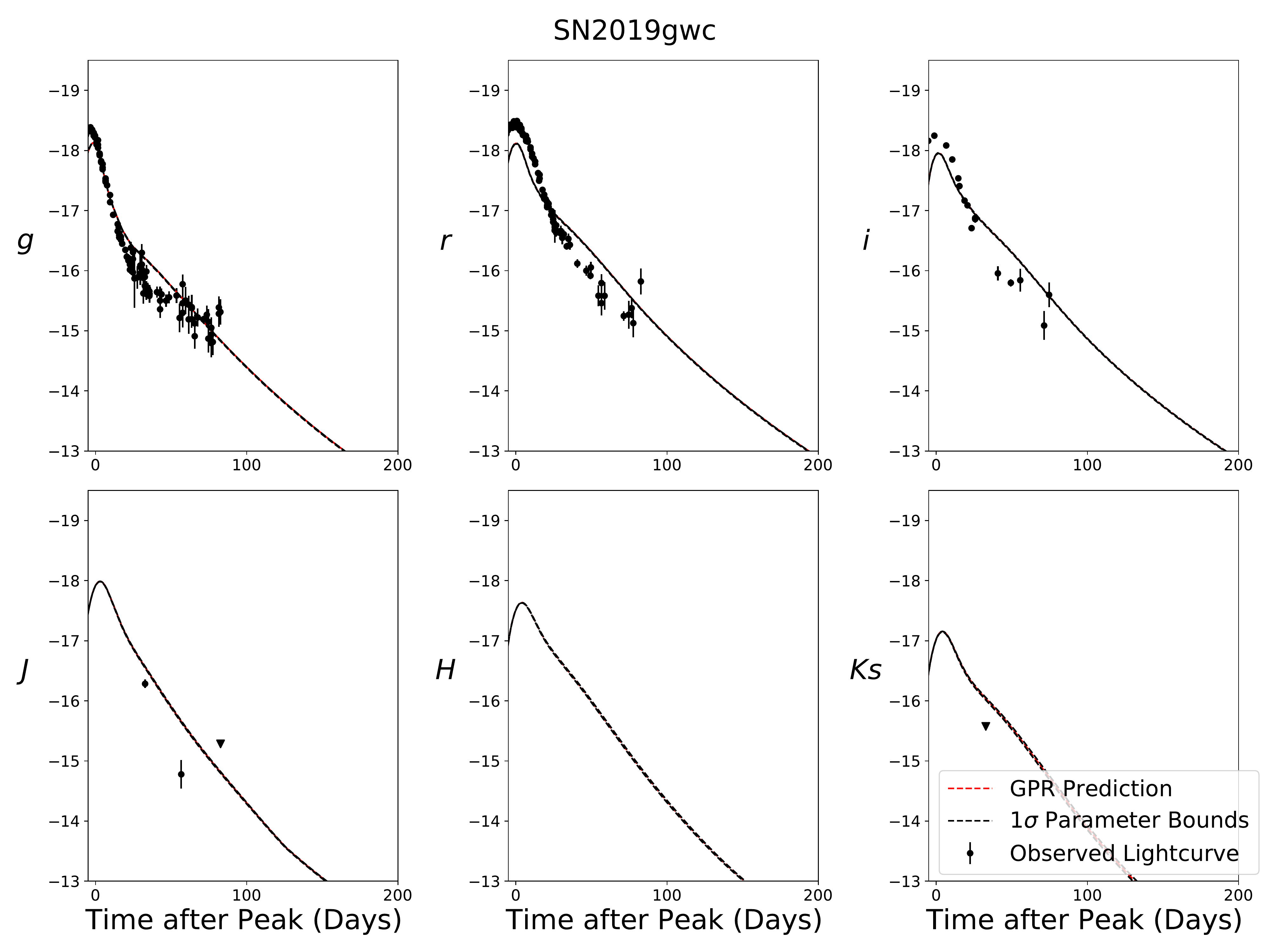}\includegraphics[width=0.5\textwidth]{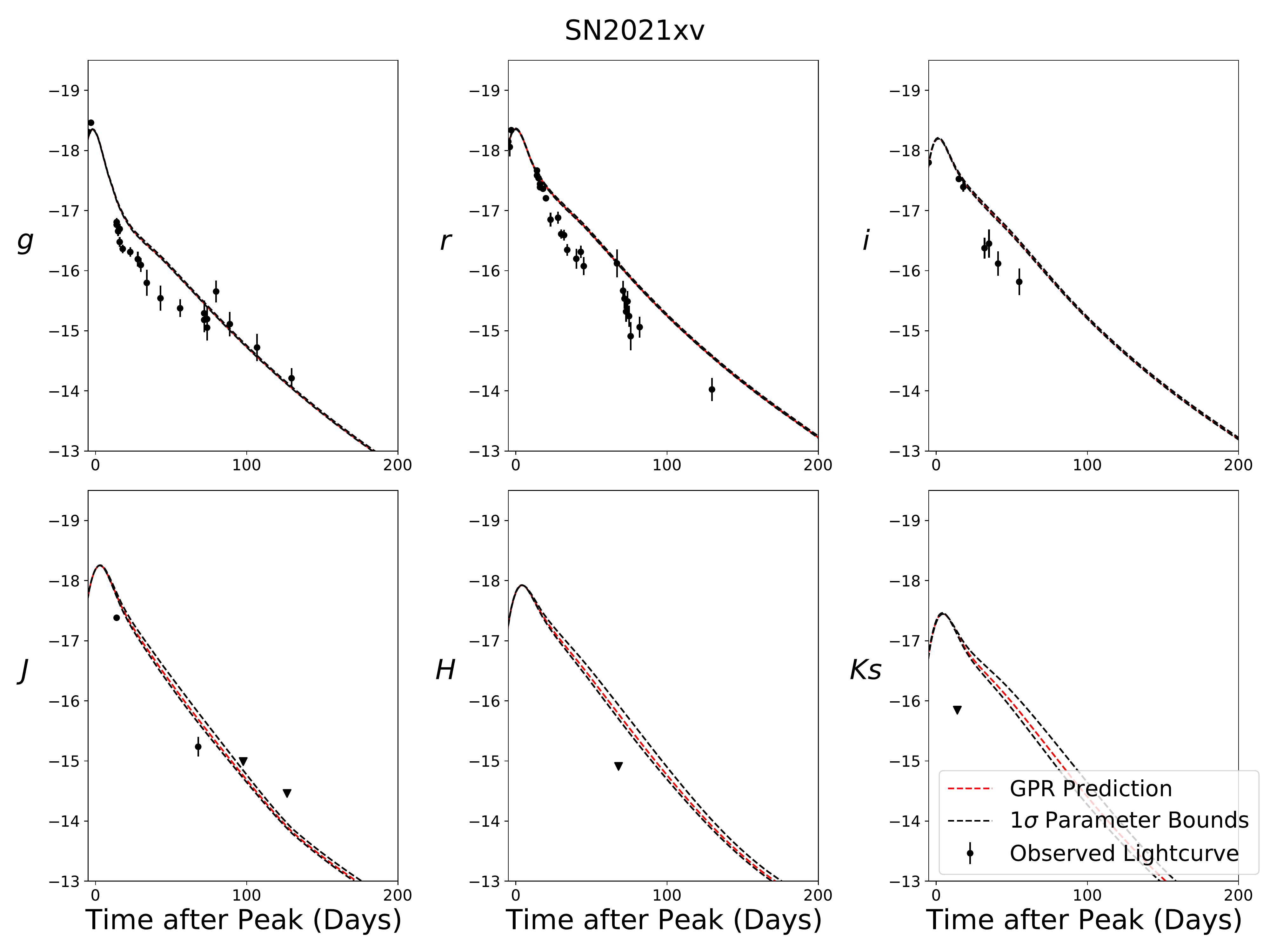}
    \end{subfigure}
    \begin{subfigure}
    \centering
    \includegraphics[width=0.5\textwidth]{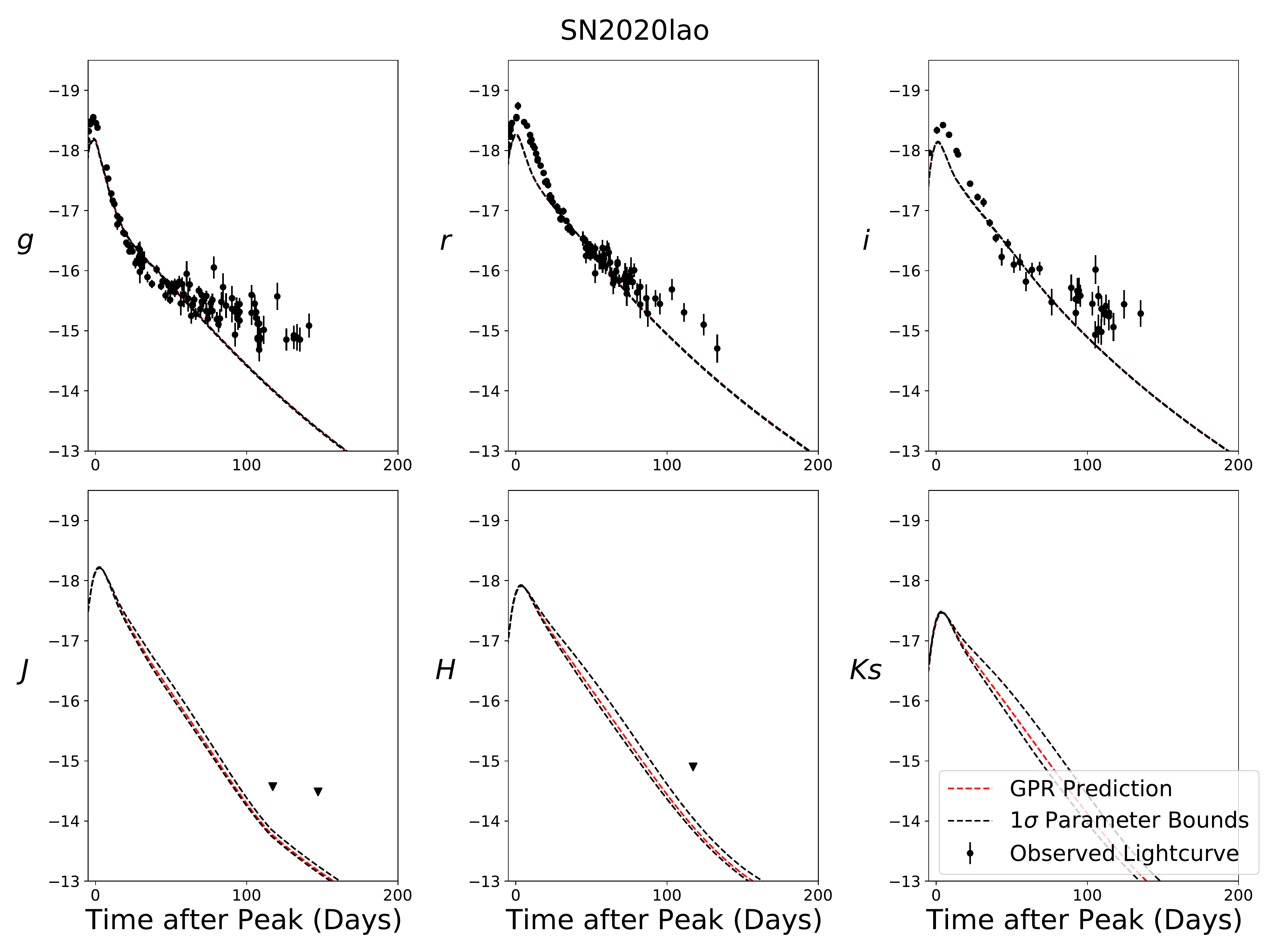}\includegraphics[width=0.5\textwidth]{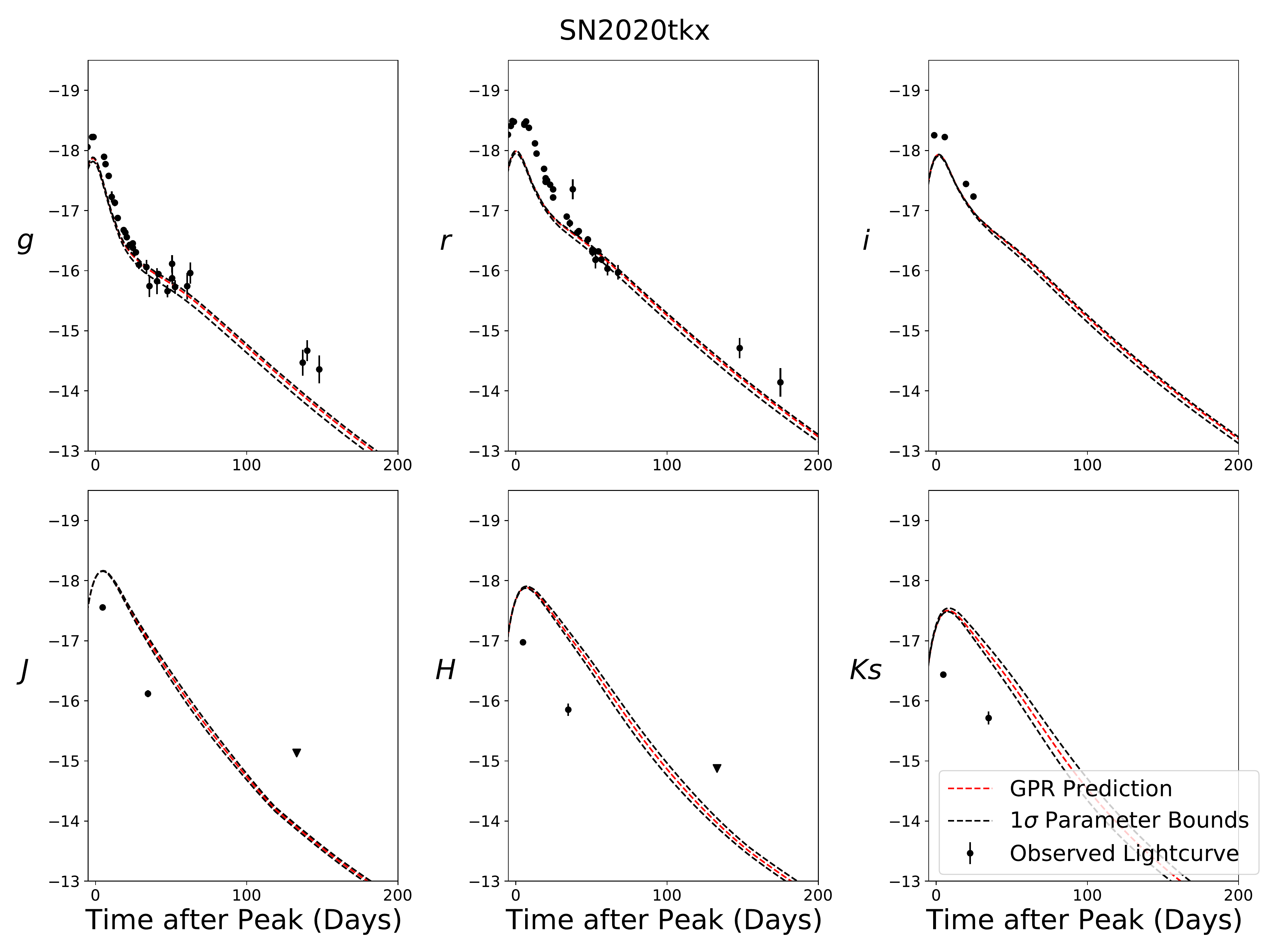}
    \end{subfigure}
    \begin{subfigure}
    \centering
    \includegraphics[width=0.5\textwidth]{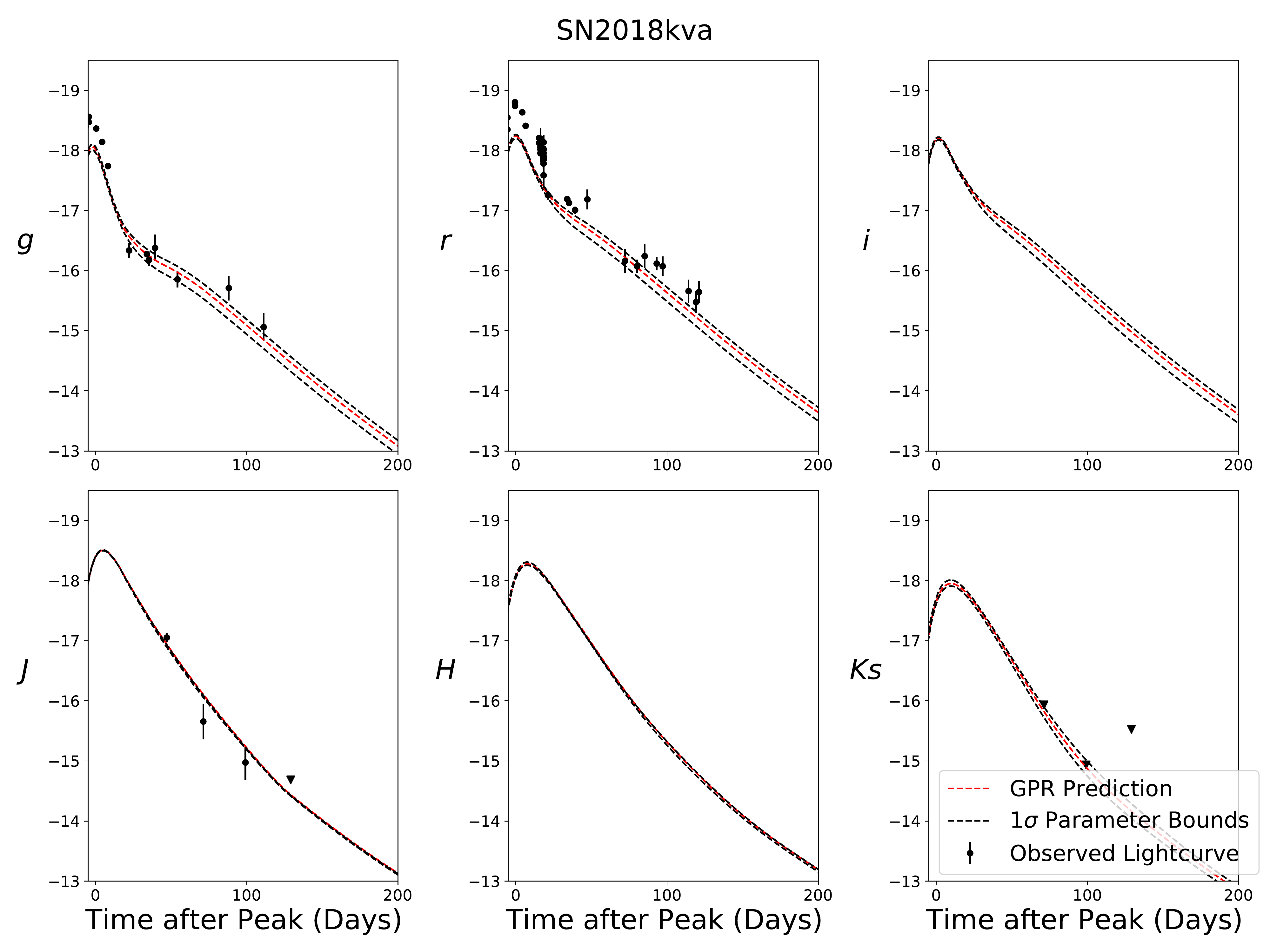}\includegraphics[width=0.5\textwidth]{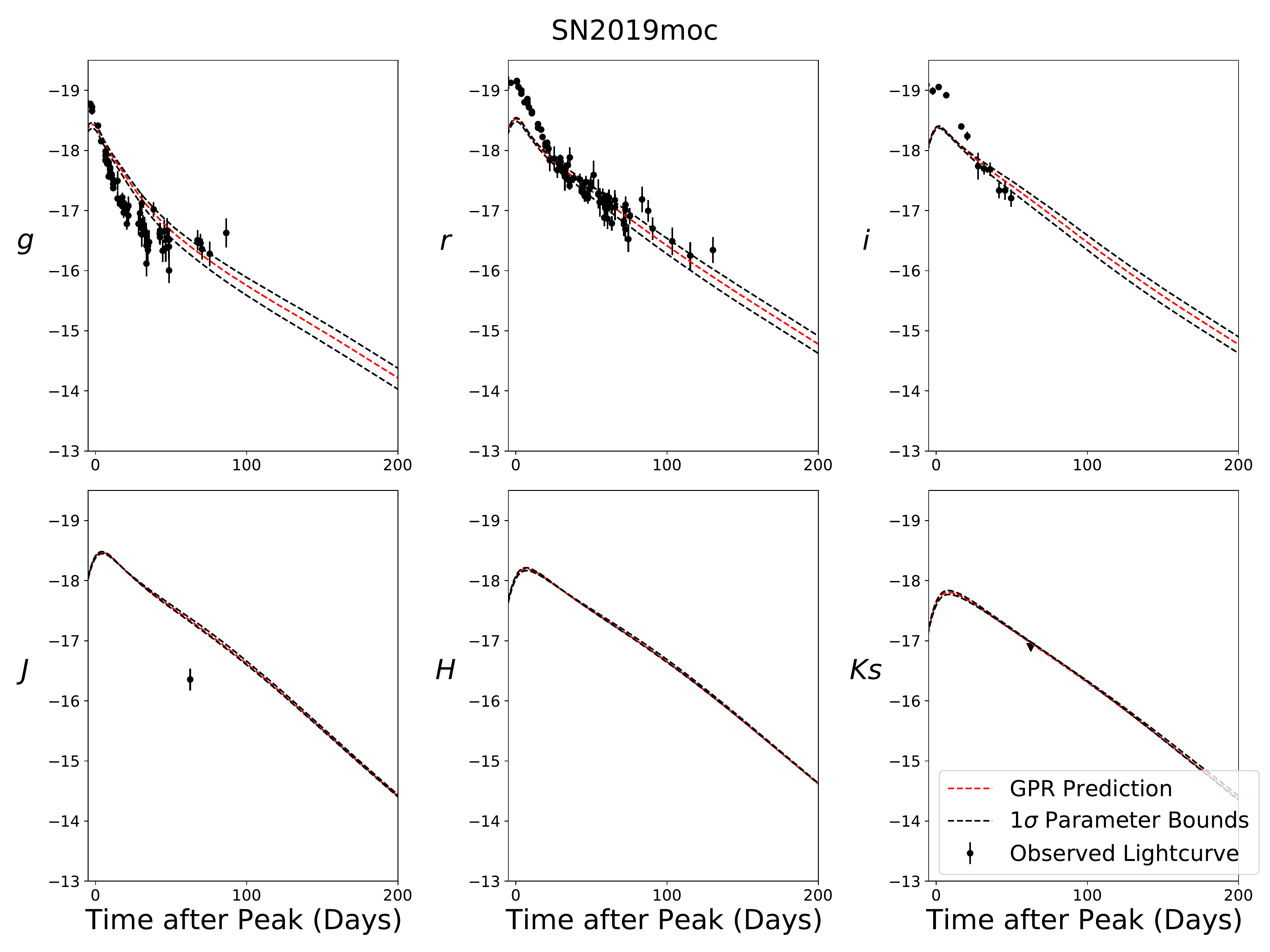}
    \end{subfigure}
\end{figure*}
\begin{figure*}
    \centering
    \begin{subfigure}
    \centering
    \includegraphics[width=0.5\textwidth]{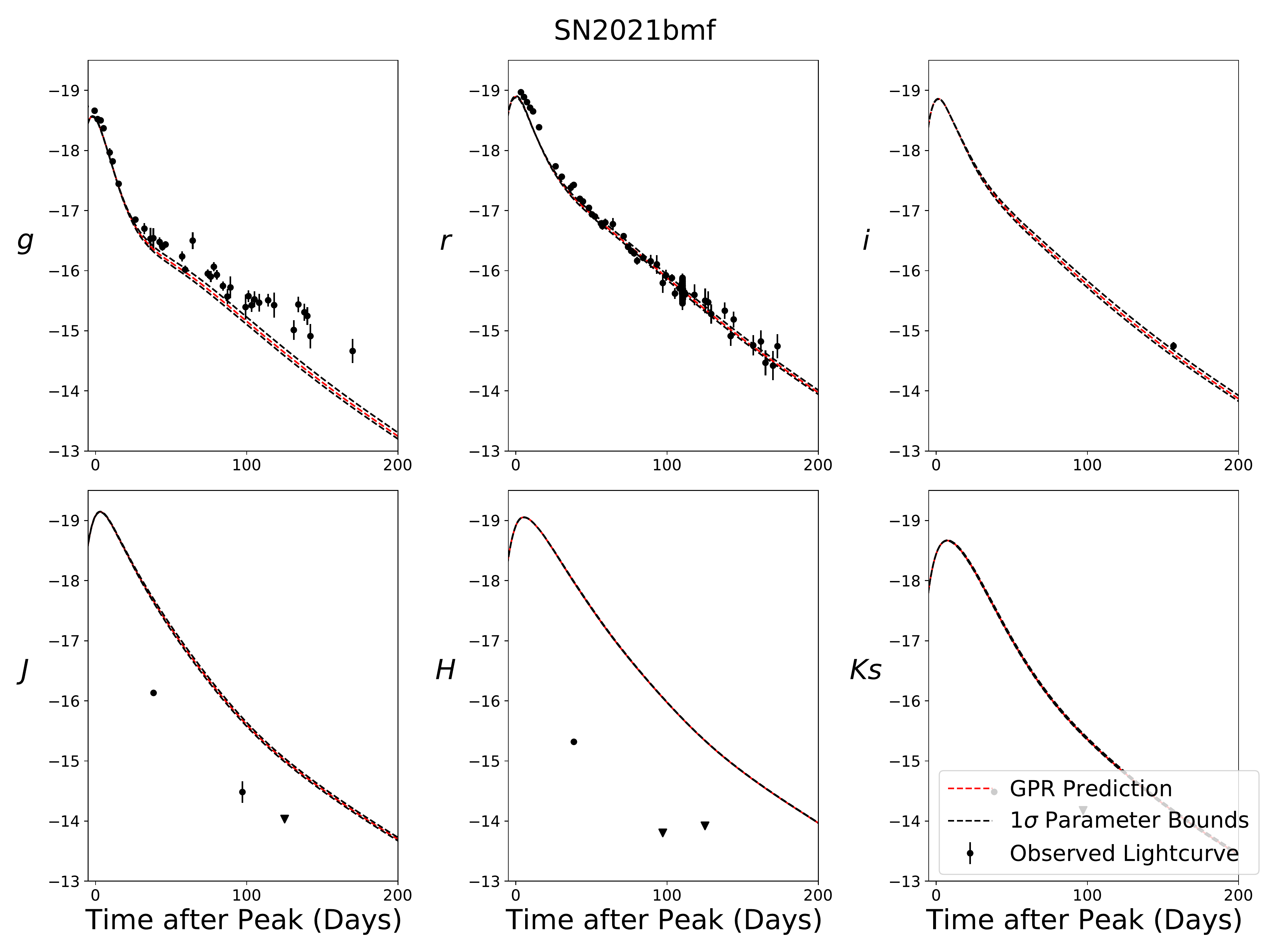}\includegraphics[width=0.5\textwidth]{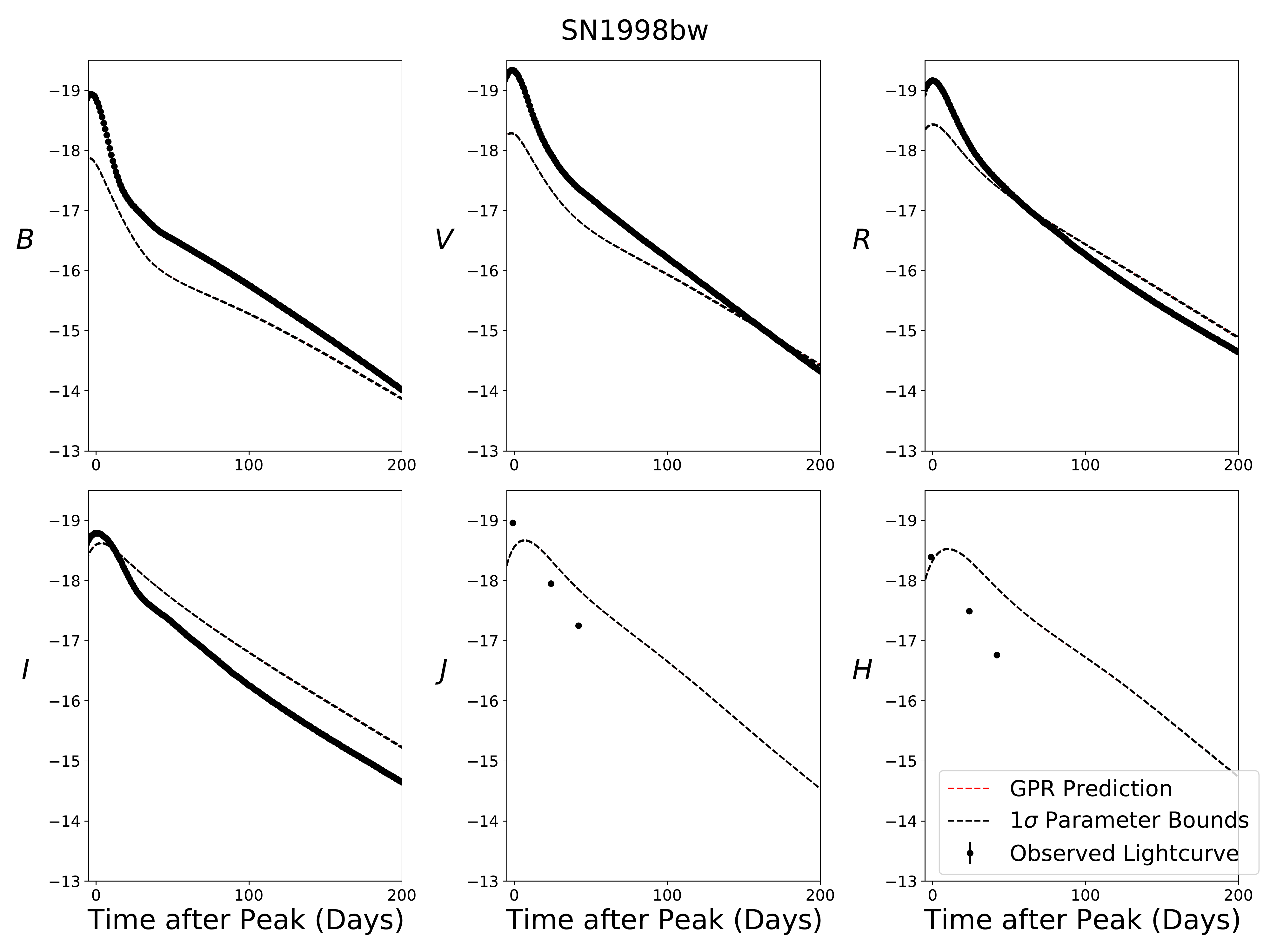}
    \end{subfigure}
    \begin{subfigure}
    \centering
    \includegraphics[width=0.5\textwidth]{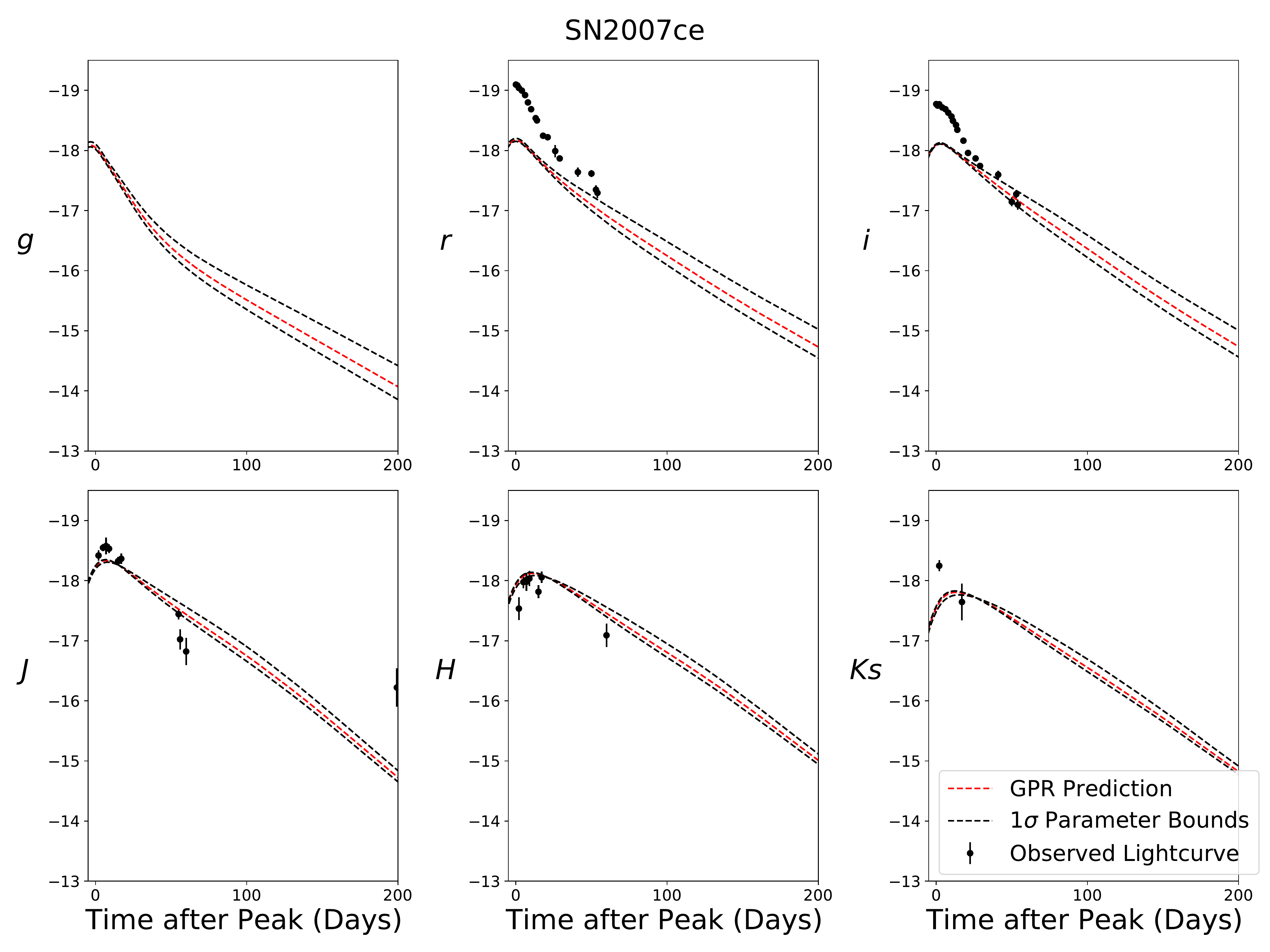}\includegraphics[width=0.5\textwidth]{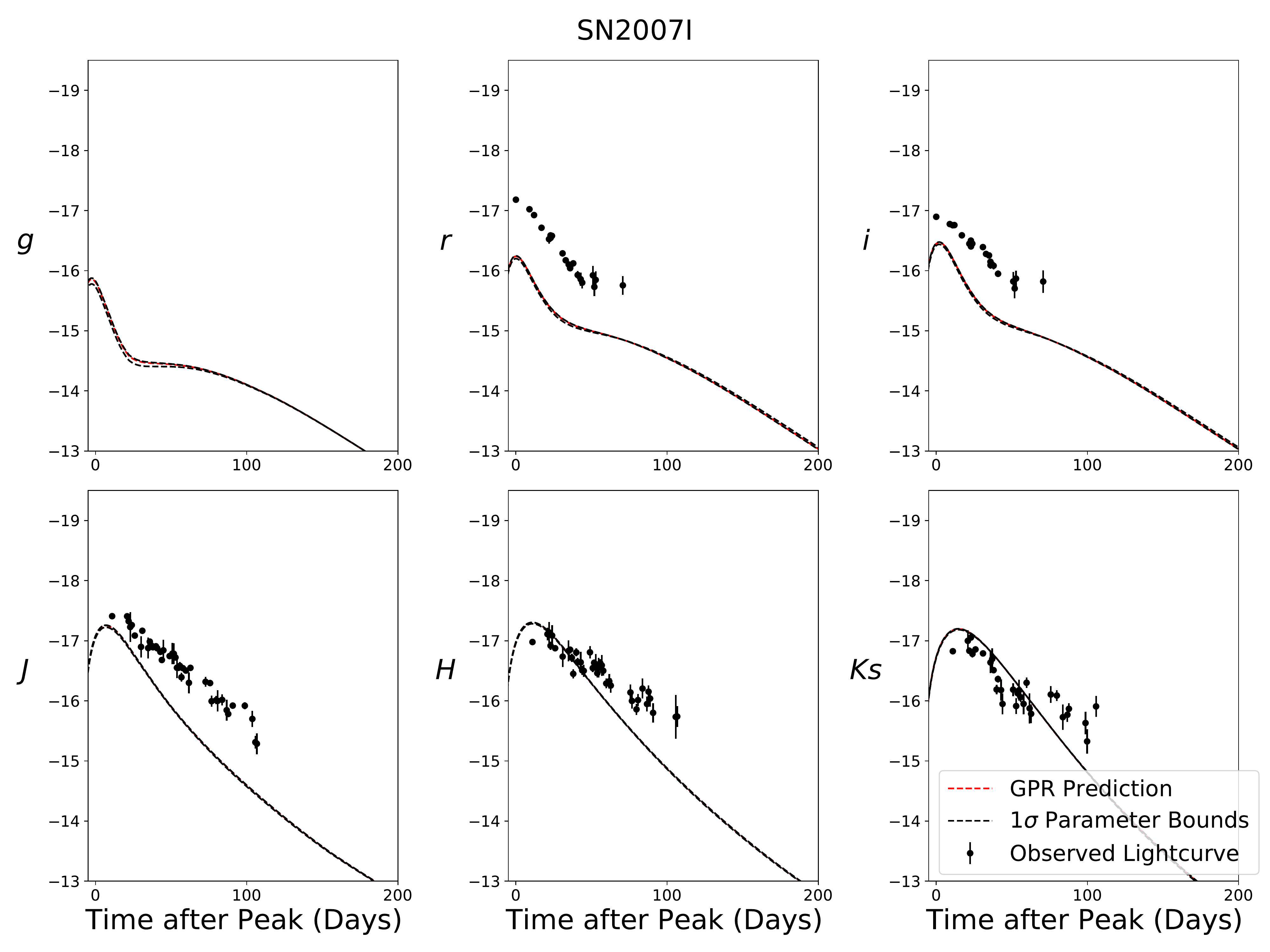}
    \end{subfigure}
    \caption{Plots of light curve models from \citet{Barnes2022} with best-fit parameters (red dotted line) and corresponding 1$\sigma$ uncertainties (black) with photometric data overplotted, for both ZTF candidates and candidates from the literature shown in Figure~\ref{fig:rp_corner} which pass our $\chi^2$ cut, ordered by inferred \mrp{}. The objects whose optical and NIR photometry are both well-described by the models are consistent with \mrp{}$\lesssim0.01\rm M_{\odot}$. The remaining objects do not show convincing fits to the \rp{} enriched models.}
    \label{fig:rp_lcs}
\end{figure*}

\subsection{\rp{} Candidate ZTF SNe} \label{sec:rp_ztfsne}

Only two of the SNe in this subset have well-constrained parameters derived from the corner plots: SN\,2020lao and SN\,2021xv. The remainder of the objects have nearly flat posteriors on \mni{} and \vej{}. For SN\,2019gwc the peaks of the posterior probability distributions for both M$_{r\mathrm{p}}$ and \xmix{} are consistent with zero. This is supported by the fact that while both the $r-$band and $i-$band light curves are slightly under-predicted by the models, the $J-$band flux is also over-predicted; the observed colors are bluer than a best-fit model with negligible \rp{}. SN\,2020lao and SN\,2021xv, in turn, have a best fit value of M$_{r\mathrm{p}}=0.01 \, \rm M_{\odot}$ and \xmix{}\,$<\,0.1$. In the case of SN\,2020lao, the optical flux is under-predicted by the models, and there are no NIR detections. On the other hand, for SN\,2021xv, the optical models provide a decent fit to the optical data, but the NIR flux is still slightly over-predicted by the models. 

SN\,2018kva, SN\,2019moc, SN\,2020tkx and SN\,2021bmf show posterior support for higher \rp{} enrichment. SN\,2020tkx and SN\,2018kva have inferred values of \mrp{}$\sim 0.03\, \rm M_{\odot}$ and \xmix{}$\lesssim 0.1$. For these two objects, the model under-predicts the peak of the optical light curve, though for SN\,2018kva the $J-$band models fit the corresponding photometry. SN\,2020tkx has two NIR detections in each of $J$, $H$, and $K_s$ filters which are well below the NIR model prediction, demonstrating that its light curve is inconsistent with the \rp{} enriched model. Finally, SN\,2019moc and SN\,2021bmf have parameter fits consistent with M$_{r\mathrm{p}} \gtrsim 0.03$ and \xmix{}$\gtrsim 0.1$. Similar to other cases, the best fit model for SN\,2019moc under-predicts its optical light curve. While the model is consistent with the $K_s-$band upper limit, it still over-predicts the $J-$band flux. SN\,2021bmf has one of the best-sampled optical light curves in our sample, and the model provides a beautiful fit to the optical bands. However, the NIR photometry is still vastly over-predicted by the same model.

\subsection{\rp{} Candidate Literature SNe}
Similarly, the two objects with $\chi^2$ fits that pass our criteria are SN\,1998bw and SN\,2007ce. In this category we also include SN\,2007I because it shows more significant photometric reddening relative to the other objects in the sample, even though it does not pass our nominal cuts.

The corner plots for these three objects show posterior distributions that are not well-constrained. However, all three objects have high predicted values for both \mrp{} as well as \xmix{}. The light curve fits show the same phenomenon that we identify for the ZTF SN light curve fits: the peak of the optical light curve is under-predicted, while the NIR data shows better agreement with the models. In the case of SN\,1998bw, the low $\chi^2$ is likely attributed to the fact that the optical data are extremely well-sampled, and the model provides a decent fit to its late-time light curve (in the $B$, $V$, and $R-$ bands), but the same model does not describe the decay in the NIR flux accurately. The best-fit model for SN\,2007ce matches the NIR bands but again under-predicts the optical. For SN\,2007I, the $riJ$-band fluxes are wholly underestimated, and in the $HK$-bands, the light curve appears to be declining much slower than predicted by the models.

\begin{figure*}
    \centering
    \includegraphics[width=0.50\textwidth]{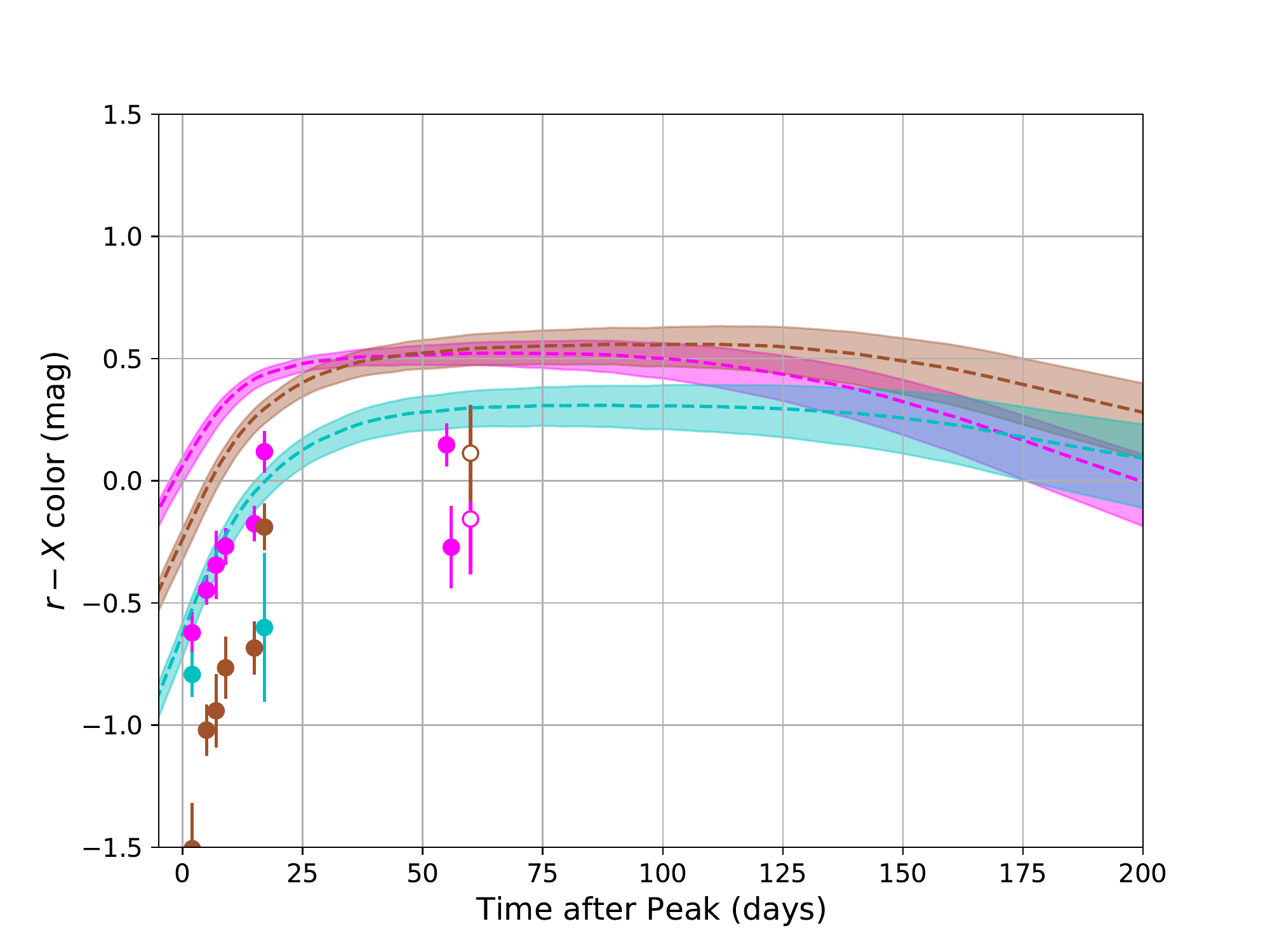}\includegraphics[width=0.50\textwidth]{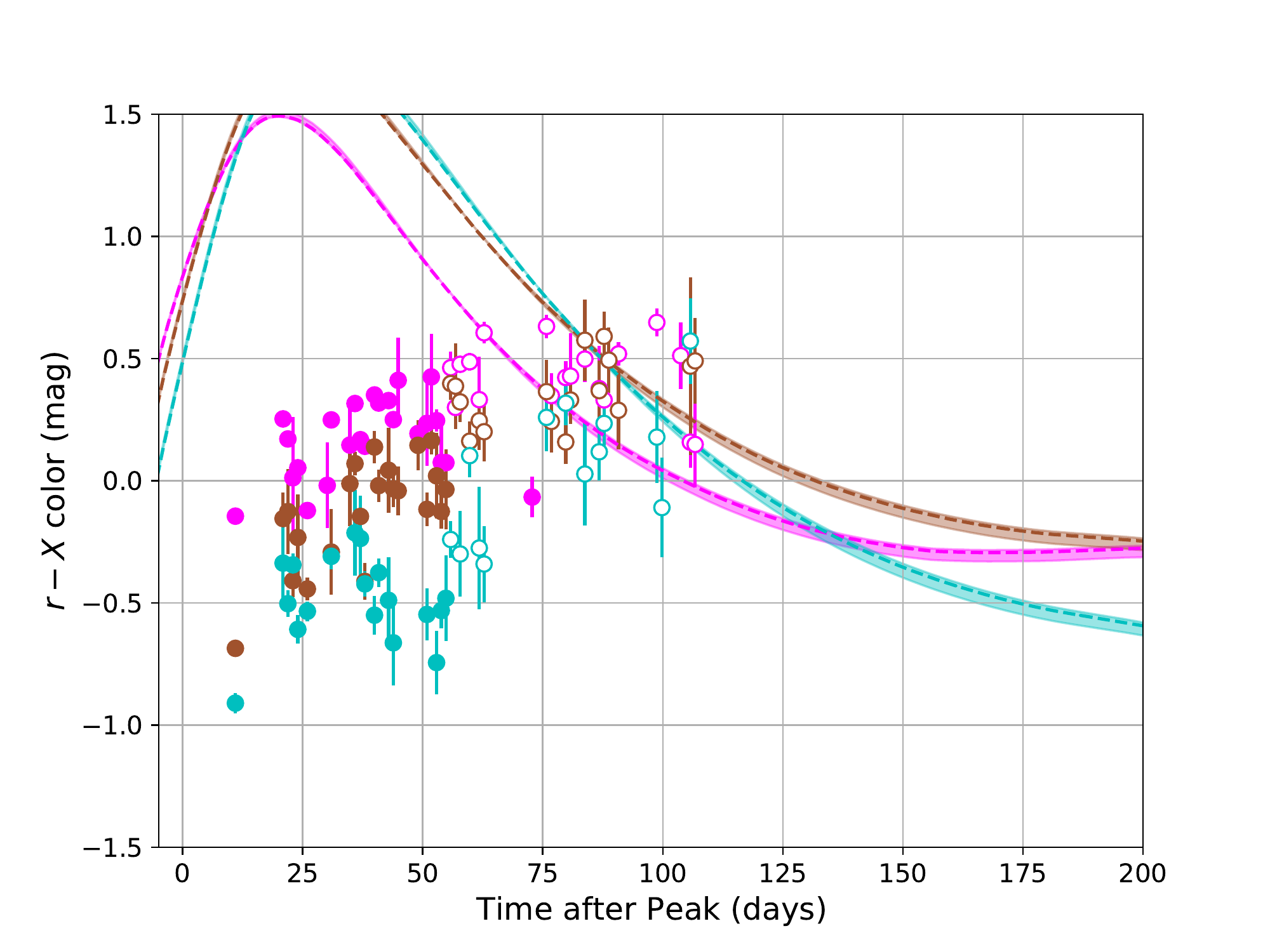}
    \caption{Color evolution as a function of time for both SN\,2007ce (left) and SN\,2007I (right). Similar to Figure~\ref{fig:color}, the filled circles with errobars represent the $r - J/H/K_s$ color estimated directly from the data, while the unfilled circles correspond to a stretched and scaled $r-$band model of SN\,2020bvc used as a proxy to estimate the color at each NIR photometric epoch, in the absence of $r-$band photometry. The dashed line represents the color evolution of the best-fit \rp{} enriched model, and the shaded regions encompass the $\pm 1\sigma$ uncertainty on the model parameters from our fits. Using the same convention as in Figure~\ref{fig:color}, magenta is $r-J$, brown is $r-H$ and cyan is $r-K_s$. As shown here, the color evolution of both SN\,2007I and SN\,2007ce appear to be inconsistent with their best-fitting \rp{} enriched model colors and associated 1$\sigma$ uncertainties.}
    \label{fig:SN2007I_SN2007ce_color}
\end{figure*}

As emphasized by \citet{Barnes2022}, color evolution can be a more powerful metric in comparison to absolute magnitude comparisons between the model light curves and data in determining whether a SN Ic-BL harbors \rp{} material. We thus plot color evolution ($r-J/H/K_s$) as a function of time for our two reddest objects, SN\,2007I and SN\,2007ce.  In Figure~\ref{fig:SN2007I_SN2007ce_color} we show their photometric colors along with their best fit \rp{}-free and \rp{} enriched models. In the shaded regions we include the $\pm 1\sigma$ uncertainty on the model parameters from our fits.
SN\,2007ce's colors appear too blue in comparison with its best-fit \rp{} model. We note that the color measurements for this object are secure because of several contemporaneous optical-NIR epochs. Given that it only attains a maximum $r-X$ color of $\sim$0.1\,mag 50\,days post-peak, we conclude that SN\,2007ce is most likely not an \rp{} collapsar.
SN\,2007I is completely inconsistent with the color evolution of its best-fit \rp{} model, even within the parameter uncertainties. However, one challenge arises from the fact that in the late-time ($\gtrsim$50\,days post-peak) SN\,2007I lacks any optical photometry. Based on our extrapolation of the $r-$band light curve of SN\,2007I we see evidence for further reddening which starts to become consistent with the \rp{} enriched model predictions in the late-times. Thus, we are unable to rule out the possibility of \rp{} production in SN\,2007I based on the \rp{} fits and the color evolution comparison alone.

% as well -- the object is systematically bluer than predicted by the models. However, based on its early color evolution ($\lesssim$50\,days post-peak, where $r$-band photometry is concurrently available), SN\,2007I could plausibly be explained by an \rp{}-free model. The challenge arises from the fact that in the late-time ($\gtrsim$50\,days post-peak) SN\,2007I lacks any optical photometry along with its NIR datapoints. Based on our extrapolation of the $r$-band light curve of SN\,2007I we see evidence for further reddening which is inconsistent with the \rp{}-free model in the late-times. Furthermore, in Figure~\ref{fig:color} this object stands out from the rest of the objects due to its photometric reddening which is consistent with a subset of the \rp{} models. Given that degeneracies between explosion properties and \rp{} properties can result in a very similar color evolution, we conclude that SN\,2007I remains as a viable candidate \rp{} collapsar.

\begin{figure*}
    \centering
    \includegraphics[width=0.5\textwidth]{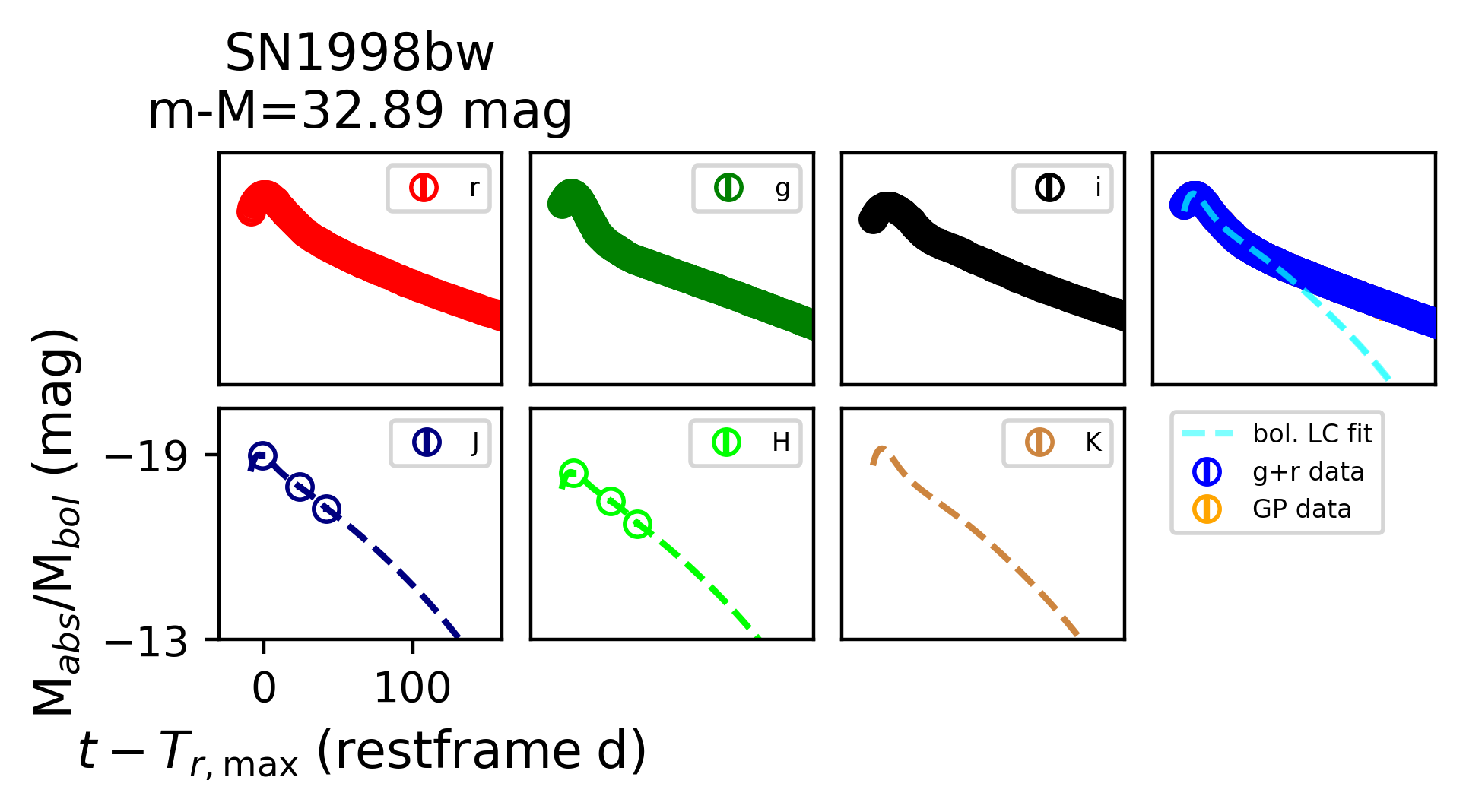}\includegraphics[width=0.5\textwidth]{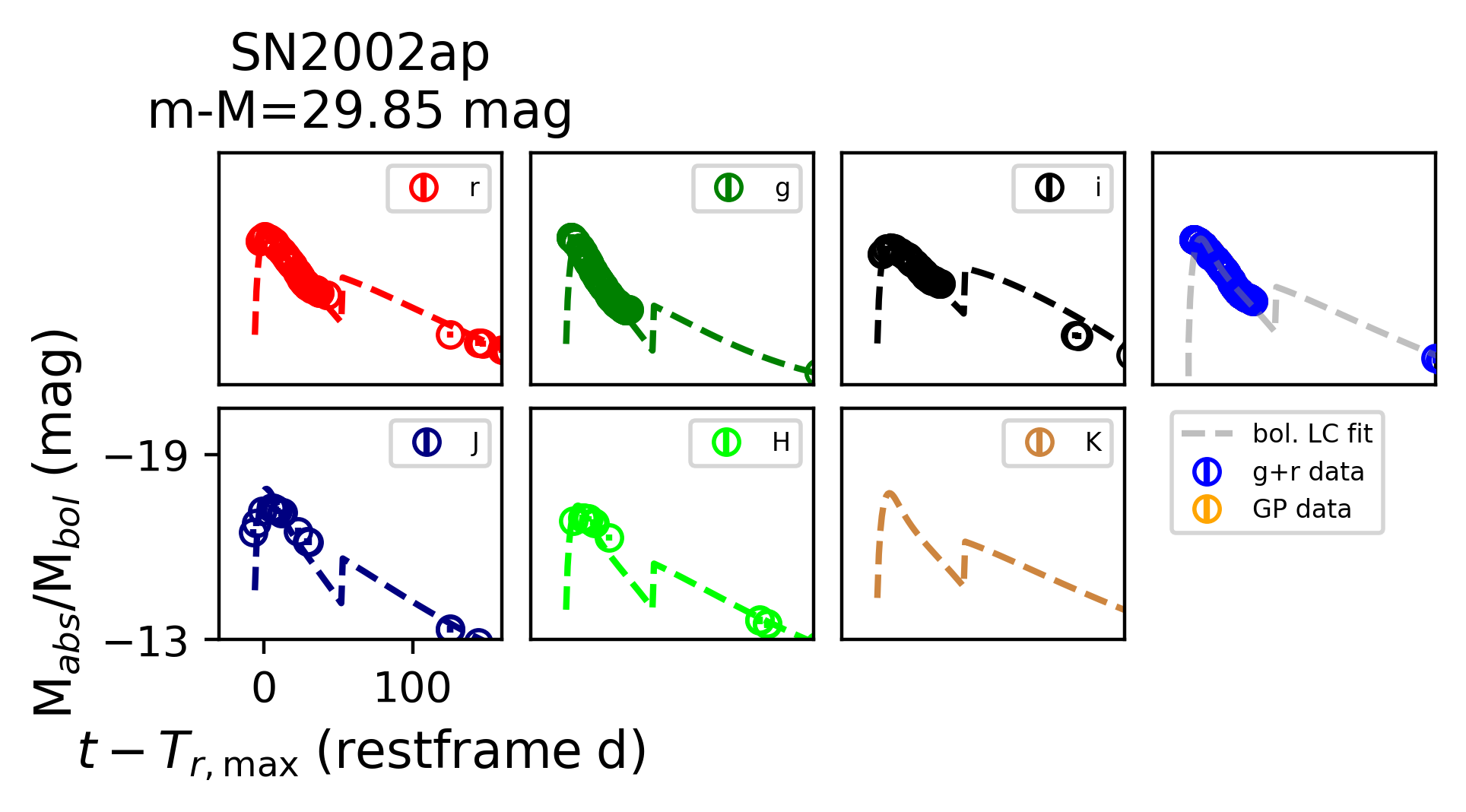}
    \includegraphics[width=0.5\textwidth]{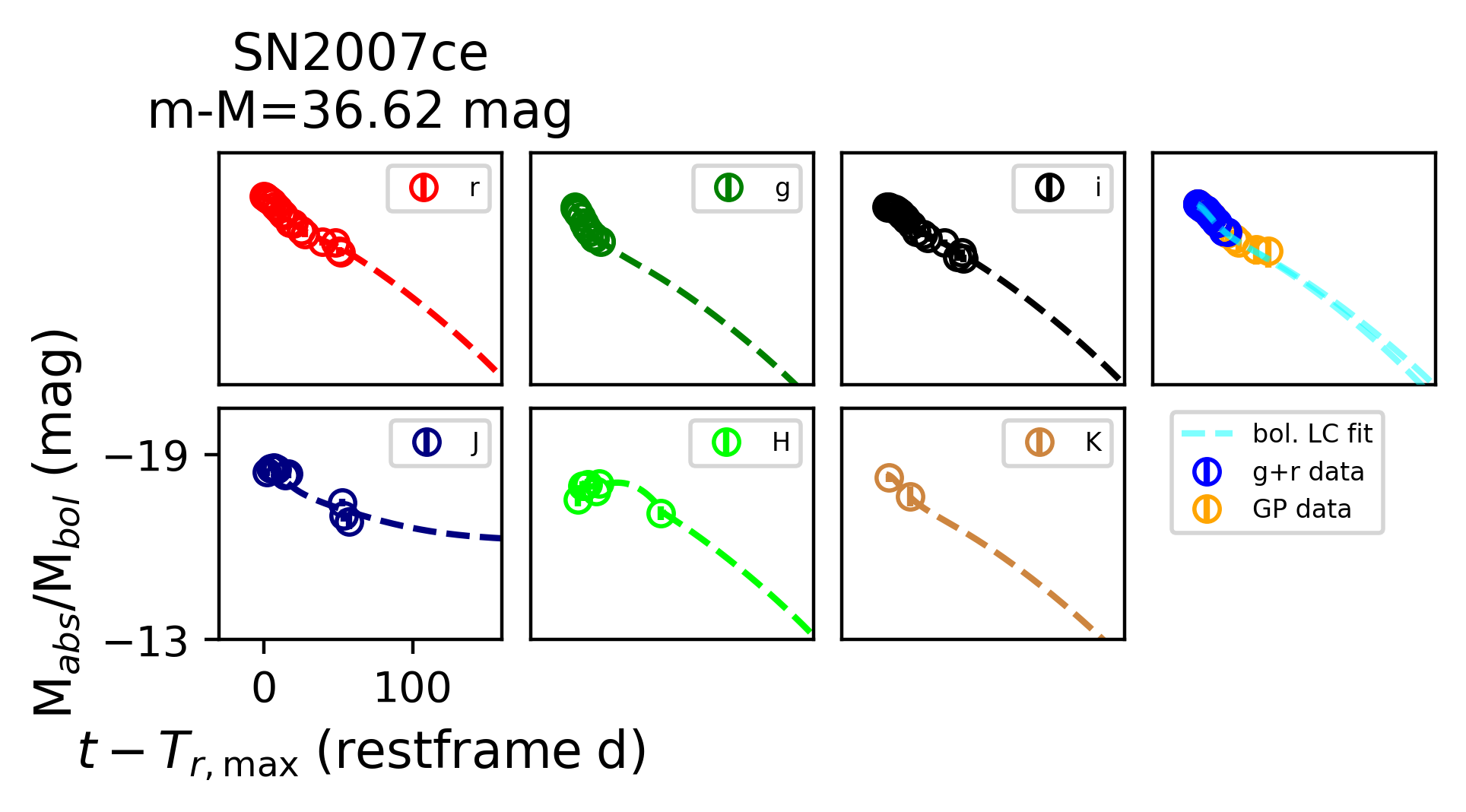}\includegraphics[width=0.5\textwidth]{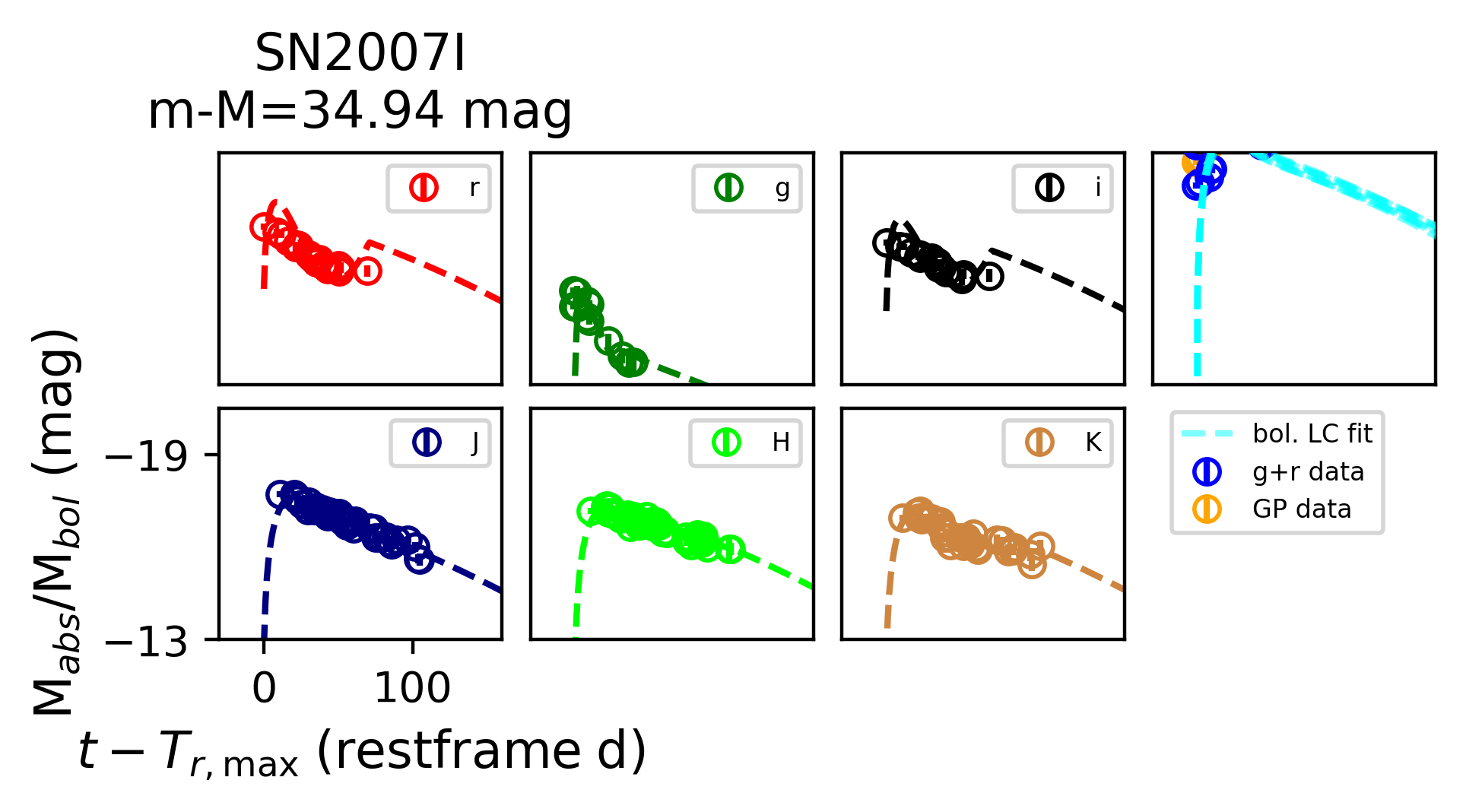}
    \includegraphics[width=0.5\textwidth]{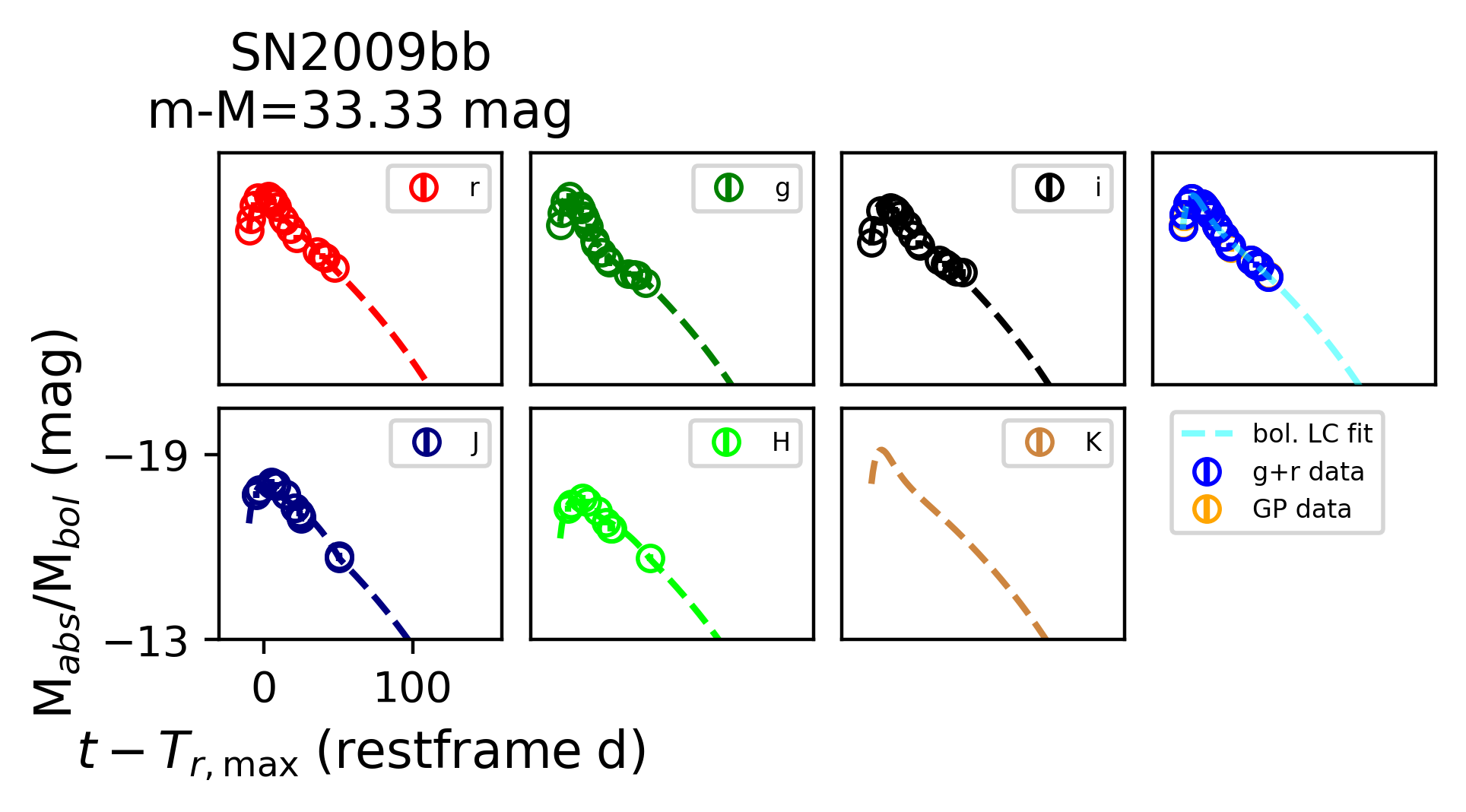}\includegraphics[width=0.5\textwidth]{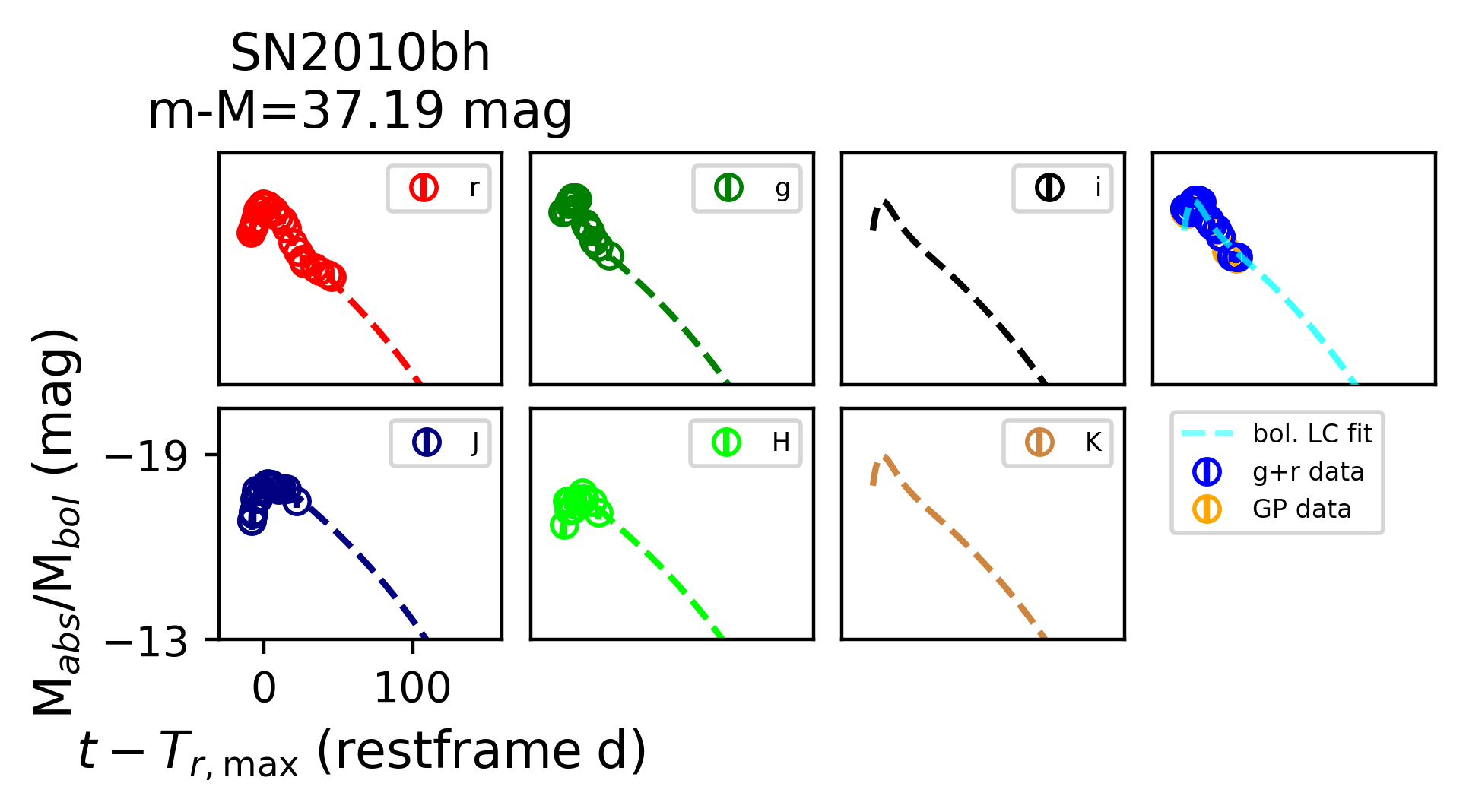}
    \includegraphics[width=0.5\textwidth]{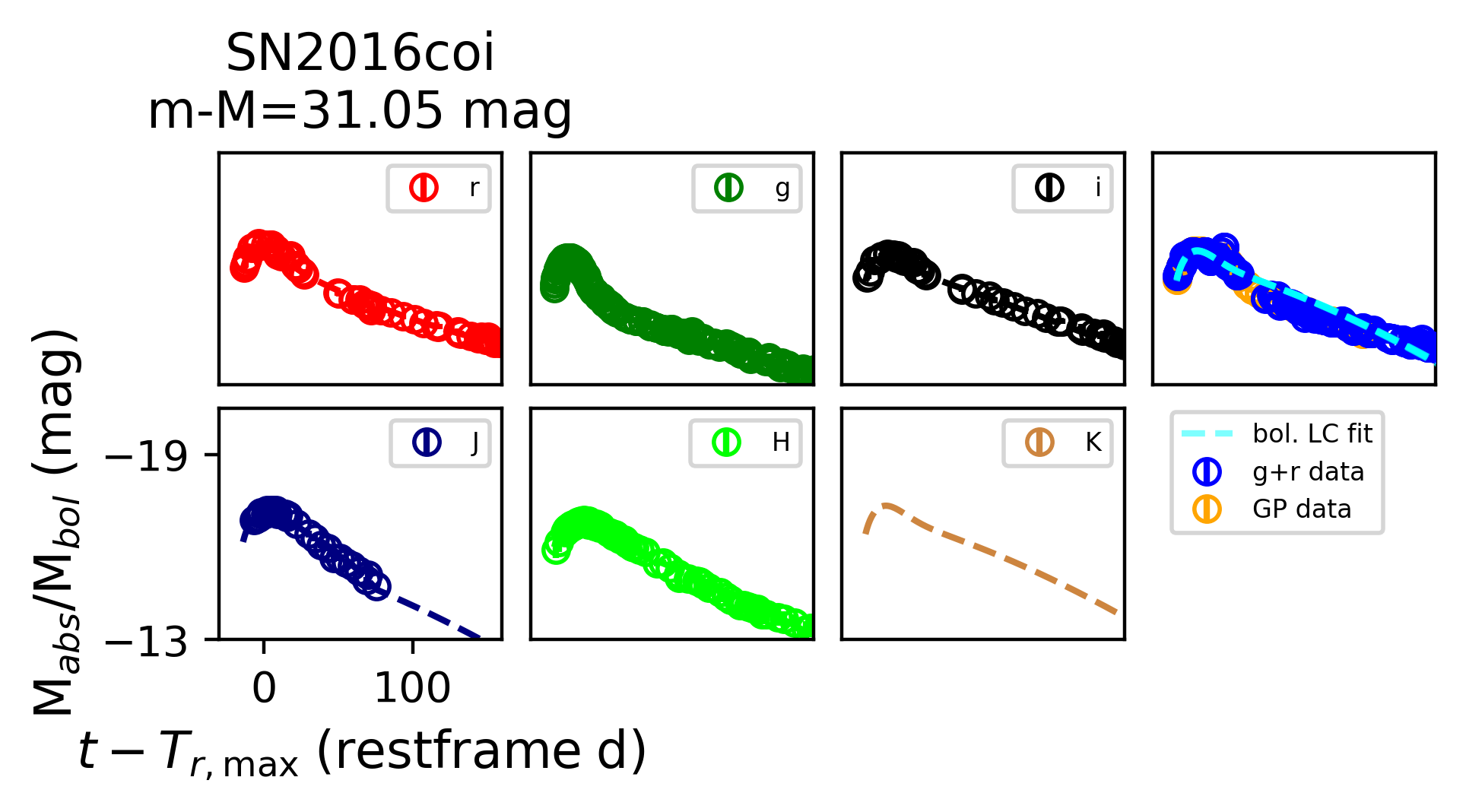}
    \caption{Light curve fits to the literature supernova events from \texttt{HAFFET}. The dashed cyan line is the best fit bolometric light curve, while the remaining dashed lines show the fits to each of the broadband light curves. Broadband light curves are calculated by fitting bolometric corrections in each band, and using these corrections to rescale the Arnett fitted bolometric light curves. The circles are the photometry for each object in $griJHK_s$ bands. In the bolometric light curve plot, the points correspond to bolometric luminosity estimated from both $g-$ and $r-$ bands, or from a single band, and using GPR to estimate the flux in the other band. For SNe with photometry in the Johnson filter system, we convert the photometry to SDSS assuming photometric conversions from \citet{Jordi2006}. We find that the \texttt{HAFFET} models are good fits to the photometric data from our sample.}
    \label{fig:BB_literature}
\end{figure*}

\begin{figure*}
    \centering
    \begin{subfigure}
    \centering
    \includegraphics[width=0.5\textwidth]{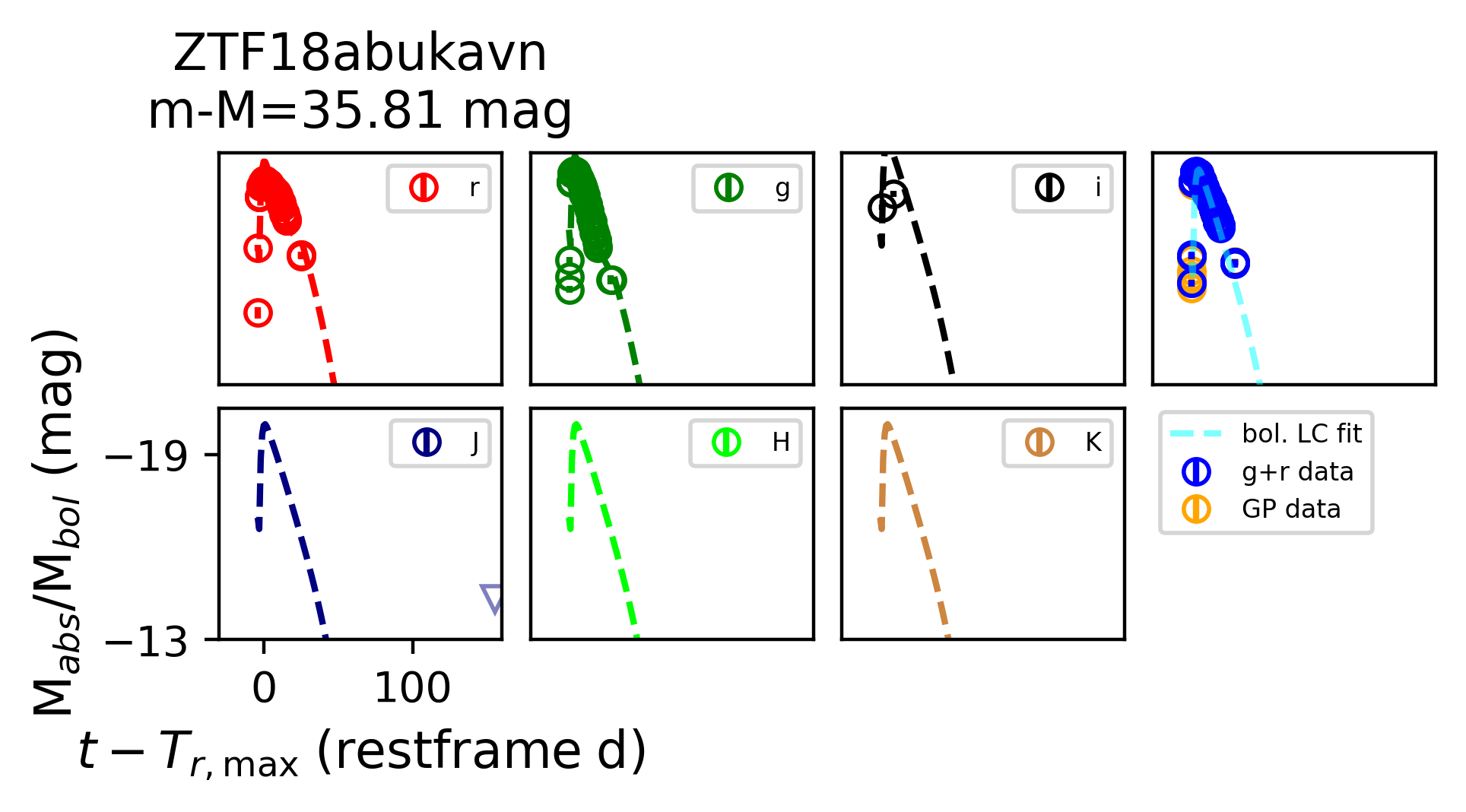}\includegraphics[width=0.5\textwidth]{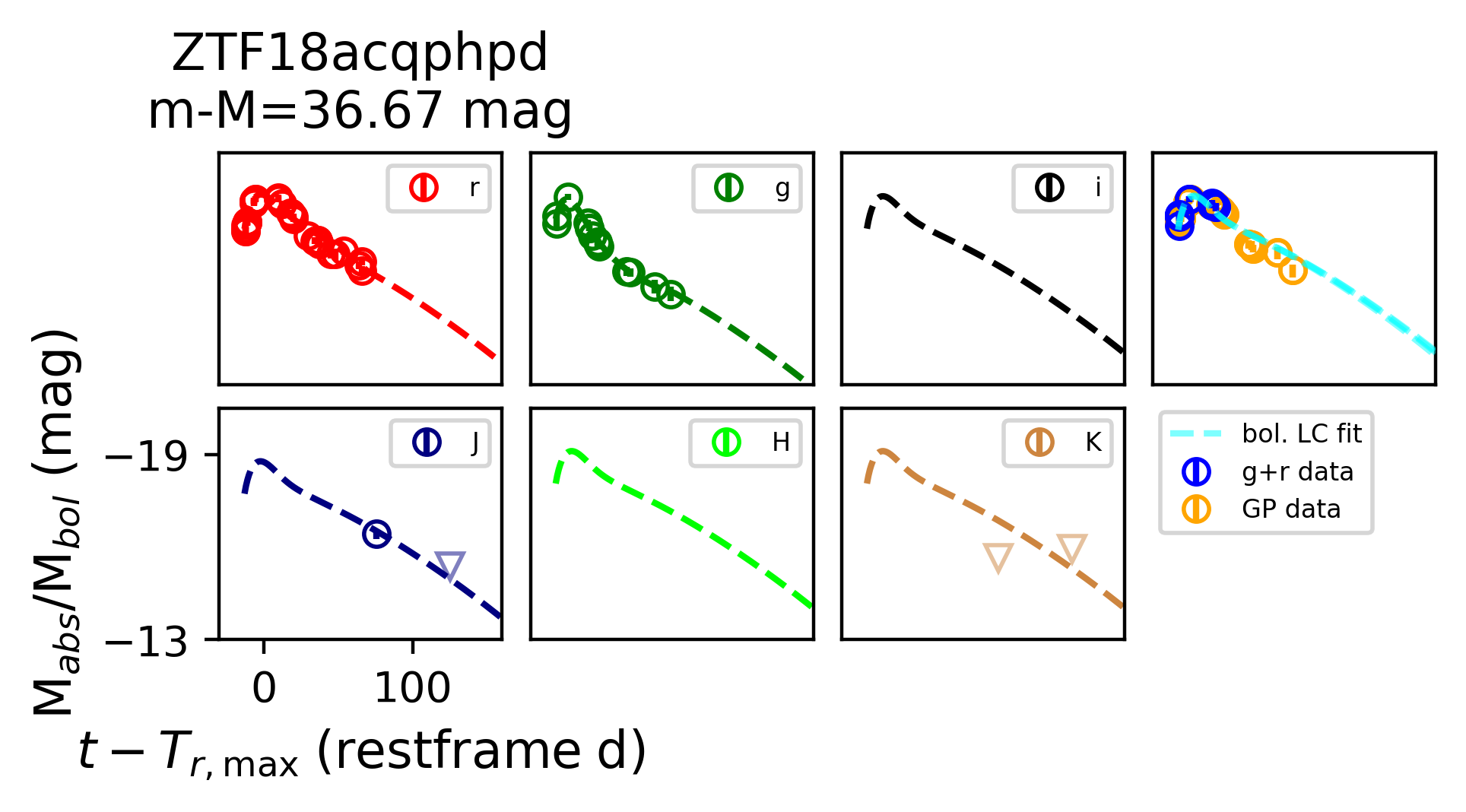}
    \end{subfigure}
    \begin{subfigure}
    \centering
    \includegraphics[width=0.5\textwidth]{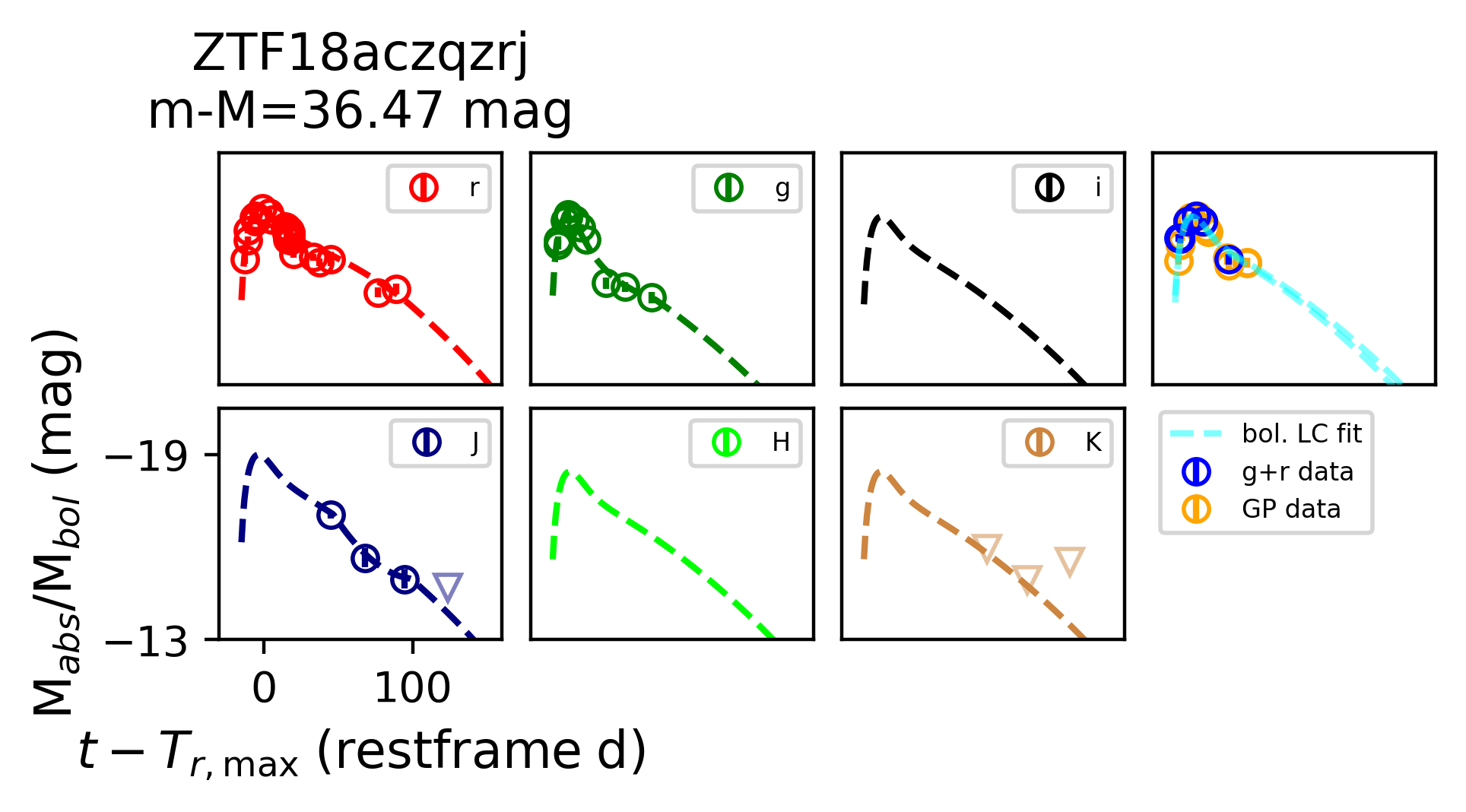}\includegraphics[width=0.5\textwidth]{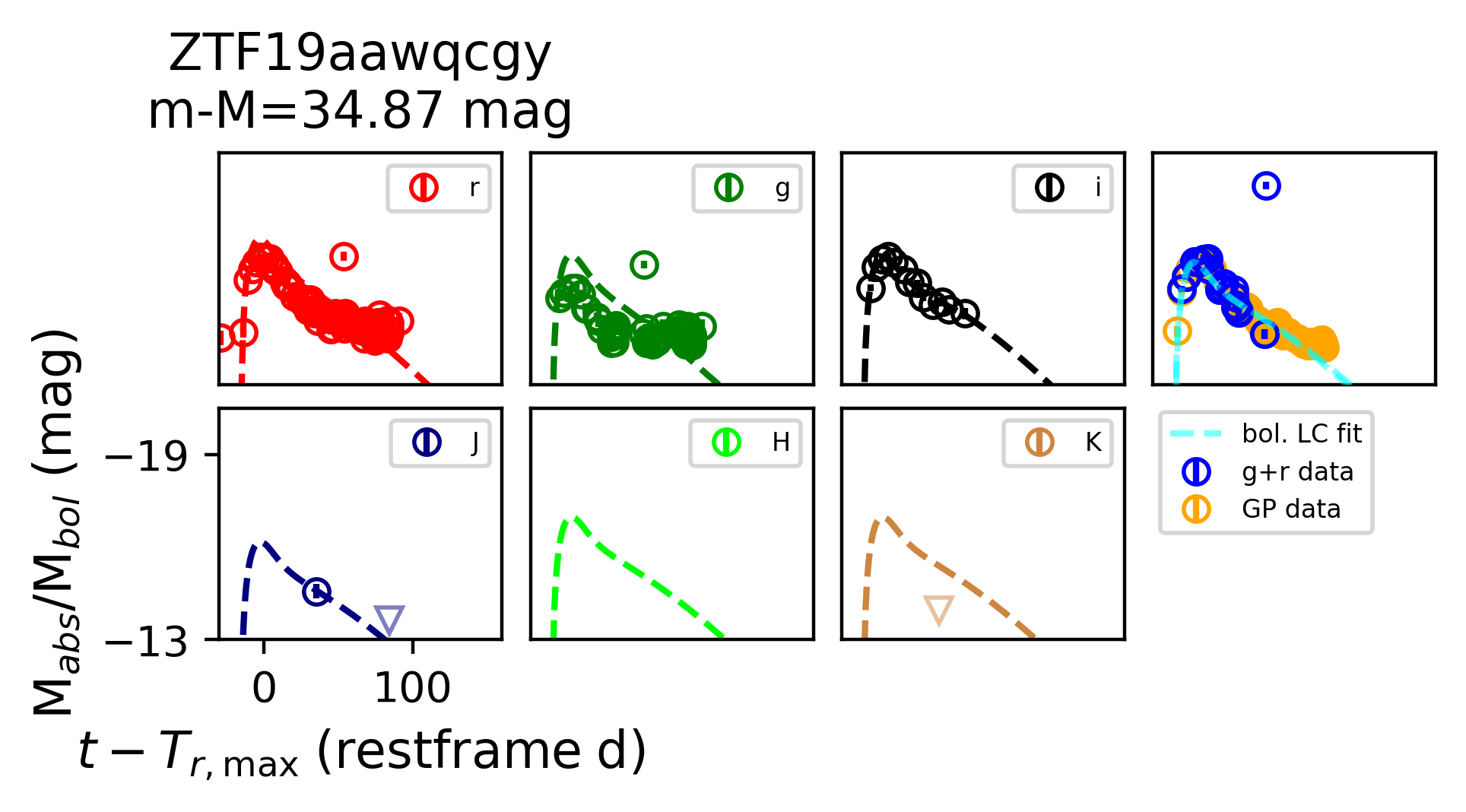}
    \end{subfigure}
    \begin{subfigure}
    \centering
    \includegraphics[width=0.5\textwidth]{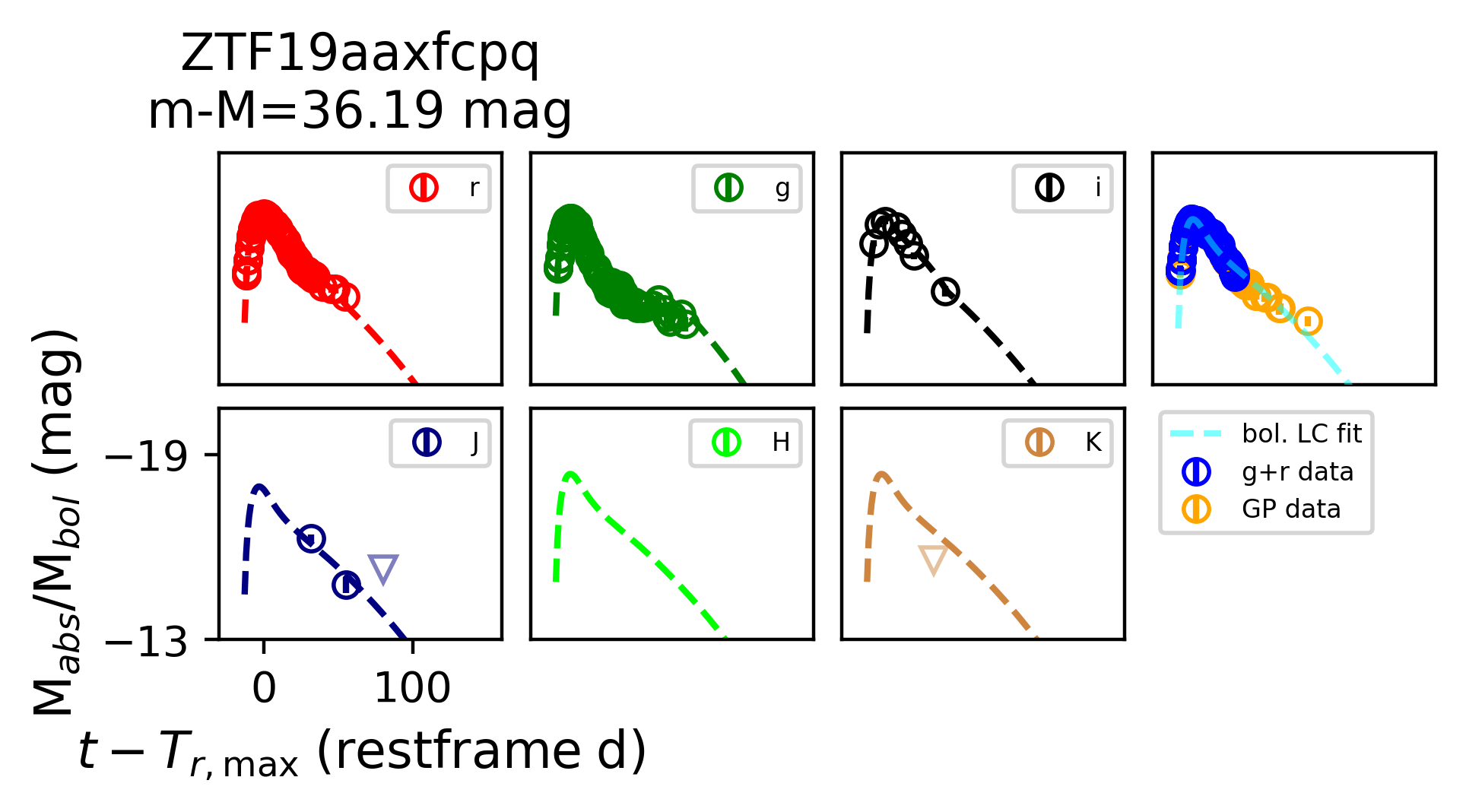}\includegraphics[width=0.5\textwidth]{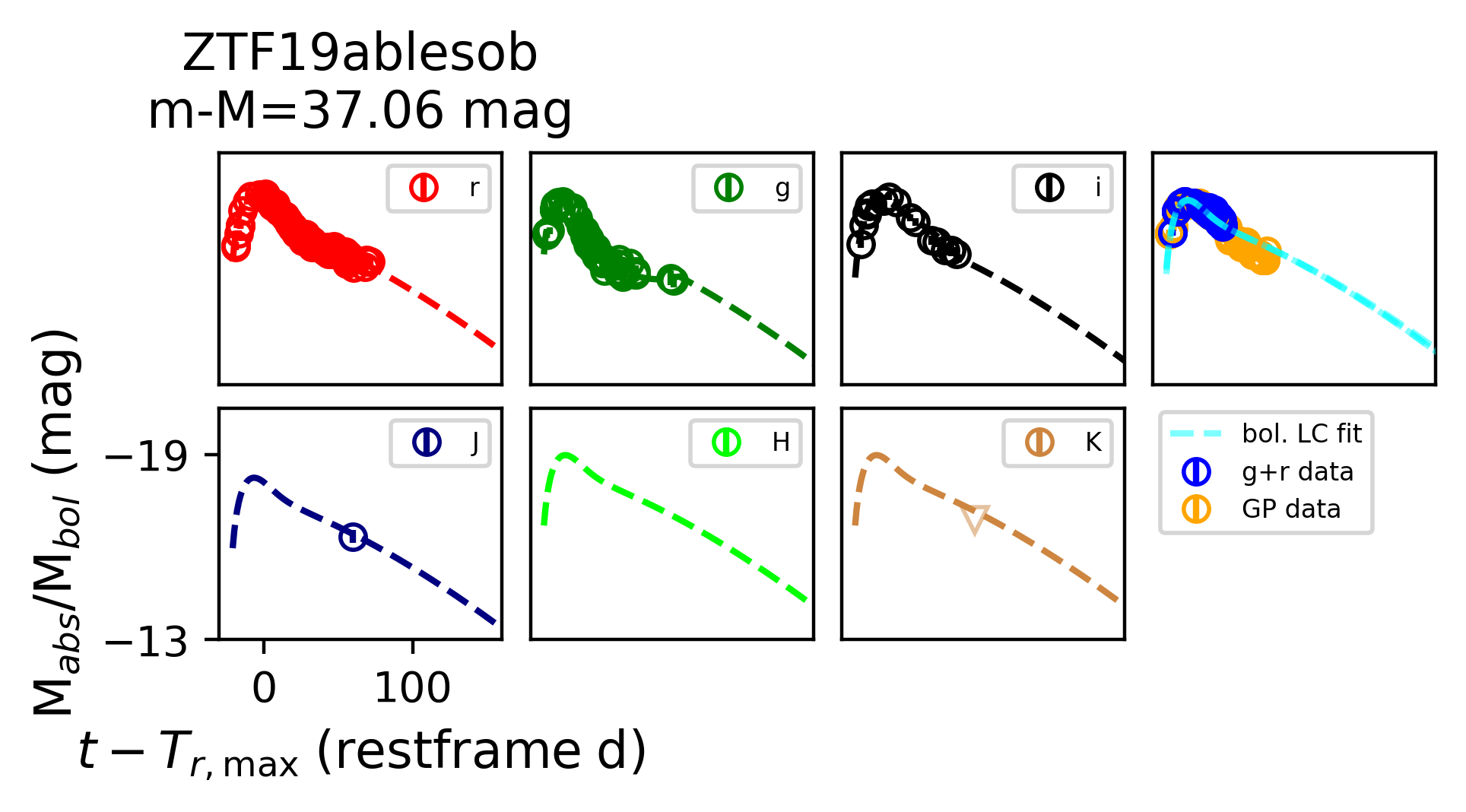}
    \end{subfigure}
    \begin{subfigure}
    \centering
    \includegraphics[width=0.5\textwidth]{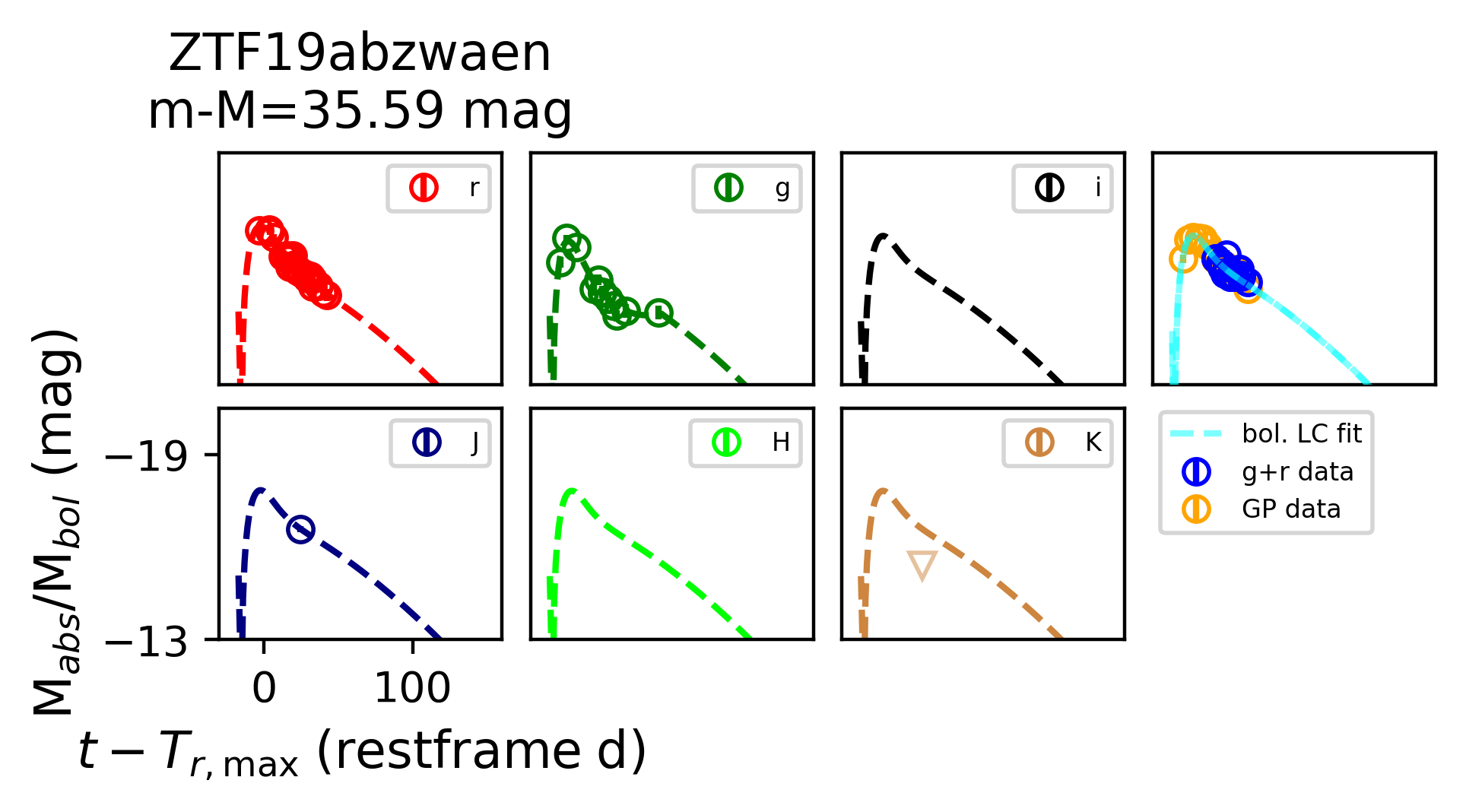}\includegraphics[width=0.5\textwidth]{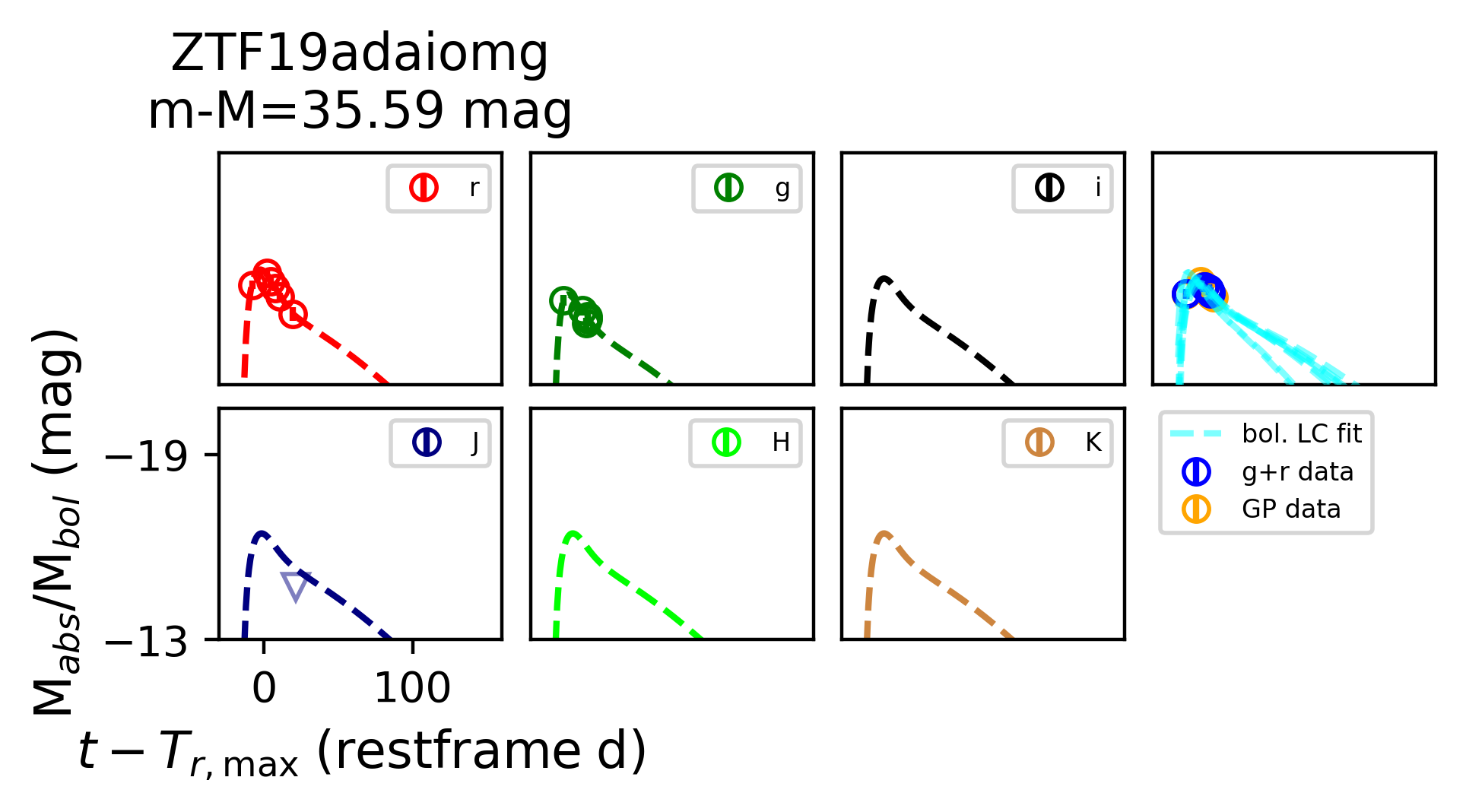}
    \end{subfigure}
    \begin{subfigure}
    \centering
    \includegraphics[width=0.5\textwidth]{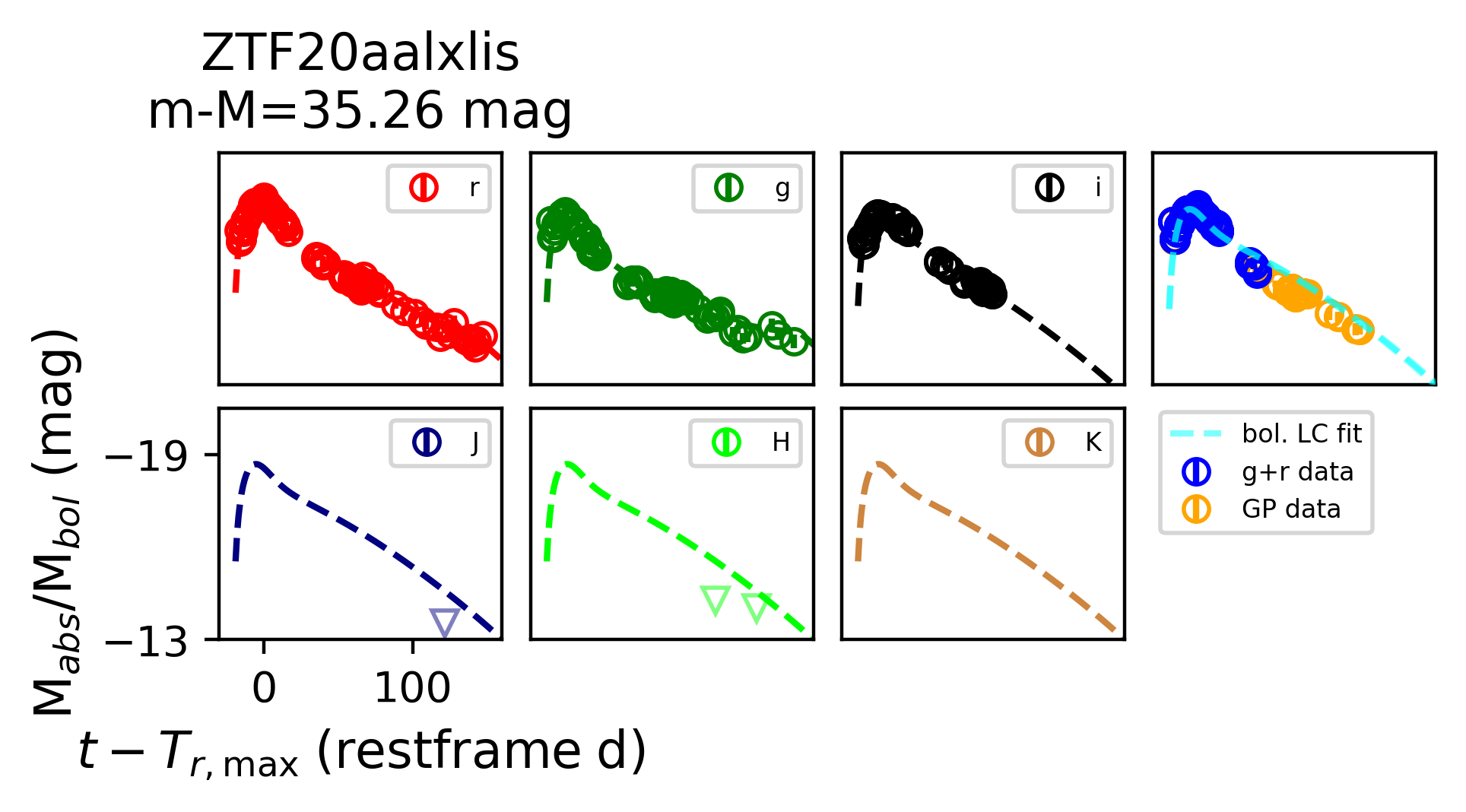}\includegraphics[width=0.5\textwidth]{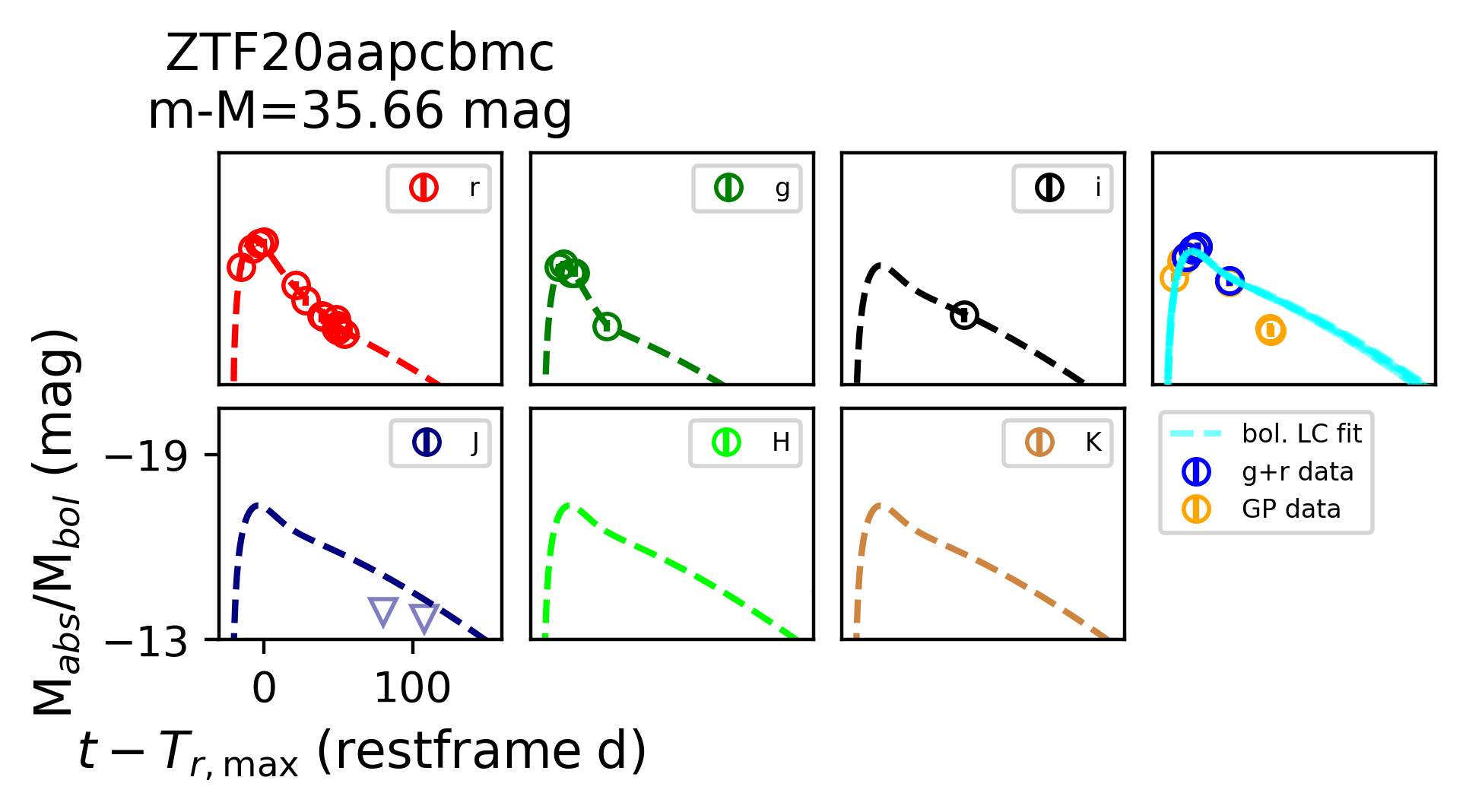}
    \end{subfigure}
\end{figure*}
\begin{figure*}
    \centering
    \begin{subfigure}
    \centering
    \includegraphics[width=0.5\textwidth]{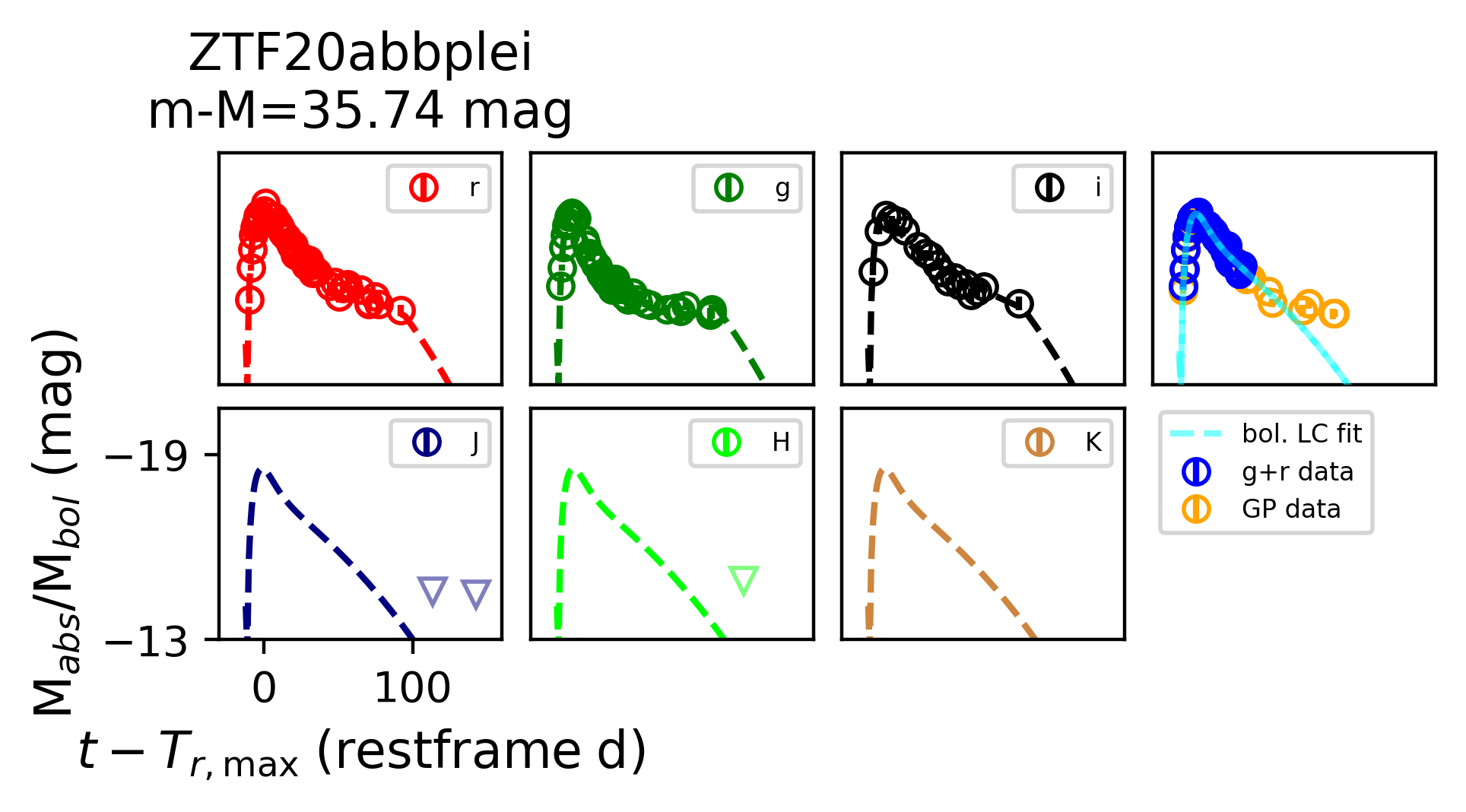}\includegraphics[width=0.5\textwidth]{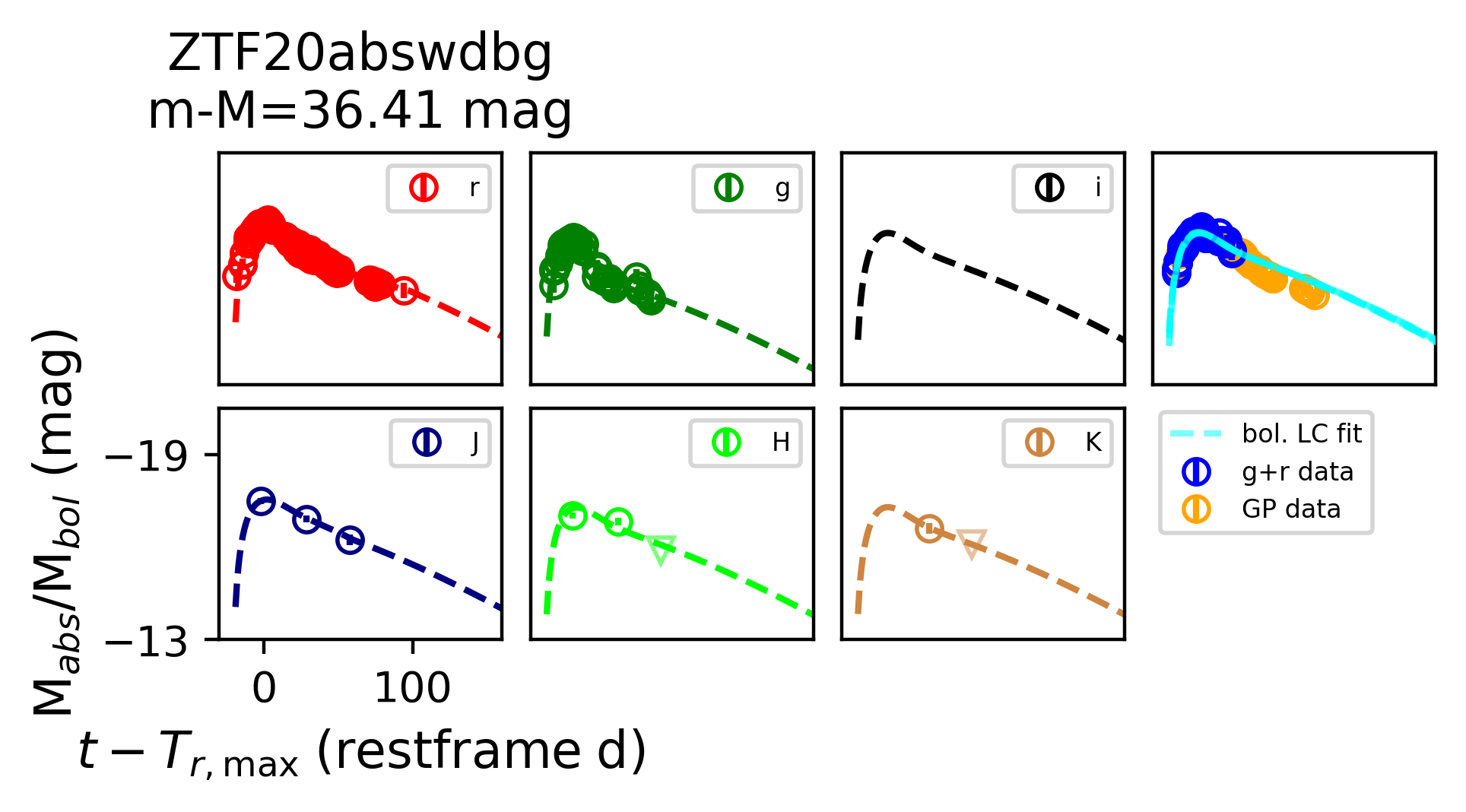}
    \end{subfigure}
    \begin{subfigure}
    \centering
    \includegraphics[width=0.5\textwidth]{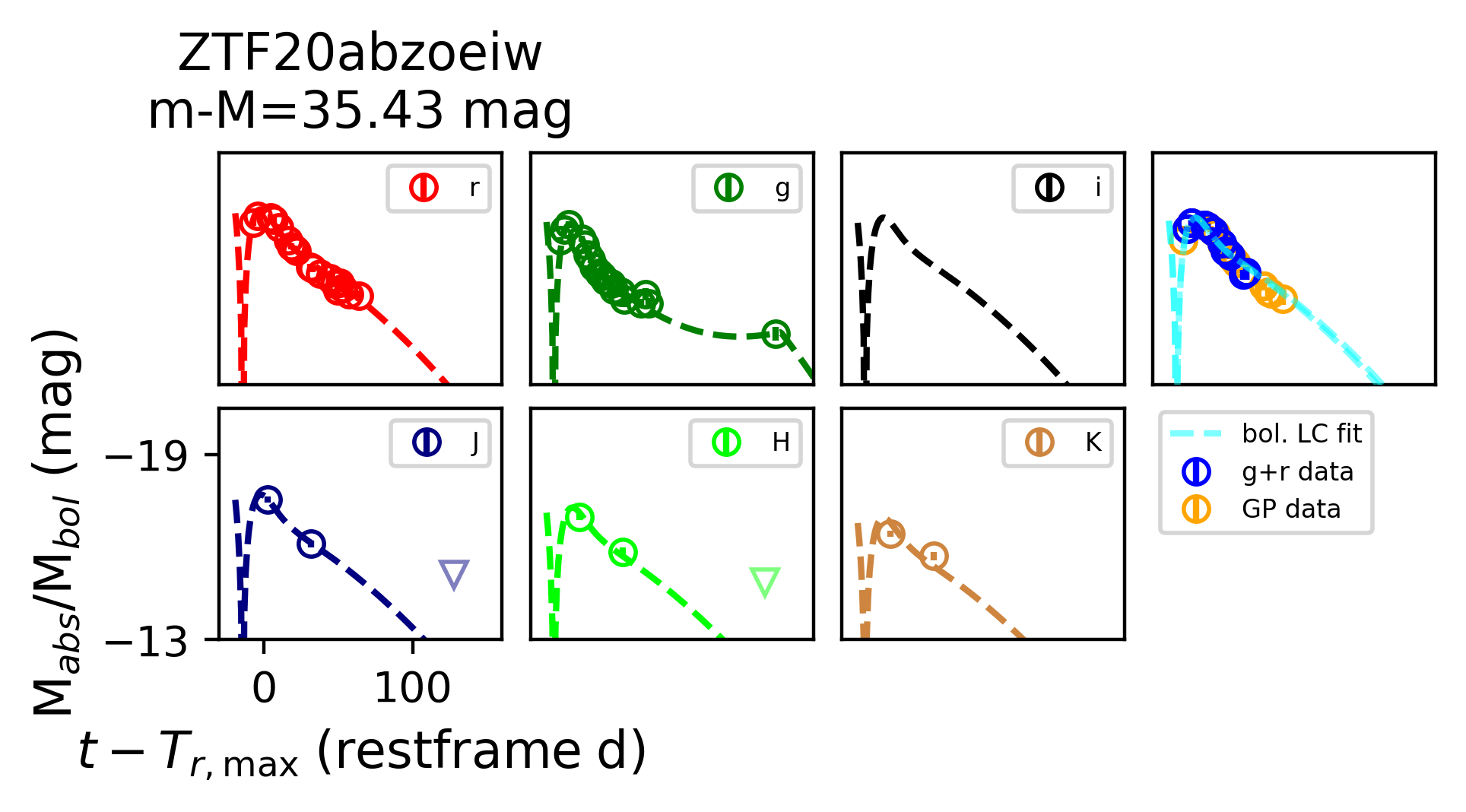}\includegraphics[width=0.5\textwidth]{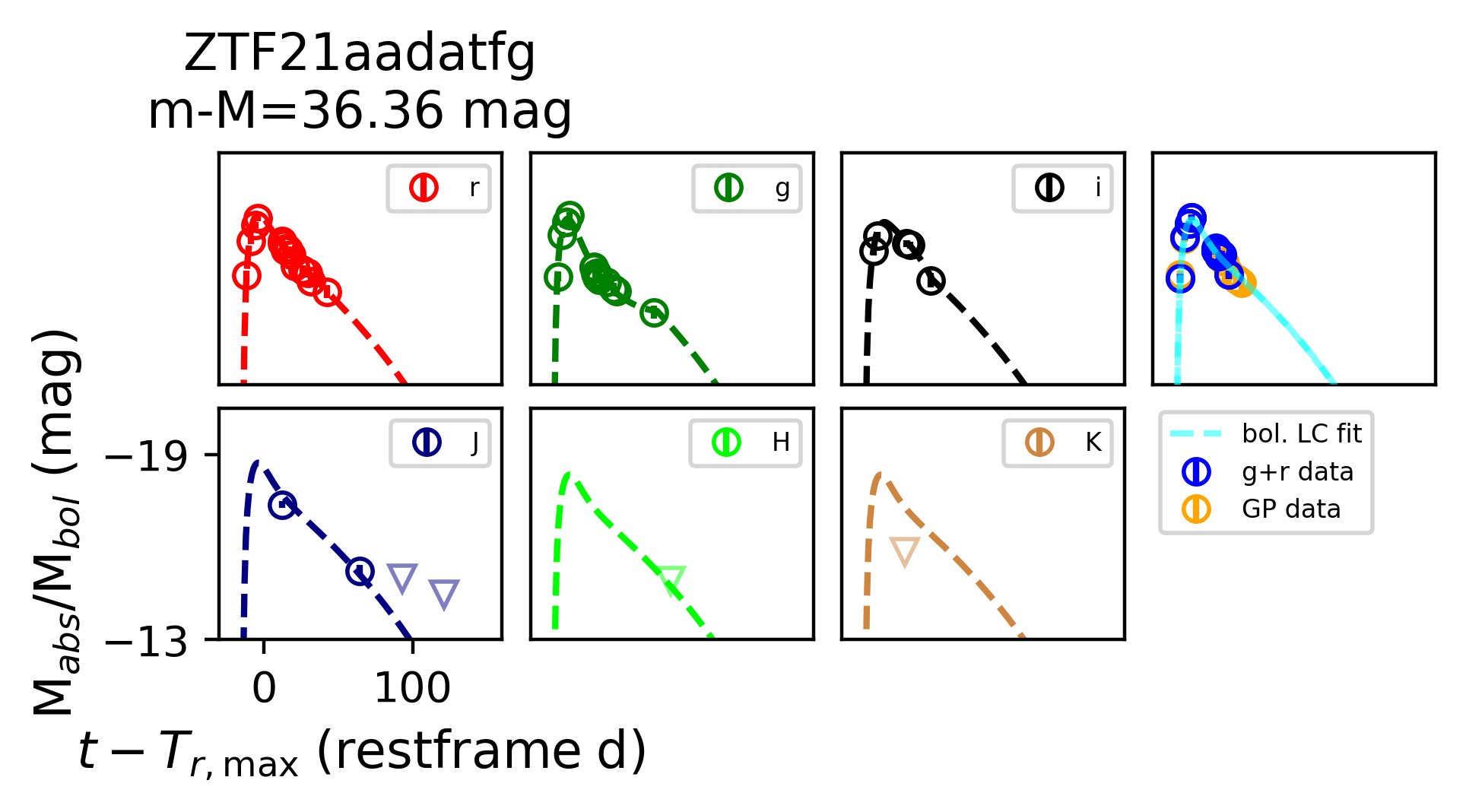}
    \end{subfigure}
    \begin{subfigure}
    \centering
    \includegraphics[width=0.5\textwidth]{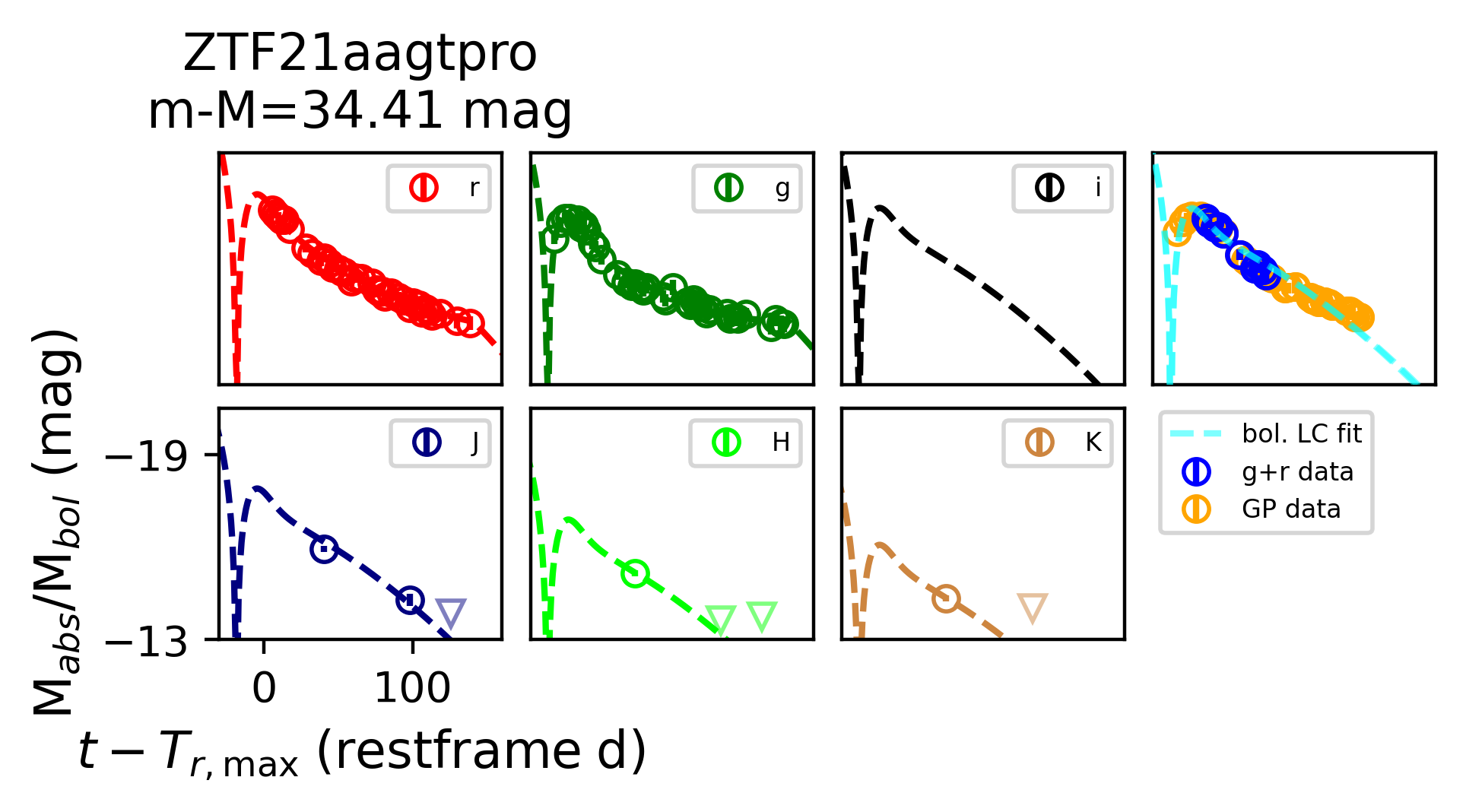}\includegraphics[width=0.5\textwidth]{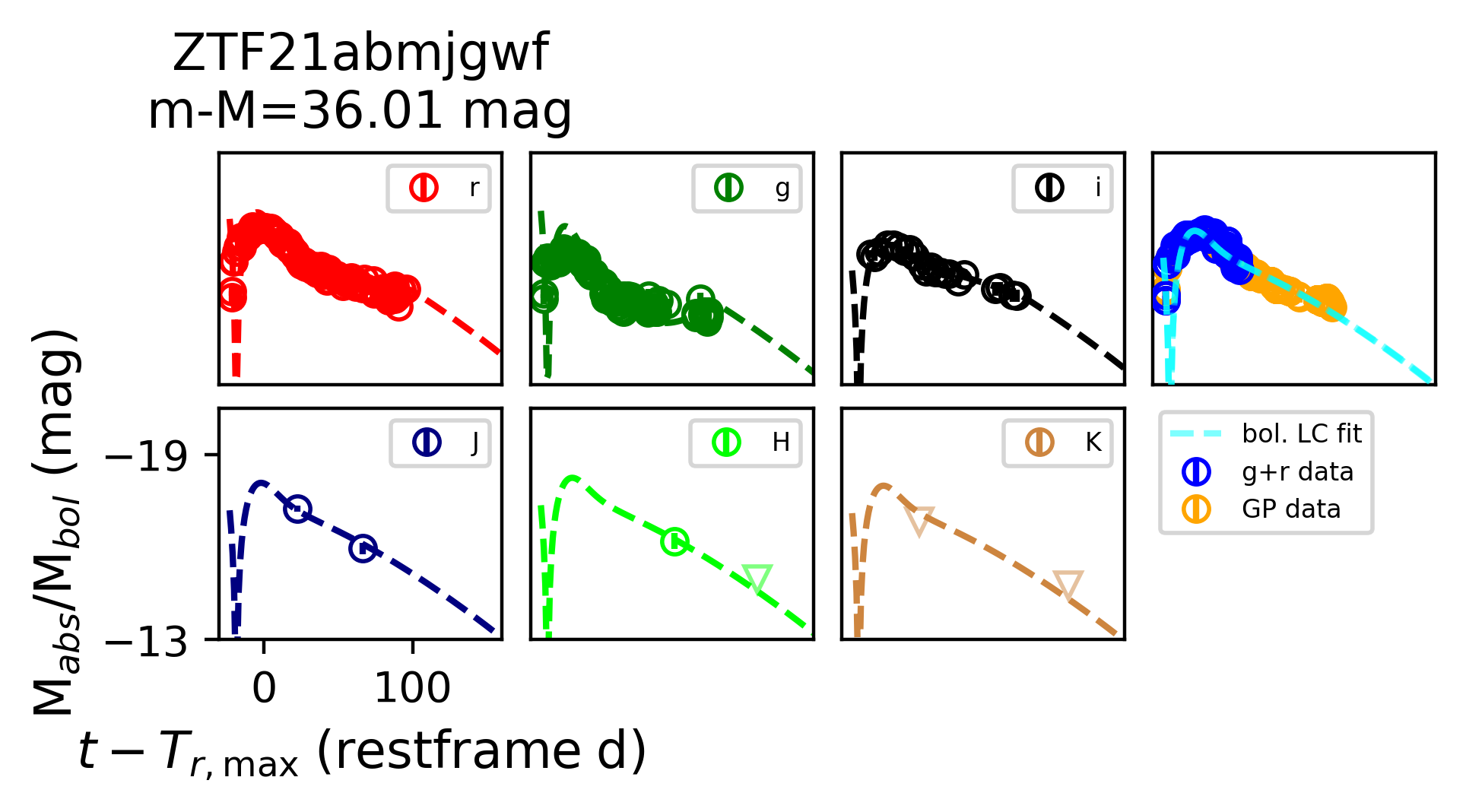}
    \end{subfigure}
    \begin{subfigure}
    \centering
    \includegraphics[width=0.5\textwidth]{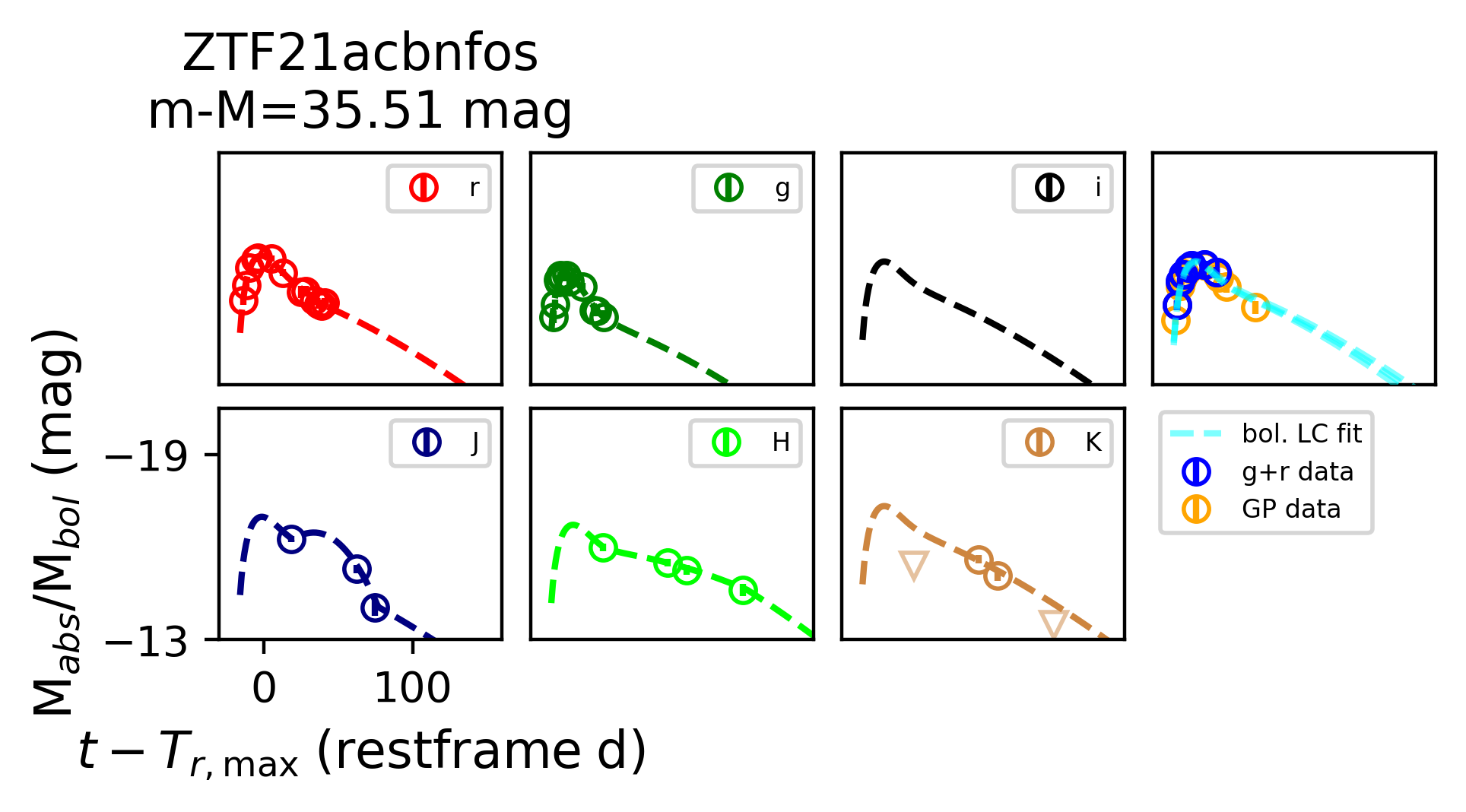}
    \end{subfigure}
    \caption{Light curve fits to ZTF SNe from \texttt{HAFFET}, similar to Fig.~\ref{fig:BB_literature}.}
    \label{fig:BB_ZTF}
\end{figure*}

\subsection{Independent Arnett Fits}
To supplement our fits to \rp{} enriched models we use \texttt{HAFFET} to construct a bolometric light curve from our optical data and fit to the standard Arnett model (as described in more detail in Sec.~\ref{sec:analysis}). We then calculate broadband light curve models by fitting bolometric corrections in each band, and using these corrections to rescale the Arnett fitted bolometric light curve models. Our fits are shown in Figures~\ref{fig:BB_literature} and \ref{fig:BB_ZTF}. 

In order to compare the two models, we compute $\chi^2$ for each of the broadband light curves in the same way as we did using the \rp{}-enriched models. We find that all of the objects, except for SN\,2021xv, have lower $\chi^2$ values with the \texttt{HAFFET} fits compared to the \rp{} model fits, insinuating that the \rp{}-free models are a better descriptor of these SN light curves. Aside from 3 objects, all other objects pass our criteria of $\chi^2 < 1.06$ (i.e. well-described by the \rp{}-free models) and none of them have $\chi^2 > 9.49$ (i.e. poorly described by the \rp{}-free models). Upon visual inspection, we find convincing fits to both the early optical light curves and the NIR light curves of these objects for the \rp{}-free models. In the case of SN\,2021xv, we note that the \rp{} parameter estimation favors little to no \rp{} mass and mixing, and the \rp{}-enriched models overestimate the NIR flux. Thus we consider SN\,2021xv to still be consistent with an \rp{}-free scenario.

Furthermore, we derive blackbody effective temperatures for the closest epoch to 30\,days post-peak where both optical and NIR photometry are available. The effective temperatures range from $4000-15000$ K; the SED colors are well-described by a single-component blackbody at this phase. Based on the quality of our Arnett fits, and the fact that the SED for these SNe in the photospheric phase is well-described by a blackbody, we conclude that no \rp{} contribution is needed to explain the color evolution of the objects in our sample, including SN\,2007I. 

Thus, we find no compelling evidence of \rp{} enrichment in any of the SNe in our sample.

\section{Discussion and Outlook} \label{sec:discussion}
From our systematic study in optical and NIR of the SNe Ic-BL associated with collapsars discovered by ZTF and reported in the literature, we do not find any evidence of \rp{} enrichment based on theoretical models which predict observable NIR excesses in the SN light curves. After constructing GPR models from the \rp{}-enriched model grid and performing fitting, the SNe that pass our nominal $\chi^2$ cuts still do not show convincing fits in both the optical and NIR to the \rp{} enriched broadband light curve predictions. On the other hand, for the \rp{}-free models, when computing broadband light curves from the bolometric corrections, we get compelling fits in both optical and NIR for each SN. Our single-component blackbody fits at $\sim$1 month after peak (see Table~\ref{tab:basic_properties}) further suggest that no additional \rp{} enrichment is required to explain the SN SED colors. 

% blurb on maximum r-process constraints (check further...)
Our use of two models, one for \rp{}-free SNe and another for \rp{} enriched cases, complicates our efforts to derive global constraints on \rp{} production in SNe. To estimate the level of enrichment our analysis is sensitive to, we take the reddest object in our sample that is consistent with the \rp{} enriched models, and compare the color measurements with the predicted color evolution from the models. To derive these global constraints, we focus on SN\,2007ce. Amongst our sample, SN\,2007ce has the highest inferred \rp{} ejecta mass of 0.07M$_{\odot}$ while passing the $\chi^2$ cut (we ignore SN\,1998bw, whose extremely well-sampled light curve could be influencing the final $\chi^2$ value). Though SN\,2007I is redder than SN\,2007ce, it shows color evolution that is completely inconsistent with the models (see Fig.~\ref{fig:SN2007I_SN2007ce_color}) making it unsuitable for deriving \rp{} constraints. In Fig.~\ref{fig:SN2007I_SN2007ce_color} we display the predicted color evolution of the best fit model bounded by its 1$\sigma$ uncertainties on the parameters, where the lower bound corresponds to a model with \mrp{}$=0.02\rm M_{\odot}$ and the upper bound corresponds to a model with \mrp{}$=0.12 \rm M_{\odot}$. SN\,2007ce's color measurements exhibit a similar shape to the model color evolution, but show a significant offset with bluer colors compared to the best-fit models. As shown in Figure~\ref{fig:color}, a model with a higher \mej{} can yield a slightly bluer color evolution for the same \rp{} mass, so it is difficult to confidently exclude the possibility that \mrp{}$=0.02 \rm M_{\odot}$ (the lower bound on the parameter inference) was synthesized in SN\,2007ce. In addition, relaxing the assumptions on the SED underlying the models could also alter the color evolution of the model. Thus, based on the the upper bound of these color curves, which corresponds to an \rp{} mass of 0.12M$_{\odot}$, we conservatively argue here that no more than 0.12M$_{\odot}$ of \rp{} material was generated in SN\,2007ce (assuming \xmix{}$=0.7$). Furthermore, since SN\,2007ce has the highest inferred \rp{} mass amongst the objects passing our $\chi^2$ cut, we suggest that \mrp{}$\lesssim0.12 \rm M_{\odot}$ represents a tentative global \rp{} constraint on all of the models in our sample, based on the observations. Future improvements in the models as well as more systematic observations will allow for tighter and more robust constraints on the \rp{} nucleosynthesis in SNe Ic-BL.

We also examine any other associated relativistic outflows to check whether that may introduce a bias. Only three objects in our full sample are counterparts to GRBs: GRB980425 (SN\,1998bw), GRB100316D (SN\,2010bh) and GRB190829A. Amongst these three, GRB980425 and GRB190829A are considered to be LLGRBs based on their peak $\gamma$-ray luminosities \citep{Galama1998, Ho2020_ZTF18aaqjovh, gcn25552}. GRB100316D is a more energetic GRB, but its emission shows a soft spectral peak, similar to other X-ray flashers \citep{Bufano2012}. While none of the other objects in our sample have any coincident $\gamma$-ray emission, some have X-ray and radio detections and upper limits. SN\,2018gep, the FBOT/SN Ic-BL, has both X-ray and radio detections that are consistent with the host galaxy emission \citep{Ho2019_SN2018gep}. On the other hand, SN\,2020bvc has mildly-relativistic X-ray and radio ejecta characteristic of LLGRBs \citep{Ho2020_ZTF20aalxlis}. \citet{Corsi2022} also obtained radio and X-ray follow-up for several ZTF SNe, a subset of which are part of the sample discussed in this work. In Table~\ref{tab:rprocess_radio} we display radio observations with the Very Large Array (VLA) and X-ray observations with the \textit{Swift} X-ray Telescope (XRT) for those SNe; the remainder of SNe which lack radio/X-ray coverage have dashes in those respective columns. Only two of the objects in the sample presented here (SN\,2020tkx and SN\,2021ywf) have a detected point-source-like radio counterpart, but their low velocities suggest that they are not the same as GRB-associated SNe \citep{Corsi2022}. 

The fact that none of these SNe are linked to standard, classical long GRBs prevents us from exploring the proposed theoretical connection between the GRB energetics and \rp{} production. If the GRB jet energy, which scales with the mass accreted by the disk, correlates with the amount of \rp{} mass produced in the disk winds, then collapsars with no GRBs may not be able to produce detectable \rp{} signatures. \citet{Siegel2019} find that for black hole accretion rates between 0.003$-$1.0 M$_{\odot}$ s$^{-1}$ needed to power relativistic outflows, the disk winds are neutron-rich and can synthesize heavy and light \rp{} elements. The association of a GRB with a SN Ic-BL could point towards a central engine that harbors high enough accretion rates to potentially generate \rp{} elements. \citet{BarnesDuffell2023} also find that hydrodynamical mixing between the \rp{} enriched and \rp{}-free layers of collapsar increases with wind mass and duration, suggesting that SNe accompanying the longest duration long GRBs may be the most promising sites to search for obvious \rp{} signatures.

It is yet unclear to what extent the populations of SNe Ic-BL and long GRBs overlap \citep{WoosleyBloom2006, Bissaldi2007, Cano2017, Barnes2018}, as some long GRBs lack SNe \citep{Fynbo2006, DelleValle2006, Tanga2018}, and most Ic-BL SNe have no associated gamma-ray emission \citep{Bianco2014, Corsi2022}. Furthermore, LLGRBs, short-duration GRBs with collapsar progenitors \citep{Ahumada2021} and long-duration GRBs from compact binaries \citep{Rastinejad2022} present evidence towards a broad diversity in collapsar central engines, ranging from mildly relativistic to ultra-relativistic explosion energies. One possibility is that a subset of SNe Ic-BL could correspond to failed GRBs with low black hole accretion rates \citep{MacFaydenWoolsey1999, Huang2002, Xu2023}. Multi-wavelength observations of SN\,2006aj suggest that another subset may be associated with a progenitor whose jet runs into a cocoon of extended stellar material \citep{Nakar2015}, even when an LLGRB is not detected, as in the case of SN\,2017iuk \citep{Izzo2019}. Yet another subset could be off-axis GRBs. 

This diversity of collapsar central engines and jet properties could lend itself naturally to a scenario where some collapsars are capable of producing \rp{} elements while others are not. However, given that only $\sim$half of the SNe in our sample have X-ray or radio observations, a more systematic NIR follow-up campaign with SNe Ic-BL associated with classical long GRBs, LLGRBs, X-ray/radio counterparts, and lacking any multi-wavelength counterparts is needed to investigate whether only those SNe that produce relativistic ejecta are able to create conditions conducive to \rp{} nucleosynthesis.

Another possibility we acknowledge is that collapsars could be a very low-yield source of \rp{} nucleosynthesis. The expected yields from the \citealt{Siegel2019} and \citealt{Barnes2022} models (0.01-0.1M$_{\odot}$) are mainly set by the joint constraints from the literature on \rp{} nucleosynthesis sites \citep[see for e.g.][]{Hotokezaka2018}. However, the discovery of minuscule amounts of Sr and Ba in an extremely metal-poor star \citep{CaseySchlaufman2017} motivates the need for core-collapse supernovae with an extremely low yield of \rp{} material whose nucleosynthesis is consistent with the Solar \rp{} abundance pattern. Due to the limitations of these models and the dataset presented here, our study only searches for enrichment levels of \mrp{}$\gtrsim0.01 \rm M_{\odot}$. Detailed analysis of the nebular-phase spectra of SNe Ic-BL would likely be required to probe such low levels of enhancement robustly.

% add some text about the possibility that collapsars could produce a negligible amount of r-process material. What if the amount is too small for us to detect this way in light curves. We are not sensitive to that level of rp in the light curves.

Despite the fact that we find no evidence for \rp{} enrichment in the SNe Ic-BL in our sample, we must also acknowledge a number of caveats to this work. 

First, we note that the \rp{} enriched and \rp{}-free models make different predictions about the relationship between nickel mass and SN luminosity. While the inferred central values from the GPR inference of both $\beta_{\rm ej}$ and \mej{} based on the \rp{} grid are generally within the 1$\sigma$ errorbar of our explosion property estimates, the nickel mass inferred shows a larger deviation from the Arnett value. Arnett-like models are constructed such that the radioactive energy-generation rate crosses the bolometric light curve precisely at peak luminosity. The \rp{} enriched models, in which energy diffuses through a series of concentric shells, do not reproduce this behavior; they generally have $L_{\rm bol}(\rm tpk) < Q_{\rm dot}(\rm tpk)$. As a result, the amounts of nickel inferred by each model for a given luminosity are inconsistent, and the Arnett-like models do not match the \citet{Barnes2022} models when \mrp{} is set to zero (see Sec.~\ref{sec:models} for other differences between the \citet{Barnes2022}-like models and the Arnett-like models). To compensate for these differences, we fit the \rp{} enriched models over a wide range of nickel masses.

% There could be two effects at play causing the discrepancy in the inferred amount of Nickel. One is that in the \rp{} enriched models, the bolometric light curve crosses the heating curve after light-curve peak, while standard Arnett models assume that the light curve peak occurs at the peak of the heating rate curve \citep{Arnett1982}, causing the \rp{} model grid to underestimate $M_{56}$. The other effect could be due to the fact that the \rp{} grid does not thermalize all of the energy from Nickel decay; the thermalization efficiency depends on the global optical depth to gamma rays. This is in contrast with Arnett models, that assume that all of the energy from Nickel decay contributes towards heating the ejecta.

Given the differences between the \rp{} enriched and \rp{}-free models we use, a more robust approach would be to conduct an apples-to-apples comparison between \rp{}-free and \rp{} enriched models from the same underlying grid. Initially, we performed fitting to both the \rp{}-free and \rp{} enriched models from \citep{Barnes2022}, but found that the colors of the \rp{}-free models were consistently much redder than the observed colors of our objects at all epochs. To construct the \rp{}-free SED, \citet{Barnes2022} uses the light curve of SN\,2007gr as it has well-sampled $B$- to $K$-band photometry up to late-times, but the detection of the CO molecule in its nebular phase NIR spectra may affect the $K-$band flux of the object \citep{Hunter2009}. Unfortunately, the semi-empirical approach to converting between bolometric and broadband light curves using \texttt{HAFFET} and the fact that the Arnett models do not allow us to define a spatial or velocity mixing coordinate renders the alternative possibility of enriching the \texttt{HAFFET} models with \rp{} material infeasible. Our inability to use \rp{} enriched and \rp{}-free models from the same grid makes our investigation to search for \rp{} production less robust. The authors are currently investigating whether varying additional parameters controlling the \rp{}-free SED may lead to better fits to the data. This will be discussed in a future work.
% A closer look at the optical-NIR colors for a systematic sample of stripped-envelope and Type II SNe may reveal whether the late-time SED of SN\,2007gr is truly representative of standard \rp{}-free SNe Ic.

Furthermore, our understanding of the emission of \rp{} ejecta in the nebular phase is quite limited. The radiation from \rp-enriched ejecta layers has a strong impact on the predictions of late-time photometry for the \rp{} grid.
\citet{Barnes2022} adopts a temperature of 2500 K for the \rp{} SED because a black-body at this temperature reproduces the optical and NIR  photometric colors of the nebular-phase kilonova model spectrum of \citet{Hotokezaka.ea_2021.MNRAS_rprocess.nebular}.
However, kilonova nebular-phase modeling is still a topic of active investigation. 
Future studies of kilonova nebulae, both observational and theoretical, may refine our understanding of nebular emission from pure \rp{} outflows.
Furthermore, differences between kilonovae and \rp-enriched SN (e.g., in their densities or their compositions) may mean that nebular-phase emission from the former is not a perfect predictor of nebular-phase emission from the latter.

Finally, we acknowledge the limitations of the dataset we present here for testing whether collapsars synthesize \rp{} elements. Due to the nature of our classical observing runs with WIRC, our NIR light curves are very sparse, and in some cases our upper limits are too shallow to be constraining. In contrast, future wide field of view NIR facilities (i.e. WINTER, DREAMS, PRIME) will enable systematic follow-up of nearby SNe Ic-BL discovered by contemporaneous wide-field optical surveys (i.e. ZTF, Pan-STARRS, ATLAS, Vera Rubin Observatory, etc) as well as counterparts to nearby long GRBs to late-times. The James Webb Space Telescope will grant the unique ability to probe the mid-infrared wavelengths and acquire IR spectroscopy to search for further signatures of \rp{} production. Higher cadence NIR photometry and nebular spectroscopy to search for the \rp{} signatures from collapsars would substantiate the results of this paper as well as determine whether the presence of a relativistic jet in the explosion is required for heavy element production. The authors plan to investigate the relative contribution of collapsars, neutron star mergers, and neutron star--black hole mergers towards the \rp{} abundance in the Universe in a future work. The next generation of optical and IR telescopes will open new windows to discoveries providing valuable insights into the open questions about \rp{} nucleosynthesis from collapsars.

\begin{table*}[]
    \centering
    \begin{tabular}{cccccccc}
    \hline
    \hline
        SN & $M_{\rm ej}$ & $E_{\rm K}$ & $M_{\rm Ni}$ & E(B-V)$_{\rm host}$ & $T_{\rm eff}$ [phase] & Ref. \\
         & (M$_{\odot}$) & (foe) & (M$_{\odot}$) & & (K [day]) & \\
         \hline
        1998bw & 10 & 50 & 0.4 & 0.06* & 5919 [24.5] & \citet{Clochiatti2011, Nakamura2001} \\
        2002ap & 2.5-5 & 4-10 & 0.07 (0.02) & 0.09 & 5126 [30.5] & \citet{Mazzali2002} \\
        2007ce & 2.90 (0.63) & 1.85 (0.89) & 0.48 (0.01) & 0.00 & 6310 [18.5] & \citet{Modjaz2008}** \\
        2007I & 6.87 (0.80) & 7.63 (1.99) & 0.10 (0.00) & 0.34 & 4064 [33.5] & \citet{Modjaz2008}** \\
        2009bb & 3.4 (0.4) & 6.2 (0.8) & 0.20 (0.02) & 0.540 & 3584 [22.6] & \citet{Taddia2018} \\
        2010bh & 2.21 (0.10) & 11.34 (0.52) & 0.21 (0.03) & 0.30 & 6102 [23.5] & \citet{Olivares2012} \\
        2016coi & 4-7 & 7-8 & 0.15 & 0.00 & 4727 [32.1] & \citet{Terreran2019} \\
        \hline\hline
    \end{tabular}
    \caption{Explosion properties and inferred \rp{} ejecta masses and mixing fractions of low-redshift SNe with contemporaneous optical and NIR imaging from our literature search. Where available, we quote the 1$\sigma$ uncertainties on the parameters in brackets. For SN\,1998bw and SN\,2002ap, we quote the ranges of explosion parameters corresponding to the best-fitting light curve models. *\citet{Clochiatti2011} already corrected for host extinction; we use the assumed host extinction to correct only the NIR photometry. **For SN\,2007I and SN\,2007ce, as explosion properties were not estimated in the literature, we conduct light curve analysis to derive the best fit properties as described in Sec.~\ref{sec:analysis}.}
    \label{tab:literature_sne}
\end{table*}

\clearpage

% \startlongtable
% \begin{rotatetable}
% \begin{center}
\startlongtable
\begin{deluxetable}{lccccccccccc}
% \rotate
% \centering
\tabletypesize{\footnotesize}
\tablecolumns{10} 
\tablewidth{0pt}
\tablecaption{Optical properties of the BL-Ic SNe in our sample. \label{tab:sn_fit_param}}
\tablehead{
\colhead{SN} &
\colhead{$t_{\rm peak}$} &
\colhead{$M_{\rm peak,r}$} &
\colhead{$t_{\rm expl}$} &
\colhead{$M_{\rm Ni}$} &
\colhead{$\tau_{m}$} &
\colhead{$M_{\rm ej}$} &
\colhead{$E_{\rm kin}$} &
\colhead{$v_{\rm ph}$} &
\colhead{$T_{\rm eff}$ [phase]}\\
\colhead{} &
\colhead{(MJD)} &
\colhead{(mag)} &
\colhead{(day)} &
\colhead{(M$_{\odot}$)} &
\colhead{(day)} &
\colhead{(M$_{\odot}$)} &
\colhead{($10^{51}$erg)} &
\colhead{(c)} &
\colhead{(K [day])}
}
\startdata
2018jaw & 58455.70 & -18.63 (0.08) & -18.74 $_{-0.66}^{+0.66}$ & 0.33 $_{-0.02}^{+0.02}$ & 13.63 $_{-1.38}^{+1.10}$ & $>$ 1.41 (0.33) & $>$ 0.40 (0.16) & 0.022 (0.004) & --\\
2018kva & 58487.05 & -18.70 (0.02) & -15.81 $_{-0.60}^{+0.49}$ & 0.29 $_{-0.01}^{+0.01}$ & 12.13 $_{-0.87}^{+1.06}$ & 2.51 (0.39) & 3.53 (0.76) & 0.051 (0.004) & 5431 [47.2]\\
2019gwc & 58650.58 & -18.48 (0.01) & -12.78 $_{-0.46}^{+0.46}$ & 0.22 $_{-0.01}^{+0.01}$ & 6.96 $_{-0.15}^{+0.12}$ & $>$ 0.60 (0.05) & $>$ 0.44 (0.08) & 0.037 (0.003) & 5953 [33.6]\\
2019hsx & 58647.07 & -17.08 (0.02) & -15.63 $_{-0.53}^{+0.38}$ & 0.07 $_{-0.01}^{+0.01}$ & 12.12 $_{-1.26}^{+1.10}$ & 1.64 (0.43) & 0.99 (0.50) & 0.033 (0.007) & 11002 [36.1]\\
2019moc & 58715.76 & -19.16 (0.03) & -20.02 $_{-3.23}^{+0.27}$ & 0.52 $_{-0.02}^{+0.01}$ & 10.60 $_{-0.30}^{+0.37}$ & 2.09 (0.50) & 3.48 (1.85) & 0.056 (0.013) & 8537 [63.0]\\
2019qfi & 58753.56 & -18.01 (0.02) & -15.09 $_{-1.40}^{+1.40}$ & 0.13 $_{-0.01}^{+0.01}$ & 10.58 $_{-1.71}^{+1.44}$ & $>$ 1.22 (0.33) & $>$ 0.70 (0.24) & 0.032 (0.004) & 5698 [25.0]\\
2019xcc & 58844.59 & -16.58 (0.06) & -10.62 $_{-2.51}^{+2.51}$ & 0.04 $_{-0.01}^{+0.01}$ & 5.04 $_{-0.95}^{+1.36}$ & 0.68 (0.30) & 2.40 (1.14) & 0.081 (0.007) & --\\
2020dgd & 58914.05 & -17.74 (0.02) & -18.03 $_{-2.50}^{+2.50}$ & 0.13 $_{-0.03}^{+0.03}$ & 13.68 $_{-2.70}^{+3.78}$ & 2.81 (1.50) & 3.07 (2.42) & 0.045 (0.013) & --\\
2020lao & 59003.92 & -18.66 (0.02) & -10.60 $_{-0.99}^{+0.99}$ & 0.23 $_{-0.01}^{+0.01}$ & 7.71 $_{-0.21}^{+0.22}$ & 1.22 (0.16) & 2.48 (0.71) & 0.048 (0.005) & --\\
2020rph & 59092.34 & -17.48 (0.02) & -19.88 $_{-0.02}^{+0.02}$ & 0.07 $_{-0.01}^{+0.01}$ & 17.23 $_{-0.89}^{+1.19}$ & 3.83 (1.59) & 3.08 (2.81) & 0.039 (0.016) & 5857 [30.5]\\
2020tkx & 59116.50 & -18.49 (0.05) & -12.77 $_{-4.54}^{+4.54}$ & 0.22 $_{-0.01}^{+0.01}$ & 10.95 $_{-0.77}^{+0.67}$ & $>$ 1.75 (0.24) & $>$ 1.82 (0.35) & 0.044 (0.003) & 7116 [32.8]\\
2021bmf & 59265.12 & -20.60 (0.04) & -23.76 $_{-5.52}^{+5.68}$ & 0.98 $_{-0.17}^{+0.16}$ & 18.08 $_{-7.94}^{+6.64}$ & 8.05 (5.37) & 23.63 (16.14) & 0.073 (0.005) & 15618 [41.4]\\
2021too & 59434.09 & -19.66 (0.02) & -23.23 $_{-0.41}^{+0.41}$ & 0.92 $_{-0.03}^{+0.03}$ & 17.67 $_{-0.66}^{+0.65}$ & 5.06 (0.78) & 6.42 (2.09) & 0.048 (0.007) & 5363 [23.5]\\
2021xv & 59235.56 & -18.92 (0.07) & -12.79 $_{-0.33}^{+0.24}$ & 0.30 $_{-0.02}^{+0.01}$ & 7.72 $_{-0.49}^{+0.66}$ & 0.89 (0.15) & 0.96 (0.23) & 0.045 (0.004) & 5969 [13.5]\\
2021ywf & 59478.64 & -17.10 (0.05) & -10.67 $_{-0.49}^{+0.49}$ & 0.06 $_{-0.01}^{+0.01}$ & 8.87 $_{-0.81}^{+0.81}$ & 1.06 (0.19) & 0.92 (0.26) & 0.040 (0.004) & 5238 [19.5]\\
\enddata
\tablecomments{Table containing the explosion properties of all of the ZTF-discovered SNe in our sample that have not yet been published. Our methods for deriving the quantities given above are described in Sec.~\ref{sec:analysis}. The velocities shown here are measured at various different phases, so do not represent the photospheric velocity of the supernova at peak. We report effective temperatures from blackbody fits around $\sim$30\,days post-peak for each of the objects that has one or more NIR detections.}
\label{tab:basic_properties}
\end{deluxetable}

\clearpage

The authors express gratitude towards Ehud Nakar, Daniel Siegel, Leo Singer, Navin Sridhar, and Peter Tsun Ho Pang for helpful discussions in preparing this work.
% Data release
All light curves presented in this work will be made public via WISeREP upon publication.
% observatory acknowledgements
Based on observations obtained with the Samuel Oschin Telescope 48-inch and the 60-inch Telescope at the Palomar Observatory as part of the Zwicky Transient Facility project. ZTF is supported by the National Science Foundation under Grants No. AST-1440341 and AST-2034437 and a collaboration including current partners Caltech, IPAC, the Weizmann Institute of Science, the Oskar Klein Center at Stockholm University, the University of Maryland, Deutsches Elektronen-Synchrotron and Humboldt University, the TANGO Consortium of Taiwan, the University of Wisconsin at Milwaukee, Trinity College Dublin, Lawrence Livermore National Laboratories, IN2P3, University of Warwick, Ruhr University Bochum, Northwestern University and former partners the University of Washington, Los Alamos National Laboratories, and Lawrence Berkeley National Laboratories. Operations are conducted by COO, IPAC, and UW.
SED Machine is based upon work supported by the National Science Foundation under Grant No. 1106171.
Based on observations made with the Nordic Optical Telescope, owned in collaboration by the University of Turku and Aarhus University, and operated jointly by Aarhus University, the University of Turku and the University of Oslo, representing Denmark, Finland and Norway, the University of Iceland and Stockholm University at the Observatorio del Roque de los Muchachos, La Palma, Spain, of the Instituto de Astrofisica de Canarias.
% co-author acknowledgments
S.~Anand acknowledges support from the National Science Foundation GROWTH PIRE grant No. 1545949.
J.~Barnes is supported by the Gordon and Betty Moore Foundation through Grant GBMF5076.
S.~Yang is supported by the National Natural Science Foundation of China under Grant No. 12303046
M.~W.~Coughlin acknowledges support from the National Science Foundation with grant numbers PHY-2010970 and OAC-2117997.
S. Schulze acknowledges support from the G.R.E.A.T research environment, funded by {\em Vetenskapsr\aa det},  the Swedish Research Council, project number 2016-06012.
L.T. acknowledges support from MIUR . A.C. acknowledges support from the NASA Swift Guest Investigator program (80NSSC22K0203).

\software{Source Extractor \citep{BertinArnouts2010}, HOTPANTS \citep{Becker2015}, SCAMP \citep{Bertin2006}, pyzogy \citep{GuevelHosseinzadeh2017}, PSFEx \citep{Bertin2011}, Pypeit \citep{Prochaska2019, Roberson2021}, LPipe \citep{Perley2019}, ForcePhotZTF \citep{Yao2019}, HAFFET \citep{Yang2023}, SNspecFFTsmooth \citep{Liu2016}, SESNspectraLib \citep{Liu2016, Modjaz2016}, PyMultinest's \citep{Buchner2014}}

\bibliography{references}

\end{document}